\documentclass[a4paper,fleqn,usenatbib,useAMS]{mnras}

\usepackage[T1]{fontenc}
\usepackage{ae,aecompl}

\usepackage{graphicx}	
\usepackage{amsmath}	
\usepackage{amssymb}	
\usepackage{times}

\title[An LBV goes SN...or not?]{SN 2015bh: NGC 2770's 4th supernova or a luminous blue variable on its way to a Wolf-Rayet star?}

\author[Th\"one et al.]{C. C. Th\"one$^{1}$\thanks{E-mail: cthoene@iaa.es}, A. de Ugarte Postigo$^{1,2}$, G. Leloudas$^{2,3}$, C. Gall$^{4}$, Z. Cano$^{5}$, K. Maeda$^{6,7}$, 
\newauthor S. Schulze$^{8,9}$, S. Campana$^{10}$, K. Wiersema$^{11}$, J. Groh$^{12}$, J. de la Rosa$^{13}$, F. E. Bauer $^{8,9,14}$,
\newauthor D. Malesani$^{2}$, J. Maund$^{15}$, N. Morrell$^{16}$, Y. Beletsky$^{16}$ 
\\
$^{1}$Instituto de Astrof\'isica de Andaluc\'ia - CSIC, Glorieta de la Astronom\'ia s/n, 18008 Granada, Spain\\
$^{2}$Dark Cosmology Centre, Niels Bohr Institute, Juliane Maries Vej 30, Copenhagen \O, D-2100, Denmark\\
$^{3}$Department of Particle Physics \& Astrophysics, Weizmann Institute of Science, Rehovot 76100, Israel\\
$^{4}$Department of Physics and Astronomy, Aarhus University, Ny Munkegade 120, DK-8000 Aarhus C, Denmark\\
$^{5}$Centre for Astrophysics and Cosmology, Science Institute, University of Iceland, Dunhagi 5, 107 Reykjav\'ik, Iceland\\
$^{6}$Department of Astronomy, Kyoto University, Kitashirakawa-Oiwake-cho, Sakyo-ku, Kyoto 606-8502, Japan\\
$^{7}$Kavli Institute for the Physics and Mathematics of the Universe (WPI), The University of Tokyo, 5-1-5 Kashiwanoha, Kashiwa, Chiba 277-8583, Japan\\
$^{8}$Instituto de Astrof\'{i}sica, Facultad de F\'{i}sica, Pontificia Universidad Cat\'{o}lica de Chile, Av. Vicu\~{n}a Mackenna 4860, Santiago, Chile\\
$^{9}$Millennium Institute of Astrophysics, Vicu\~{n}a Mackenna 4860, 7820436 Macul, Santiago, Chile\\
$^{10}$INAF, Osservatorio Astronomico di Brera, via E. Bianchi 46, 23807, Merate, Italy\\
$^{11}$Department of Physics and Astronomy, University of Leicester, University Road, Leicester, LE1 7RH, United Kingdom\\
$^{12}$ School of Physics, Trinity College Dublin, Dublin 2, Ireland\\
$^{13}$ Department of Physics and Astronomy, University of Texas at San Antonio, San Antonio, Texas, 78249, USA\\
$^{14}$Space Science Institute, 4750 Walnut Street, Suite 205, Boulder, CO 80301, USA\\
$^{15}$Department of Physics and Astronomy, University of Sheffield, Hicks Building, Hounsfield Road, Sheffield S3 7RH, UK\\
$^{16}$Las Campanas Observatory, Carnegie Observatories, Casilla 601, La Serena, Chile
}

\date{Accepted XXX. Received YYY; in original form ZZZ}

\pubyear{2016}

\begin{document}
\label{firstpage}
\pagerange{\pageref{firstpage}--\pageref{lastpage}}
\maketitle

\begin{abstract}
{Very massive stars in the final phases of their lives often show unpredictable outbursts that can mimic supernovae, so-called, ``SN impostors'', but the distinction is not always straigthforward. Here we present observations of a luminous blue variable (LBV) in NGC\,2770 in outburst over more than 20 years that experienced a possible terminal explosion as type IIn SN in 2015, named SN 2015bh. This possible SN or ``main event'' was preceded by a precursor peaking $\sim$ 40 days before maximum. The total energy release of the main event is $\sim$1.8$\times$10$^{49}$ erg, which can be modeled by a $<$\,0.5\,M$_\odot$ shell plunging into a dense CSM. All emission lines show a single narrow P-Cygni profile during the LBV phase and a double P-Cygni profile post maximum suggesting an association of this second component with the possible SN. Since 1994 the star has been redder than during a typical S-Dor like outburst. SN 2015bh lies within a spiral arm of NGC 2770 next to a number of small star-forming regions with a metallicity of $\sim$ 0.5 solar and a stellar population age of 7--10 Myr. SN\,2015bh shares many similarities with SN 2009ip, which, together with other examples may form a new class of objects that exhibit outbursts a few decades prior to ``hyper-eruption" or final core-collapse.  If the star survives this event it is undoubtedly altered, and we suggest that these ``zombie stars" may evolve from an LBV to a Wolf Rayet star over a very short timescale of only a few years.  The final fate of these types of massive stars can only be determined with observations years after the possible SN.}
\end{abstract}



\section{Introduction}
Massive stars towards the end of their short life-span are unstable and can lose large amounts of mass in the form of eruptions. Luminous blue variables (LBVs) are a type of very massive stars known to show different types of variabilites: During S-Dor outbursts, which occur over time scales from months to decades, the star drops in temperature while staying constant or decreasing in luminosity \citep{2009ApJ...698.1698G} after which it goes back to its original state. Those variations are now thought to be caused by a change in the hydrostatic radius of the star, hence the S-Dor variations would be irregular pulsations of the star \citep{2009ApJ...698.1698G, 2011ApJ...736...46G, 2012A&A...541A.146C}. Another form of eruptive mass loss are the more rare ``giant eruptions'', the most famous of which are the historic eruptions of Eta Carinae in 1861 \citep[e.g.][]{2008Natur.455..201S} and P Cygni in the 17th century \citep{1999PASP..111.1124H} in the Milky Way (MW). During those eruptions, the star reaches both higher luminosities and lower temperatures as the outer layers detach from the rest of the star for reasons still poorly understood. The mass loss in those eruptions can be considerable, e.g. Eta Carinae has been estimated to have lost already $>$40\,M$_\odot$ in the last thousand years by means of eruptions and stellar winds \citep{2010MNRAS.401L..48G}. 

More evidence for eruptive mass loss comes from the observations of ``Interacting supernovae'' such as SNe~IIn \citep{1990MNRAS.244..269S} and SNe~Ibn \citep{2008MNRAS.389..113P}. In these events the fast SN shock-wave interacts with previously ejected material, creating the characteristic narrow ('n') emission lines. The mass of the circumstellar material (CSM) required to power the luminosity of these types of SNe ranges from 0.1 (P Cygni) to 20--30 M$_\odot$ (Eta Car and AG Car, \citealt{2006ApJ...645L..45S}) which is too large to be explained by mass loss through stellar winds \citep[e.g.][]{Smith06tf, 2013A&A...555A..10T}. Smaller, periodic mass-loss events such as S-Dor variabilities have been suggested to explain the periodic pattern in the radio data of some Type IIn SNe \citep{2006A&A...460L...5K}.

Eruptive mass loss is also associated with several SN ``impostors''  that resemble SNe~IIn spectroscopically but are typically fainter and are non-terminal explosions where the progenitor star survives. SN impostors could be LBV giant eruptions, although some SN impostors do seem to behave differently from the giant eruptions of e.g. Eta Carinae \citep[for an extensive discussion see][]{2015arXiv151008050S} and not all of them have been associated to LBVs \citep[e.g. SN 2008S and NGC 300-OT,][]{0004-637X-741-1-37} The line between SN impostors (non-terminal event) and a ``true'' SN Type IIn (terminal explosion) is often unclear and largely debated, see e.g. SN 1961V which, despite its luminosity comparable to a SN IIn is considered by some to be a non-terminal event (\citealt{1995AJ....110.2261F, 2002PASP..114..700V} but see also \citealt{2011ApJ...737...76K})

There is growing evidence that LBVs can directly explode into SNe IIn, but not all type IIn originate from LBV progenitors. Previously it was thought that LBV stars have not yet evolved enough to develop an iron core, and classical models predict that an LBV will first go through a Wolf-Rayet (WR) phase before exploding as a SN  \citep{2000A&A...361..159M}. Based on the observations of LBV outbursts, the lifetime of this phase had been estimated to last only a few 10$^4$ yr \citep{1994PASP..106.1025H}, however, many mass-loss outbursts may be too faint to be directly observed, and the lifetime may up to 10$^5$ yr \citep{2015MNRAS.447..598S, 2014ARA&A..52..487S}. Two main issues now put in doubt the role of the LBV phase in stellar evolution: SNe IIn require interactions with an circumstellar medium (CSM) with masses and densities only achievable with giant eruptions, stellar winds alone would be too weak for this purpose \citep{2014ARA&A..52..487S}, although this has been questioned as well \citep{2011MNRAS.412.1639D}. On the other hand, several likely cases of LBV progenitors of Type IIn and Ibn SNe have now been imaged prior to explosion with \emph{HST}: SN~2005gl  \citep{2009Natur.458..865G}, SN 2009ip \citep{2011ApJ...732...32F} and SN~2010jl \citep{2011ApJ...732...63S}, although there is ample debate about the nature of all of these events. 

A very curious event in this context is SN~2009ip: Initially classified as an LBV SN impostor \citep{2010AJ....139.1451S, 2011ApJ...732...32F}, it was observed by \cite{2013ApJ...767....1P}, until September 2012, when  \cite{2012ATel.4412....1S} proposed that it had evolved into a real SN, after experiencing a major outburst in August 2012. Despite the many studies that have been published on SN~2009ip \cite[e.g.][]{2013MNRAS.430.1801M,2013ApJ...767....1P,2013MNRAS.433.1312F,2014ApJ...780...21M,2014ApJ...787..163G,2015arXiv150206033F} there is no clear consensus on whether the progenitor star has survived the 2012 outburst. 

Several SNe have also shown outbursts just before the main event, as was observed for 2009ip (see above). \cite{2007Natur.447..829P} was the first to report on a massive eruption that preceded the explosion of Type~Ibn SN~2006jc that had been recorded in images obtained by amateur astronomers. A similar event, SN~2010mc, showing an outburst 40 days before the main explosion, was presented by \cite{2013Natur.494...65O}. In a subsequent statistical study, based on the pre-explosion observations of 16 SNe IIn detected by the Palomar Transient Factory (PTF), \cite{2014ApJ...789..104O} estimated that at least 50\% of SNe~IIn experience a major outburst in the last 100 days prior to explosion. The physical mechanism of these outbursts is still unknown and some argue that the ``pre-curser'' is the actual explosion event while the brighter ``main'' explosion is the interaction of the SN shell with the CSM. Recently, precursor explosions have been identified also in the case of superluminous SNe \citep{2012A&A...541A.129L,2015arXiv150501078N, 2016MNRAS.457L..79N, 2016ApJ...818L...8S} although it remains debatable whether these objects are powered by circumstellar interaction.

\begin{figure*} 
   \centering
   \includegraphics[width=18 cm]{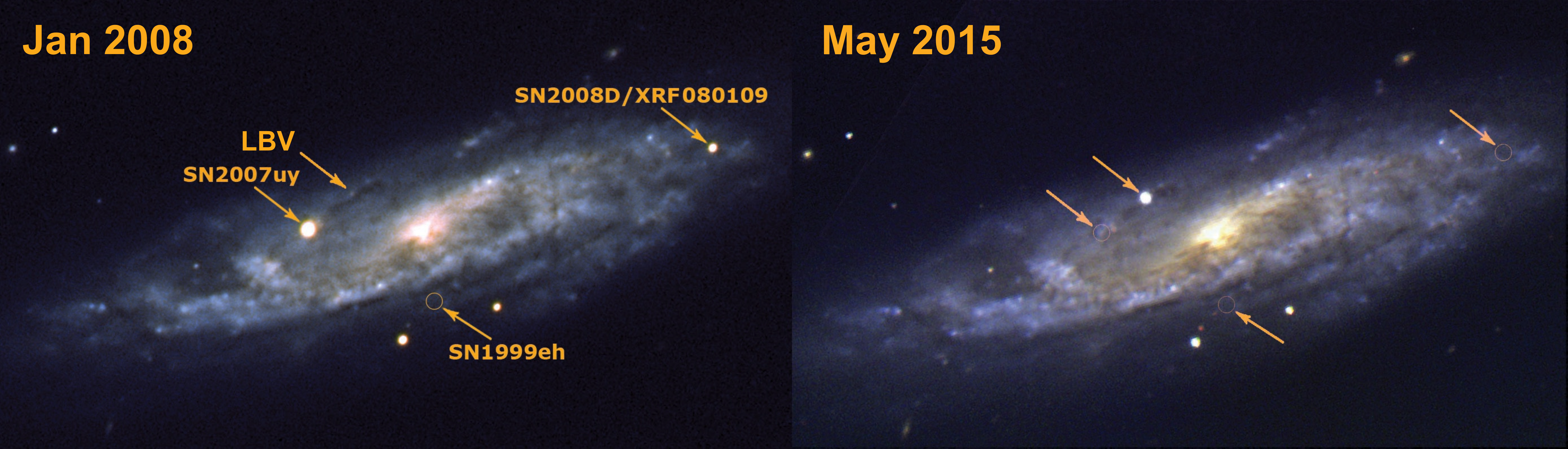}
   \caption{Left: NGC2770 observed with VLT (+FORS1) in January 2008, while there were two active SNe. An earlier SN from 1999 is indicated by a circle. Right: In May 2015 the SNe are gone in this image from GTC (+OSIRIS), but there is a new object, SN 2015bh, an LBV that potentitally suffered a terminal explosion as a SN IIn.} \label{Fig:FC}
            
    \end{figure*}

In this paper, we present an extensive dataset of an LBV star in the galaxy NGC~2770  that has possibly exploded after experiencing multiple outbursts over the last 21 years. The galaxy (also known as ''SN Ib factory`` \citealt{2009ApJ...698.1307T}) had previously hosted 3 type Ib SNe, 1999eh, 2007uy (later classified as a Ib pec \citet{2014AJ....147...99M}) and 2008D, the last two of which happened almost simultaneously (see Fig. \ref{Fig:FC}). Our collaboration has extensively studied SN 2008D \citep{2009ApJ...692L..84M,2009ApJ...705.1139M,2010A&A...522A..14G} and SN 2007uy \citep{2013MNRAS.434.2032R} as well as the host galaxy \citep{2009ApJ...698.1307T}. Another detailed study of the host using resolved spectroscopic information is in preparation (Th\"one et al.). 

Sporadic monitoring of the galaxy shows that the LBV had several outbursts as well as quiescent periods between 1994 and 2015 but none of them had been reported previously. iPTF discovered an outburst in December 2013 but did not publish the discovery until May 2015 \citep{2015ATel.7515....1D}, which has now been discussed in a recent paper by \cite{2016arXiv160502450O}. In February 10, 2015, a SN candidate was reported at that position on the CBAT/TOP page by Stan Howerton as ``SNhunt275'', and was classified as a SN impostor by \cite{2015ATel.7042....1E}. Since then, the event was monitored more closely by several groups as well as amateur astronomers. In April 2015, we obtained photometry and spectra \citep{2015ATel.7409....1P} and started monitoring SNhunt275 weekly until on May 15 we noticed that the LBV had brightened to $M \sim-16.4$,  wo magnitudes brighter than our previous observation one week before. This behavior was similar to SN~2009ip and we considered the possibility that the progenitor star had exploded. We immediately reported the discovery \citep{2015ATel.7514....1P} and initiated a dense follow-up campaign on the source, including ground-based imaging and spectroscopy and photometry from \textit{Swift} \citep{2015ATel.7517....1C}. The source disappeared behind the Sun in mid June 2015 and reappeared at the end of September 2015 in a significantly faded state.

In Sect.\ref{sec:obs} we present our extensive imaging and spectroscopy campaign as well as LBV/SN environment observations. In Sect. \ref{sec:evol1} and \ref{sec:evol2} discuss the different episodes of the LBV/SN in detail as well as its physical properties. In Sect.\ref{sec:prog} and \ref{sec:env} we determine the properties of the possible progenitor star and its environment, respectively. Finally in Sect. \ref{sec:comp} we compare SN 2015bh with other events and discuss the evidence of whether or not the star might have actually experienced a terminal explosion.  Throughout the paper we use a flat lambda CDM cosmology as constrained by Planck with $\Omega_m$$=$0.315, $\Omega_\Lambda$$=$0.685 and H$_0$$=$67.3. For NGC 2770 we apply a luminosity distance of 27\,Mpc corresponding to a distance modulus of 32.4\,mag.

\section{Observations}
\label{sec:obs}

\subsection{Imaging}
Archival imaging data of the LBV are available at irregular intervals since 1994. Two epochs in 1994 and 1996 were observed with the 1m Jacobus Kapetyn Telescope (JKT), the 2.5m Isaac Newton Telescope (INT) in 1994, 1995, 2002 and 2012 using the Prime Focus Cone Unit (PFCU) and the Wide Field Camera (WFC) as well as with the WHT in 1997. The LBV is detected in most of these observations except for a few data taken under poor conditions. Archival data are also available from the SDSS taken in March 2004 in $u'g'r'i'z'$ bands.

During the observations of SN 2008D from January to March 2008, the LBV was within the field-of-view (FoV), therefore frequent imaging is available from NOT/ALFOSC and FORS1+2/VLT. Further observations were used from the \emph{HST} archive for 5 epochs in 2008 and early 2009. Acquisition images in r$^\prime$ from a drift-scan spectroscopy program at GTC were obtained in Nov. 2013, serendipitously covering the onset of one of the outbursts (episode 2013). Finally, we used images during the episode 2015A from Asiago in $ugri$ and from a UVOT/Swift programs (PI. P. Brown) in Feb. 2015. 

On March 27 and April 9 2015 we observed NGC 2770 with OSIRIS/GTC using narrow continuum filters to complete a dataset on NGC 2770 during which we discovered that the source was still bright, and we decided to undertake a dedicated follow-up of the event. We then obtained weekly $r$-band observations through a DDT programme (PI A. de Ugarte Postigo) at the Observatorio de Sierra Nevada (OSN) in Granada, Spain, using both the 1.5 m and the 0.9 m telescopes where we discovered a sudden brightening on May 16, 2015, which we call the ``main event'' or episode 2015B. On May 9, we also obtained one epoch with the nIR imager Omega2000 at the 3.5 m telescope at Calar Alto Observatory (Spain) in $J, H$ and $K_S$ bands.

Subsequently, we requested ToO observations by {\it Swift}/UVOT (PI S. Campana) that were first executed daily until the end of May 2015 and then at larger intervals until the object was too close to the Sun in mid June. During the UVOT observations, XRT also observed the source in photon counting mode at 0.3--10\,keV to look for possible X-ray emission from the supernova. Dense photometric monitoring of the main event was done from the ground with OSIRIS at the 10.4 m Gran Telescopio Canarias, Spain (PI A. de Ugarte Postigo), the 0.5 m telescope at the University of Leicester, UK (UL50), the 0.6 m Rapid Eye Mount telescope (REM) at La Silla Observatory, Chile and further observations with the 0.9m OSN telescope. Together, these observations provide us with an almost daily coverage of the LC from May 16 to June 20. NIR observations were performed at several epochs during episode 2015B with the REM telescope.

After the object became visible again by the end of September 2015 we continued with weekly monitoring of the object until the submission of this paper. Photometry in $B, V, R I$ were obtained weekly using the 1.5m OSN telescope, but the detection proved to be increasingly difficult, especially in the blue bands. On Nov. 26 we also obtained IR observations in J-band using the PAnoramic Near Infrared CAmera (PANIC) at the 2.2 m telescope in Calar Alto Observatory during its science verification phase. The observation consisted of 9$\times$82\,s under non optimal conditions and the object was detected at low S/N.

The optical and NIR images were reduced using IRAF with standard routines and pipelines developed for the different instruments or self-made procedures. After registering the images with \texttt{alipy} v2.0 and \texttt{nshiftadd.cl} they were coadded and astrometrised with \texttt{astrometry.net}\footnote{\href{http://obswww.unige.ch/~tewes/alipy/}{http://obswww.unige.ch/\~{}tewes/alipy/}} \citep{2010AJ....139.1782L}.. Aperture photometry was done for the different datasets using apertures of 1 -- 1.5 $\times$ FWHM depending on the instrument and seeing conditions. Photometric calibrations were performed using SDSS for those images taken in Sloan filters, 2MASS for images in nIR and reference stars from \cite{2009ApJ...692L..84M} for images in John-Cousin filters. Typically photometry was done using 3--11 reference stars. Once an instrumental magnitude was established, it was photometrically calibrated against the brightness of a number of field stars measured in a similar manner. Photometry was tied to the SDSS DR8 \citep{2011ApJS..193...29A} in the optical filters ($griz$) and 2MASS \citep{2006AJ....131.1163S} in the NIR ($JHK_s$).

The result of all the photometric observations can be found in Tab. \ref{Tab:photlog} in the Appendix. The light curve (LC) in R-band from 2008 to 2015 indicating the different episodes mentioned in the text is plotted in Fig. \ref{Fig:LC}.

\subsection{Spectroscopy}

During the onset of episode 2013 (Episode 2013A) we serendipitously obtained spectraon 11 November 2013 with OSIRIS at the 10.4 m GTC during a drift-scan spectroscopy run to observe NGC 2770 and its satellite galaxy NGC 2770B. Episode 2013A was also detected by iPTF in December 2013 and spectroscopy obtained with GMOS-N, but none of the observations had been published until recently \citep{2016arXiv160502450O}. Following the detection of the precursor in Feb. and March 2015 we performed spectroscopy on April 14, close to the maximum of the 2015A episode. Early data of this episode from February 2015 were published by the Asiago collaboration which we included in our dataset.  

From May 16 onwards, after the onset of episode 2015B/SN explosion, we started an intense spectroscopic monitoring, first daily, then ever second day and in larger intervals from mid June. Data were taken primarily with OSIRIS at the 10.4 m GTC using grism R1000B (covering 3700--7800\,\AA{}), on a few occasions supplemented by grism R1000R (covering up to 10,000\,\AA{}). On June 4 and 6 we also obtained high resolution observations from 5580 to 7680 \AA{} using the R2500R grism. Further visible spectra were obtained with TWIN and CAFOS, at the 3.5 m and 2.2 m CAHA telescopes of the Calar Alto Observatory (Almer\'ia, Spain). High resolution spectra were also obtained using HDS at Subaru (R$\sim$40,000). Three more late time spectra were obtained after the Sun gap at the end of September 2015, November 2015 and January 2016 from OSIRIS/GTC and a single observation on Dec. 5, 2015 from FOCAS/Subaru. The log of observations is detailed in Table~\ref{Tab:speclog}, the spectra are plotted in Fig. \ref{Fig:specs}.

Spectra were reduced with standard procedures in IRAF and flux calibrated with standard stars taken the same night. The TWIN spectra show some issue in the flux calibration altering the shape of the spectrum, the reason for which is unknown and we hence exclude these data from any analysis requiring flux calibrated spectra. The Subaru data were reduced using the procedures described in \cite{Maeda2015} for the HDS data and \cite{Maeda2016} for the FOCAS data. 

\begin{table}
\caption{Log of the spectroscopic observations of SN\,2015bh and its precusors.}
\label{Tab:speclog}
\centering                          
\begin{tabular}{c c c c}        
\hline\hline                 
Date 		& Telescope 	& Exposure	& Grism	\\    
\hline                        
20131112.191   	& OSIRIS/10.4m GTC    		& 830	   	& R1000B 		\\ 
20150209.93 	& Asiago  					& ---	  		& 3400-8200 	\\
20150414.96    	& OSIRIS/10.4m GTC     		& 3$\times$450	& R500B	 	\\
20150516.856 	& TWIN/3.5m CAHA			& 2$\times$900	& B		 	\\
20150516.856 	& TWIN/3.5m CAHA			& 2$\times$900	& R		 	\\
20150516.885  	& OSIRIS/10.4m GTC  		& 3$\times$180	& R1000B	 	\\
20150516.893  	& OSIRIS/10.4m GTC  		& 3$\times$180	& R1000R	 	\\
20150517.856 	& TWIN/3.5m CAHA			& 2$\times$900	& B		 	\\
20150517.856 	& TWIN/3.5m CAHA			& 2$\times$900	& R		 	\\
20150518.905  	& OSIRIS/10.4m GTC  		& 3$\times$180	& R1000B		\\
20150519.892  	& OSIRIS/10.4m GTC  		& 3$\times$180	& R1000B 		\\
20150520.906  	& OSIRIS/10.4m GTC  		& 3$\times$180	& R1000B		\\
20150522.927  	& OSIRIS/10.4m GTC  		& 3$\times$180	& R1000B		\\
20150522.934  	& OSIRIS/10.4m GTC  		& 3$\times$180	& R1000R		\\
20150525.860  	& CAFOS/2.2mCAHA  		& 600		& Grism4		\\
20150525.874  	& CAFOS/2.2mCAHA  		& 600		& Grism7		\\
20150524.894  	& OSIRIS/10.4m GTC  		& 3$\times$180	& R1000B		\\
20150526.921  	& OSIRIS/10.4m GTC  		& 3$\times$180	& R1000B		\\
20150529.918  	& OSIRIS/10.4m GTC  		& 3$\times$180	& R1000B		\\
20150530.251		&HDS/Subaru				& 2$\times$1200	& (R=40000)	\\
20150531.887  	& OSIRIS/10.4m GTC  		& 4$\times$180	& R1000B		\\
20150602.882	& OSIRIS/10.4m GTC  		& 3$\times$180	& R1000B		\\
20150604.890	& OSIRIS/10.4m GTC  		& 3$\times$180	& R1000B		\\
20150604.896	& OSIRIS/10.4m GTC  		& 300		& R2500R		\\
20150606.889	& OSIRIS/10.4m GTC  		& 3$\times$180	& R1000B		\\
20150606.894	& OSIRIS/10.4m GTC  		& 300		& R2500R		\\
20150610.897	& OSIRIS/10.4m GTC  		& 3$\times$180	& R1000B		\\
20150614.888	& OSIRIS/10.4m GTC  		& 3$\times$180	& R1000B		\\
20150617.903	& OSIRIS/10.4m GTC  		& 3$\times$180	& R1000B		\\
20150619.891	& OSIRIS/10.4m GTC  		& 5$\times$180	& R1000B		\\
20150927.248	&  OSIRIS/10.4m GTC		& 3$\times$300 & R1000B		\\
20151121.244	&  OSIRIS/10.4m GTC		& 3$\times$400 	& R1000B	 	\\
20151121.258	&  OSIRIS/10.4m GTC		& 3$\times$400 	& R2500R		\\
20151205.465 & FOCAS/Subaru			& 2$\times$600	& B300		\\
20160121.168	&  OSIRIS/10.4m GTC		& 4$\times$400 	& R1000B	 	\\
\hline                                   
\end{tabular}
\end{table}

\subsection{Tuneable filters and driftscan spectroscopy}

An ongoing project within our collaboration is analyzing spatially resolved spectroscopic data throughout the host galaxy of NGC 2770. Two of the datasets also cover the position of SN\,2015bh: 1) Tuneable narrowband filter (TF) observations of [NII], H$\alpha$, [SII] lines as well as several continuum images across the galaxy were taken between 2013 and 2015 using the Fabry-Perot tunable filters at OSIRIS/GTC. 2) A data cube obtained by drift-scan spectroscopy with the longslit spectrograph of OSIRIS/GTC in November 2013 at the onset of episode 2013A. The spectra cover the wavelength range from 3600 to 7900 \AA{} at a spectral resolution of 2.6 {\AA} and a spatial sampling of 1.2 arcsec. 

The tunable filter data cover the strong emission lines of H$\alpha$, [NII] and [SII] in different steps, although [SII] is not used in this paper. To disentangle H$\alpha$ and [NII] we used filter widths of 12\,\AA{} in steps of 8\,\AA{} and a continuum filter of width 20\,\AA{} red wards of [NII]. In order to improve the continuum subtraction and to obtain a coverage in a blue filter we performed observations in three order separating filters (width 13 -- 45 \AA{}) roughly corresponding to $g'$, $r'$ and $i'$ filters in March 2015. The images were reduced with standard imaging techniques. To account for the wavelength shift across the field in the emission line scans, we assigned the true wavelength to each step in the scan using the wavelength dependency from the optical axis as described in \cite{2011PASP..123.1107M}. After this we shifted the profile in each pixel to a common rest frame before integrating over a certain wavelength range for each emission line. This shift was not applied to the continuum filters. No absolute flux calibration was done for the TF data.

The driftscan data were reduced as 2D longslit spectra and wavelength calibrated. For each OB a template trace was extracted using a bright, fairly isolated region in the galaxy. We then extracted 50 regions per 2D spectrum, shifting the template trace by equal steps. The resulting 1D spectra were flux-calibrated using a single sensitivity function derived from a standard star observation during a photometric night and combined into a 3D data cube.

In both datasets the site of SN 2015bh is contaminated and/or dominated by emission from the LBV itself, either during a more quiescent state (part of the TF dataset) or even during one of the pre-explosion outbursts (drift-scan dataset). Hence, we cannot use the data to measure any properties at the site of SN 2015bh directly, instead we can only infer properties from nearby SF regions. Here we use the part of the datasets described above in the vicinity of SN 2015bh. The full dataset as well as supplementing data from other regions in the host supersede the scope of this paper and will be presented in a separate paper (Th\"one et al. in prep.)

\section{Evolution of the LBV 2008-2015}\label{sec:evol1}

   \begin{figure*}
   \centering
   \includegraphics[width=17.5cm]{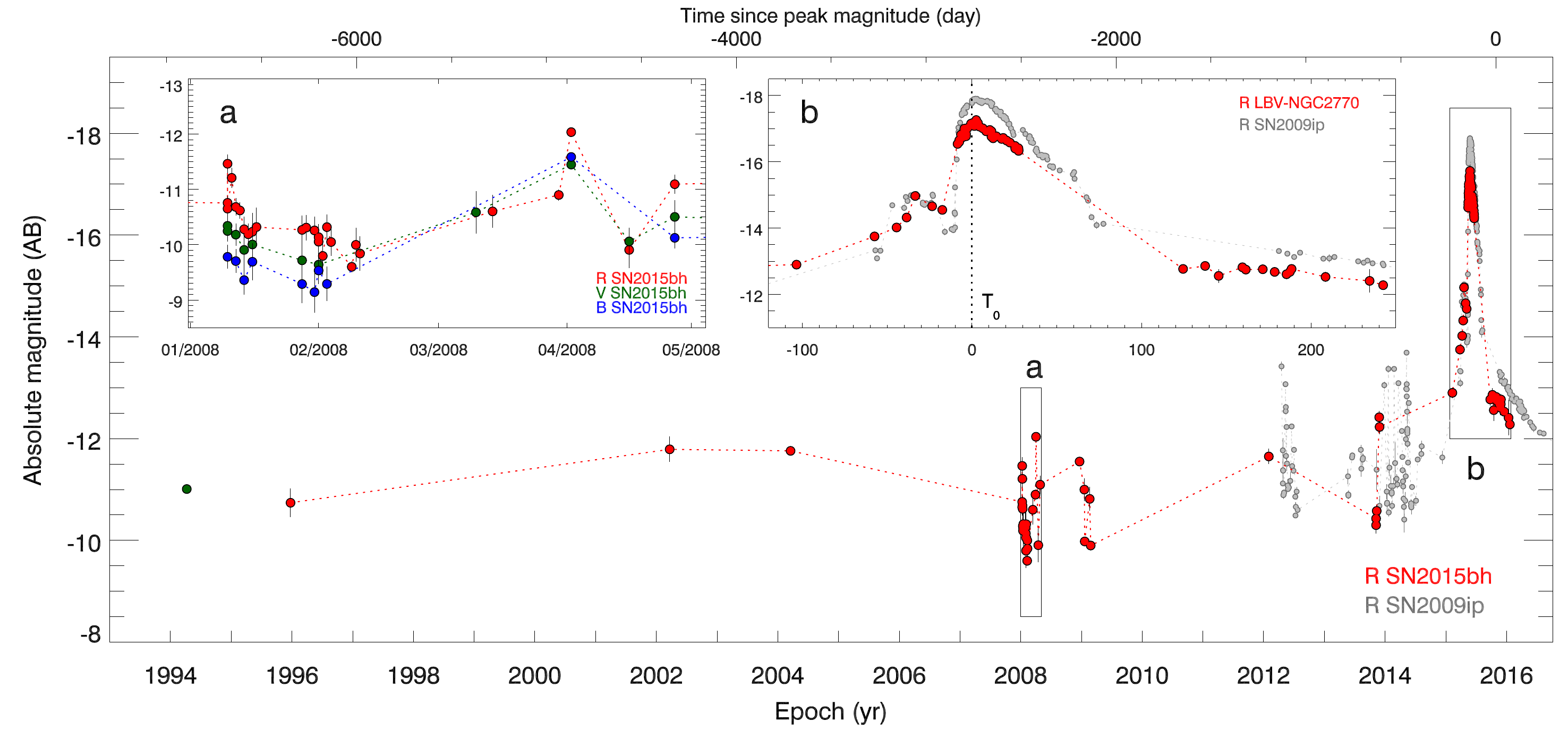}
      \caption{Top: Light curve of the progenitor of SN 2015bh from 2004 to the explosion in 2015, the insets are blow-ups of episode 2008 in different colors and episodes 2015A\&B (main event and precusor). SN 2009ip is displayed for comparison, as it shows a very similar behaviour in all stages of the evolution.
              }
         \label{Fig:LC}
   \end{figure*}

\subsection{Photometric monitoring}
Archival imaging dating back 21 years before the main event indicate that the progenitor of SN 2015bh had been in an active state for at least two decades (see Fig.~\ref{Fig:LC}). Until the discovery in 2013 by iPTF, or officially, the detection of SNhunt275 in Feb. 2015, the LBV progenitor had falsely been identified as an H{\sc ii} region in the same spiral arm of NGC 2770 as SN 2007uy and hence there was no monitoring of the object during these years.

Already in 1994, the object was likely in a non-quiescent state with an absolute magnitude of --11 in $V$, similar to the outbursts at later times. Further single detections in R-band in 1996, 2002 and 2004 ranging between --11 and --12 mag are further indicative of ongoing outbursts. During the observations in 2008 (see Fig.~\ref{Fig:LC}, inset a), significant variability was observed with variations in brightness of more than 2 mag between $\sim$--9.8 mag and --12.1 in $R$-band: At the onset of SN 2008D on Jan. 10, 2008 and the subsequent $\sim$ 30 days, the LBV was in a declining phase after a potential earlier outbust, fading to an absolute magnitude of $\sim$ -10 mag (episode 2008A).  Another episode was observed around the turn of the year 2008/2009 (episode 2008B) with HST, showing a rebrightening of less than 2 mag and a quick decline afterwards. The color of the object usually turned redder during  outbreaks after which it became slightly bluer again.

The LBV was in the FoV of the Westerbork Synthesis Radio Telescope during observations of SN 2007uy and SN 2008S on Jan 19 and 20, 2008, Nov 15 and 16, 2008 and Sep 12 2009  \citep{2011ApJ...726...99V}. The object was not detected in either of the epochs, which were performed during an LBV outburst phase, with a limit of $\sim$0.1\,mJy or 8.7$\times$10$^{26}$ erg\,s$^{-1}$\,Hz$^{-1}$ at 4.8\,GHz (A. van der Horst, priv. comm.). 

During the following years, only narrow-band TF images in H$\alpha$ were obtained, including some narrow continuum filters around H$\alpha$ that likely contain still part of the asymmetric H$\alpha$ emission from the LBV and are hence not reliable for any study of the LBV. A single observation from INT in Feb. 2012 showed the LBV at R$=$19.8 mag (--12.6 mag) probably indicating another outburst phase.

In Nov. 2013 we serendipitously obtained imaging data and a spectrum while observing NGC 2770B (a the satellite galaxy of NGC 2770) with driftscan spectroscopy covering also the central part of NGC 2770 and the site of SN 2015bh. By chance, we caught the object at the onset of another outburst (episode 2013A) where it showed an increase in brightness from $R=$--\,10.2 to --\,12.5\,mag. This outburst was detected later by iPTF, classified as an LBV outburst and named ``iPTF13evf''. On Dec. 21, 2013 the object had been faded to the level of comparison images taken between Jan. and Apr. 2014 \cite{2016arXiv160502450O}.  \cite{2016arXiv160502450O} searched the PTF and Catalina Real Time Survey (CRTS, \citet{2009ApJ...696..870D}) archives for other possible outbursts. No detections are made with PTF up to $\sim$ mag 21 (or --11.4 mag) up to 240\,d before the 2013 episode and no outbursts brighter than magnitude 19 (or --13.4 mag) from CRTS until the precursor 2015A. Possibly no other major outburst happened at that time, although our detection from INT in 2012 would have been missed by those surveys.

\subsection{Features in the spectra during the LBV phase}
\begin{figure*}
   \centering
   \includegraphics[width=15cm]{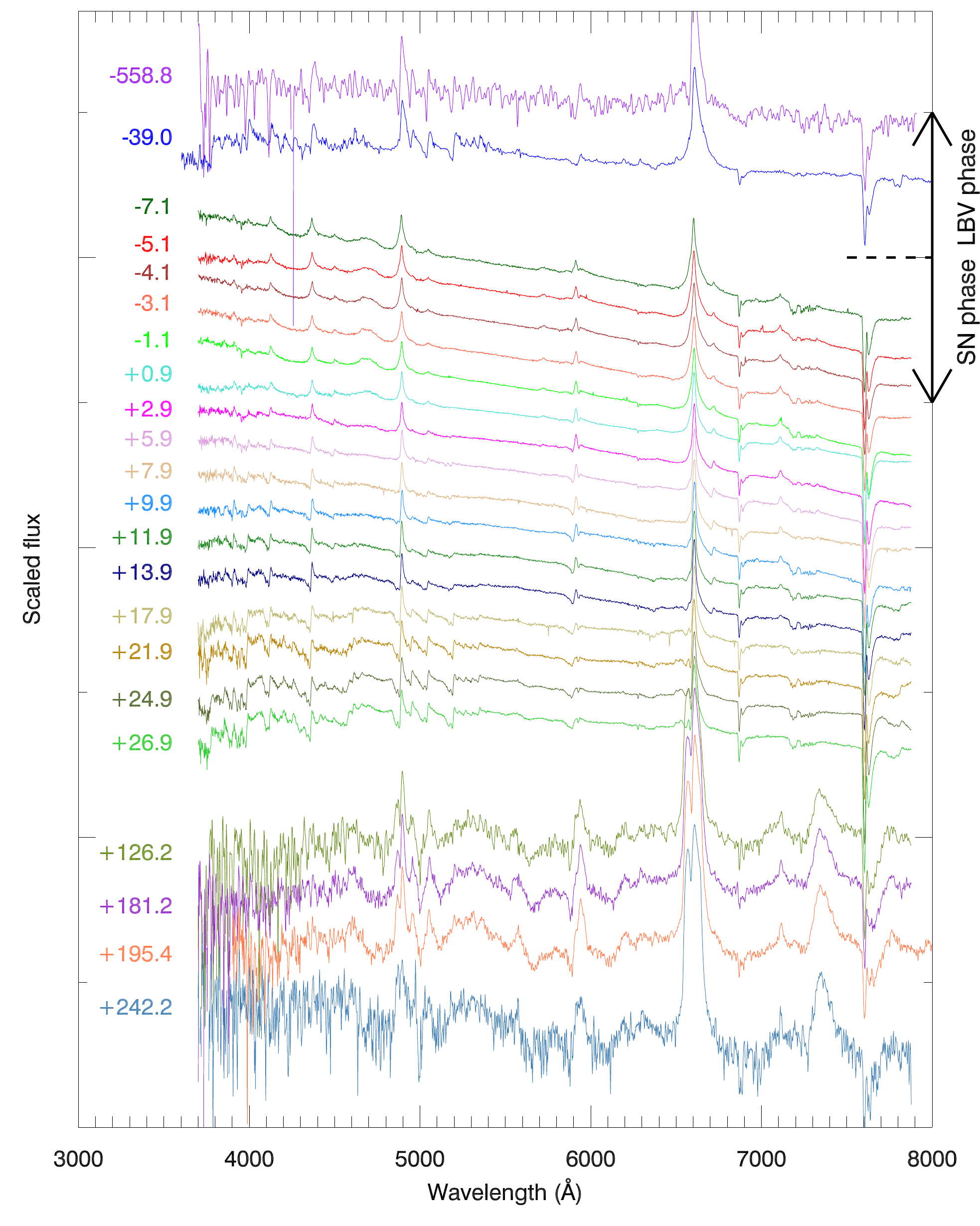}
      \caption{Spectroscopic evolution of the LBV and SN, including two epochs during the LBV outbursts (episode 2013, -559\,d and episode 2015A at --39\,d) and a closely sampled time-series during the possible SN explosion from May 16 to June 19 (-7 to +27\,d). After a gap in the observations due to sun constraints we obtained four new spectra in late September (+126\,d), November 2015 (+181\,d), December (+195.4\,d) and January 2016 (+242\,d).}
         \label{Fig:specs}
   \end{figure*}

\begin{figure*}
   \centering
   \includegraphics[width=16cm]{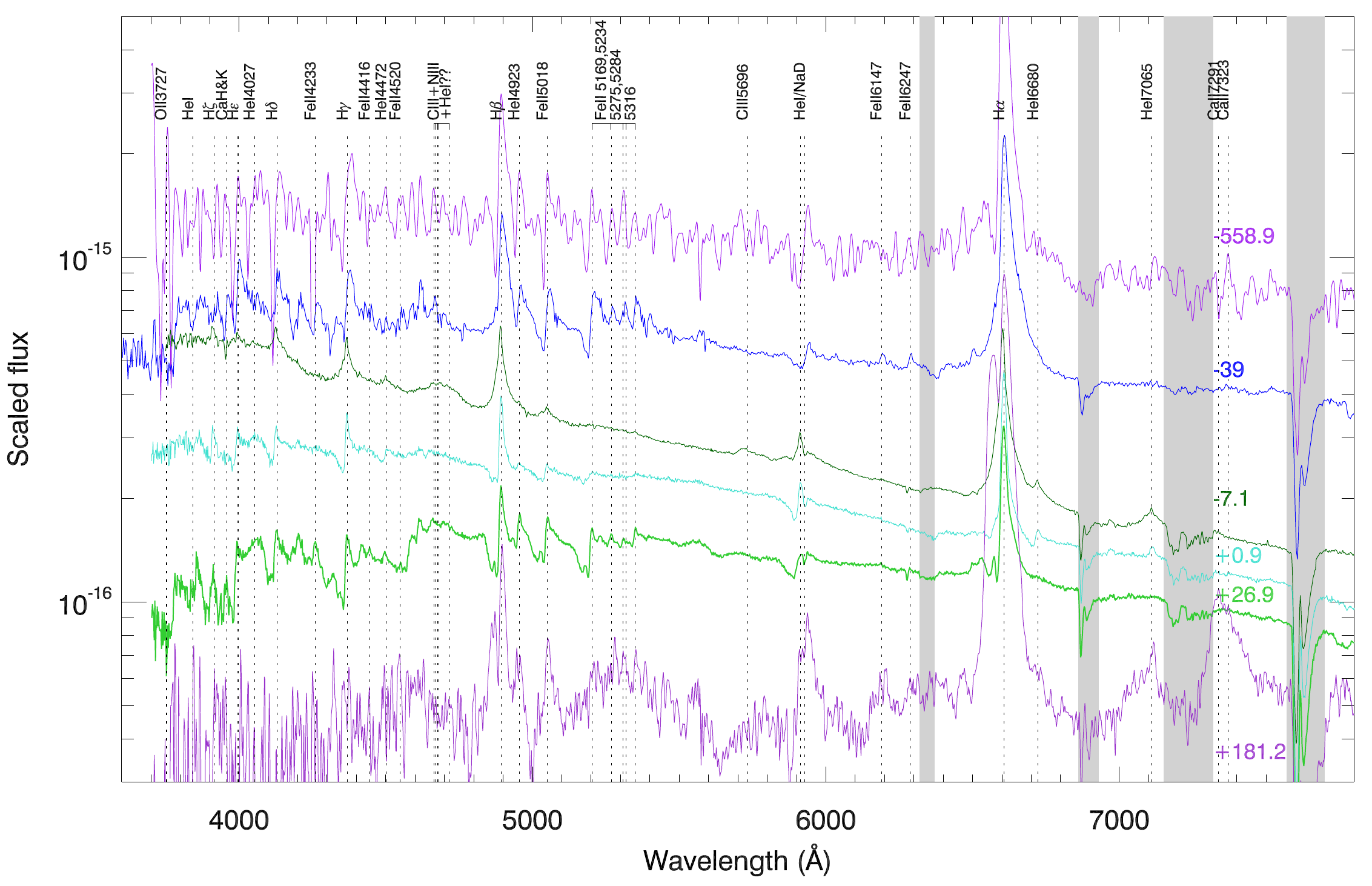}
      \caption{Optical spectrum at 6 representative epochs and line identifications. Not all lines are visible in all epochs. The grey bars indicate atmospheric lines. Color coding of the spectra is the same as in Fig. \ref{Fig:specs}}
         \label{Fig:spectra_ids}
   \end{figure*}

Our serendipitous spectra from the onset of episode 2013 are the only public spectra of SN 2015bh before the precursor event 2015A. The complete set of spectra from the outburst through the main event is plotted in Fig.~\ref{Fig:specs}, in Fig.~\ref{Fig:spectra_ids} we furthermore plot distinct epochs together with line identifications. 

The spectrum from Nov. 2013 is characterized by intense narrow P-Cygni profiles of the Balmer series, He I and Fe II. The Balmer emission lines are very narrow (FWHM $\sim$ 600 km\,s$^{-1}$) and show a very steep drop in the blue wing towards the absorption part of the P-Cygni profile. This single, narrow P-Cygni profile basically stays unchanged from 2013A throughout the main event and is probably reminiscent of material ejected prior to the 2013A event. 

The spectrum of that time looks very similar to an LBV in quiescence or at least free of CSM interaction, which is not available for any other SN or impostor so far. We find a good match between the Nov. 2013 spectrum and a model for an LBV with L$=$1$\times$10$^{6}$ L$_\odot$, T$_\mathrm{eff}$=8500\,K, a mass loss rate of $\dot{M}=$5$\times$10$^{-4}$M$_\odot$\,y$^{-1}$ and a wind terminal speed of 600 km\,s$^{-1}$ (see Fig.\ref{Fig:specsmodel} in the Appendix). This is similar to Eta Carinae today with the stellar radiation propagating through a dense wind.

\cite{2016arXiv160502450O} publish spectra taken with GMOS/Gemini-N on Dec. 13, 2013, one month after our spectrum and during the maximum of the 2013A event. The spectra are of lower resolution, S/N and cover a smaller wavelength range and the only feature clearly detected in H$\alpha$ together with some weak tentative detections of various He I transitions also present in later spectra of SN 2015bh. These authors measure the velocity of the absorption component of the P-Cygni profile at --1,300 km\,s$^{-1}$ which is slightly higher than what we observe in the Nov. 12 spectra. The most curious feature in their spectra is an absorption trough blue wards of H$\alpha$, also noted in their work, which, if associated to H$\alpha$ would have a velocity of $\sim$--15,000 km\,s$^{-1}$. However, it remains unclear whether this feature is real or due to some unknown issue in the reduction of the spectra.

\section{The 2015A event and possible SN explosion}\label{sec:evol2}  
    
   \subsection{Lightcurve evolution}
   
        \begin{figure}
  \centering
   \includegraphics[width=\columnwidth]{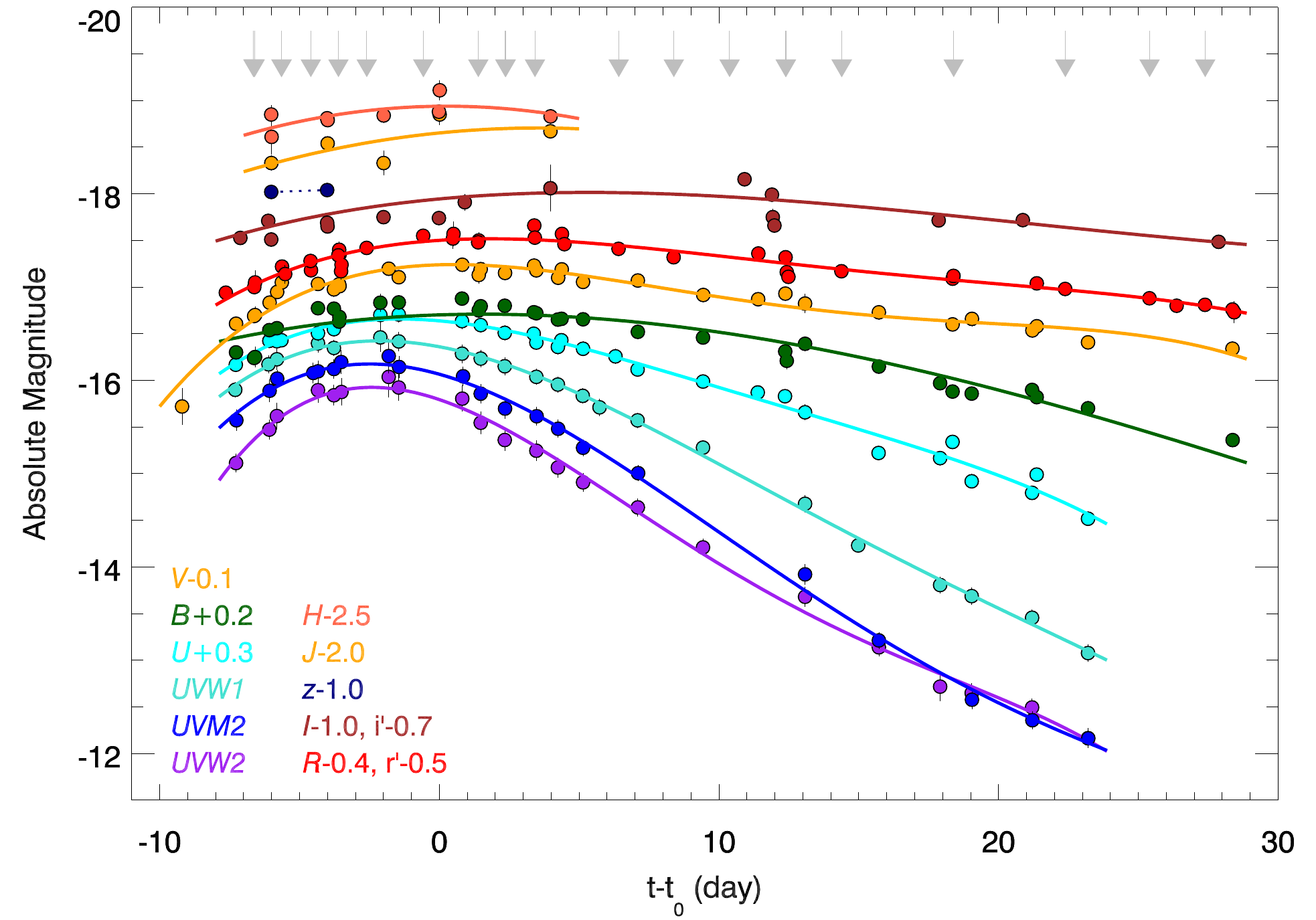}
      \caption{Lightcurve in different bands during the main episode (2015B) and polynomial fits to the LCs. Grey arrows indicate the dates where we obtained spectroscopy.}
         \label{Fig:LCbands}
   \end{figure}

There are no data public of SN\,2015bh  between the end of 2013 and Feb. 2015. On Feb. 7 2015, SN hunt\footnote{http://crts.caltech.edu} reported the discovery of a ``new transient'' which they named ``SNhunt275''  with an absolute magnitude of --13, slightly higher than during the last observed outburst in 2013. Spectra taken on Feb. 10 \citep{2015ATel.7042....1E} reported asymmetric Balmer lines and classified this event as a ``SN impostor''. We obtained our own photometric data in late March and early April 2015, which showed a further increase in brightness. During our subsequent weekly monitoring, the object rose to a maximum of $R=$--13.9\,mag (around 25 days before the SN explosion). After this, its brightness declined down to $R=$--13.6\,mag on May 12, 2015. We call this episode the ``precursor'' or 2015A (see Fig. \ref{Fig:LC}, inset b). Both the shape and magnitude difference of the precursor resemble that observed for SN 2009ip. 

Instead of a further decline, observations from the 1.5m OSN telescope on May 16 showed that the event suddenly brightened by $\sim$2 mag, and the brightness continued to rise in the days thereafter. The maximum was reached around May 24 which we subsequently define as T$=$0 of the main event 2015B, after which the LC started a linear decline. The main event of 2009ip was slightly brighter but had similar rise and decay times. In addition, it showed a small ``bump'' during the decline post maximum which we did not observe for SN 2015bh; however, such a bump could have occurred during the observational gap starting in late June 2015. Both SN 2009ip and SN 2015bh were affected by observational gaps of $\sim$ 3 months due to sun constraints, so the exact evolution between $\sim$ 30--120 days and 50--180 days for SN\,2015bh and SN 2009ip, respectively, is not well determined. At late times ($>$100\,d), the LCs of both events flatten and subsequently decay very slowly. As of June 2016, SN 2015bh remains in this shallow decline phase.

During the UVOT observations at the main event (epoch 2015B), we also took observations with XRT onboard Swift from May 16 to June 7, 2015. We do not detect any X-ray emission from the SN at any epoch.  Single observations of $\sim 2$ ks reach a limiting 0.3-10 keV unabsorbed flux of $3\times 10^{-13}$ erg cm$^{-2}$ s$^{-1}$ assuming a power law spectrum with $\Gamma=2$ and the Galactic column density of $2\times 10^{20}$ cm$^{-2}$ \citealt{2005A&A...440..775K}). At a distance of 27 Mpc this converts to an X-ray luminosity $L_X<3\times 10^{40}$ erg s$^{-1}$. Integrating the full set of X-ray observations we collected 48\,ks and derive an upper limit of $5\times 10^{-14}$ erg cm$^{-2}$ s$^{-1}$ ($L_X<4\times 10^{39}$ erg s$^{-1}$).

\subsection{Bolometric LC and energy release}

Using our photometry during episode 2015A and B we construct a pseudo-bolometric LC following the standard method outlined in \cite{2014MNRAS.438.2924C}: (1) we corrected all magnitudes for foreground and rest-frame extinction, (2) then we converted the AB UV/optical/NIR magnitudes into monochromatic fluxes using a flux zeropoint of 3.631 ($\times10^{-23}$ erg~cm$^{-2}$~s$^{-1}$~Hz$^{-1}$, e.g. \citealt{1995PASP..107..945F}) to obtain flux densities (units of mJy).  For epochs where there were no contemporaneous observations, we linearly interpolate the flux LCs to estimate the missing flux.  Then, for each epoch of multi-band observations, and using the effective wavelength of each filter \cite{1995PASP..107..945F, 2006MNRAS.367..454H, 2008MNRAS.383..627P} we (3) linearly interpolated between each datapoint, (4) integrated the SED over frequency, assuming zero flux at the integration limits, and finally (5) corrected for filter overlap.  The linear interpolation and integration were performed using a program written in {\sc Pyxplot}\footnote{http://pyxplot.org.uk}.

During the main event and the late epochs, there are not enough multi-band observations for us to determine the bolometric luminosity for a given epoch using the method described above. In order to construct a bolometric LC at those times  we apply a different method by fitting the Planck function to each epoch of multi-band photometry to determine the black-body temperature and the radius of the blackbody emitter. Then, we integrated the fitted Planck function between the filter limits of our data (UV to $I$-band) in order to consider the same frequency range as that used when constructing the filter-integrated bolometric LC.  The fitting and integration were performing using custom  \textsc{python} programs.  It was seen that both the shape and peak luminosity of both bolometric LCs agree very well.

From the bolometric LC we derive estimates of the total energy release during the different events: The 2015A event released 1.8 $\times$10$^{47}$ erg, though we must consider the caveat that this estimate only includes data obtained over three epochs, i.e. days -145 to -45. The first epoch of observations of the main event until the object disappeared behind the Sun counts with a total release of 1.2$\times$10$^{49}$ erg. Adding the luminosity of the observations after the object became visible again does not significantly increase the total energy release (the total energy between day 121 and 242 is 9.9$\times$10$^{46}$ erg). Assuming a linear interpolation during the time when the object was not observable yields a total energy release of the main event (2015B) of 1.8$\times$10$^{49}$ erg. \cite{2016arXiv160502450O} estimated a total energy release of $>$ 2.4$\times$10$^{46}$ erg for the LBV outburst in 2013.

  \begin{figure}
   \centering
   \includegraphics[width=\columnwidth]{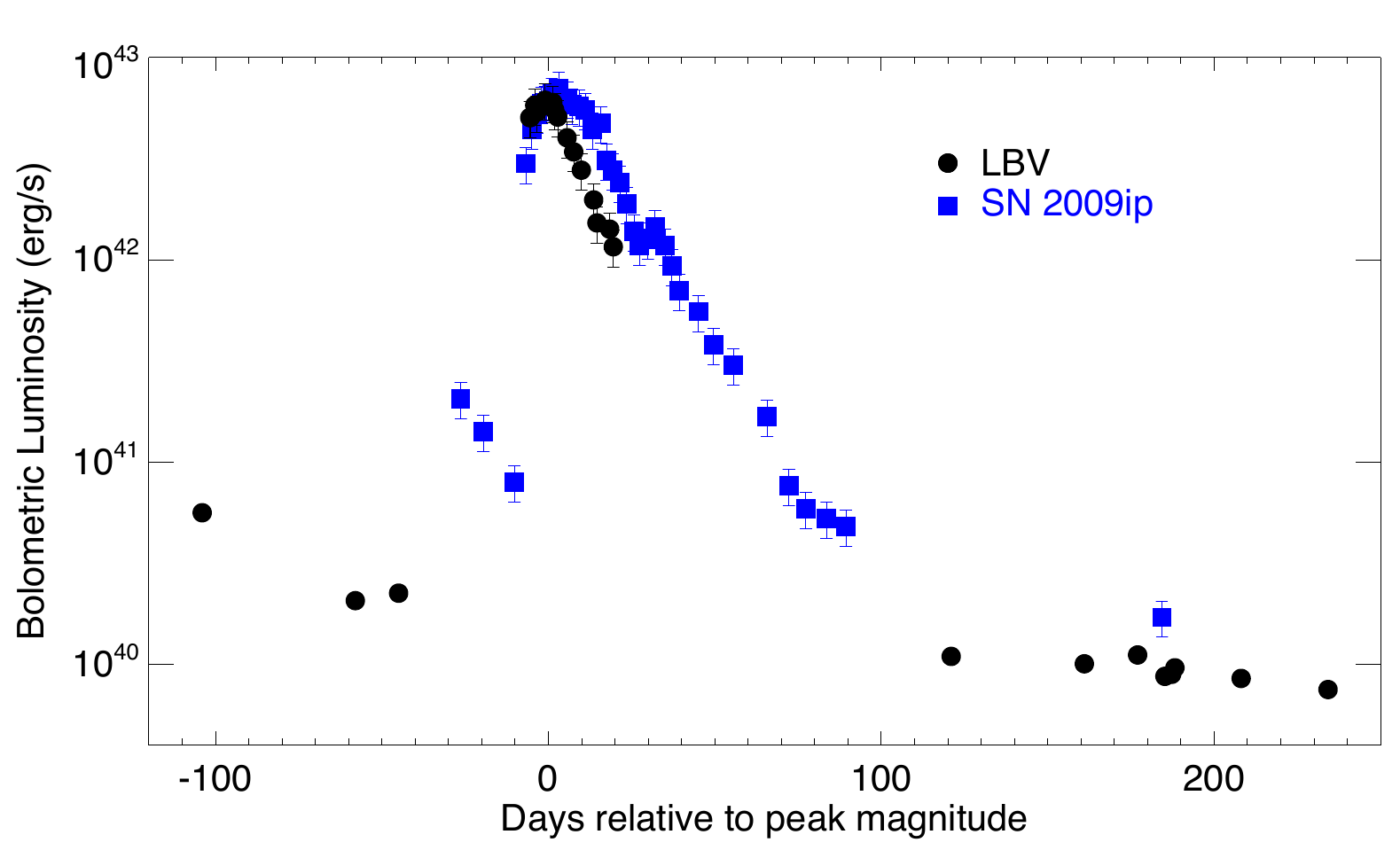}
      \caption{Bolometric LC of the precursor (episode 2015A) and the main explosion event of SN 2015bh (epoch 2015B). As comparison we also plot the bolometric LC of SN 2009ip obtained in the same way as for 2015bh.}
         \label{Fig:bolo}
   \end{figure}

\subsection{Spectral energy distribution, temperature and radius}
During the main event, starting May 16, we have sufficient data in each epoch to fit the SED of the event from UVW1  ($\sim$ 1500 \AA{}) to $z$/$H$-band (IR data are only available up to day 6 post maximum). We further have an early SED on May 9 during the pre-explosion event using the $J,H,K$ observations from Omega2000 and a late epoch on Nov. 27 (184 days post maximum) using the $J$-Band observations from PANIC. 

\begin{figure*}
   \centering
   \includegraphics[width=14cm]{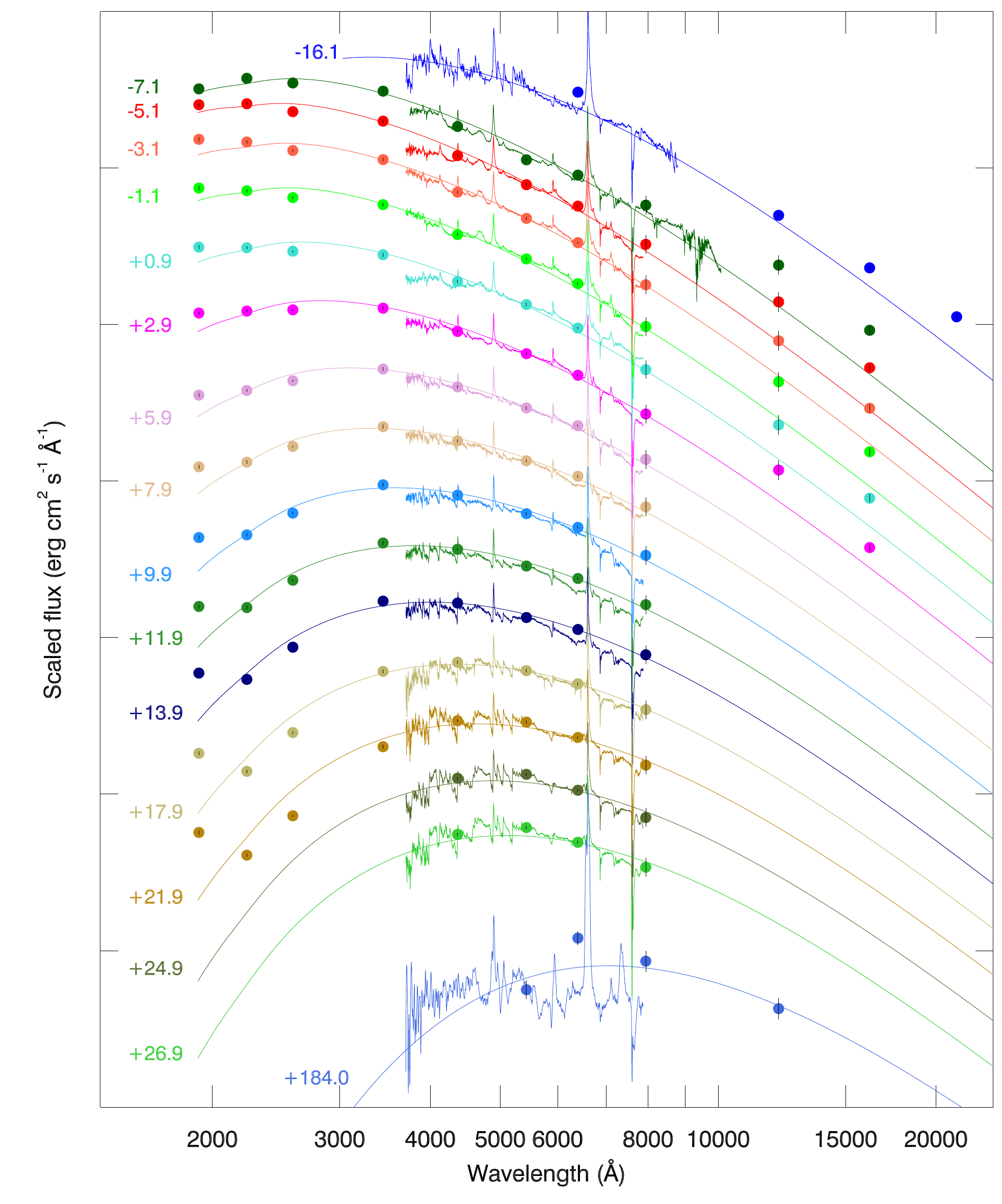}
      \caption{SED from UV to IR from the pre-explosion bump (one epoch), during the main explosion event until 180 days post maximum. The fit shown is a simple BB fit with varying temperatures. For the VIS bands we derive magnitudes using the flux calibrated spectra. In the last spectrum/SED, the R-band is clearly affected by the strong relative H$\alpha$ line flux at that time.}
         \label{Fig:SED}
   \end{figure*}

In Fig. \ref{Fig:LCbands} we show the multi band LC during the main event, in Fig. \ref{Fig:SED} we plot the complete SED evolution and a fit with a single expanding and cooling BB. The data are observed flux densities, the BB has been corrected for extinction in the host galaxy. We determine the extinction using the resolved NaD doublet from the high-resolution spectrum on June 4. The NaD$_1$ and D$_2$ lines have EWs of 0.45 and 0.57 \AA{}, respectively, applying the relation between NaD and extinction as described in \cite{2012MNRAS.426.1465P} we get an extinction of $E(B-V)=$0.21\,mag $^{+0.08}_{-0.05}$. The high-resolution data from HDS/Subaru on May 30 give very similar values.

  \begin{figure}
   \centering
   \includegraphics[width=\columnwidth]{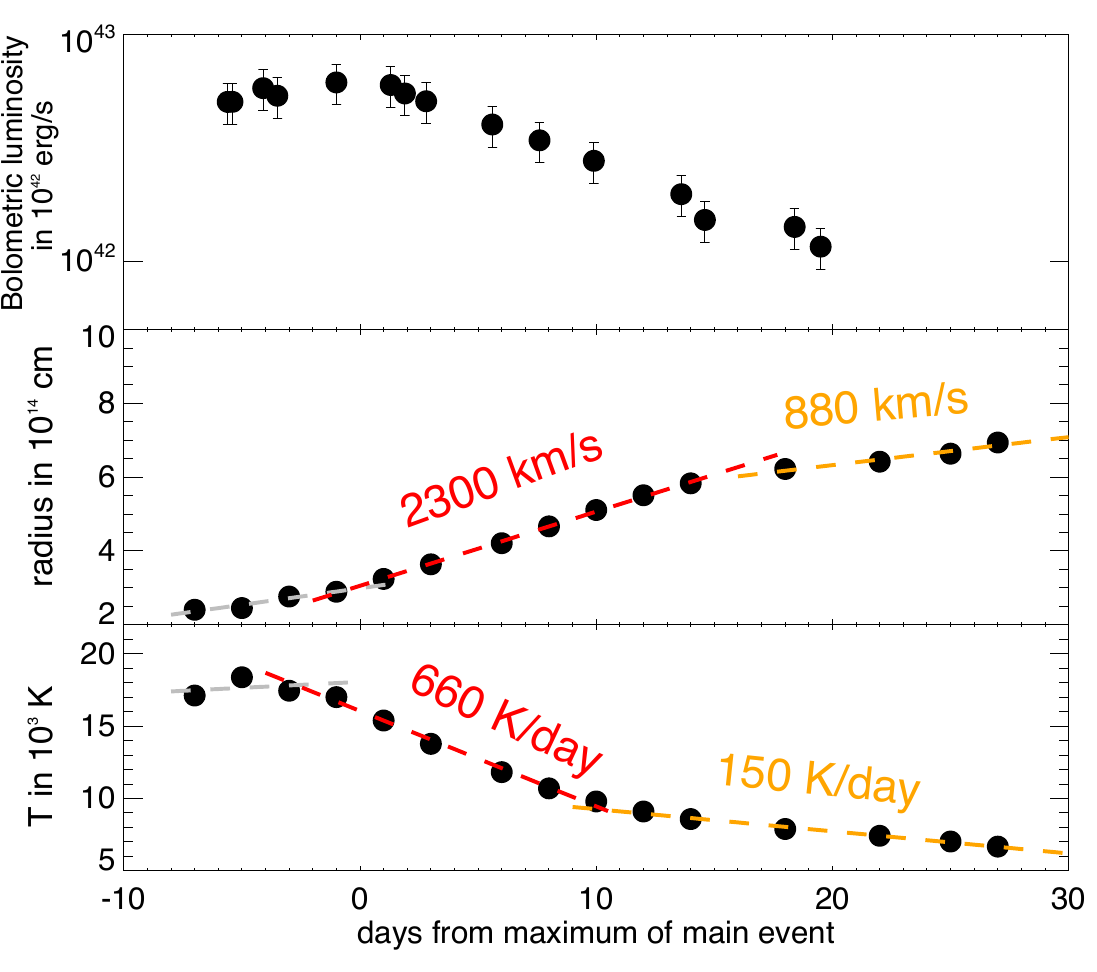}
      \caption{Temperature and radius evolution of the BB fitted to the SED during the main event (see Fig. \ref{Fig:SED}). In the top panel we plot the bolometric LC for comparison.}
         \label{Fig:Tevol}
   \end{figure}

During the first days, coincident with the rise in the LC, the BB gets hotter while there is very little expansion visible (see Fig. \ref{Fig:Tevol}). Past maximum, the radius increases with a rate of $\sim$ 2300 km\,s$^{-1}$ and about 15 days past maximum its rate decreases to 880 km\,s$^{-1}$. The temperature also seems to change during this evolution, first cooling faster until $\sim$10 days post maximum after which it cools down at a considerably slower rate. A similar change in the behavior of the temperature and radius evolution was also seen in the last episode of SN 2009ip \citep{2014ApJ...780...21M}, although the initial expansion velocity was slightly higher with $\sim$ 4400 km\,s$^{-1}$ but slower cooling rates (initially $\sim$ 350 K\,day$^{-1}$). Note also that the BB radius is significantly smaller than the one in SN 2009ip, although one has to be careful to associate the radius derived from the SED fit with a physical radius of the SN photosphere or interacting shell.

The fits in Fig. \ref{Fig:SED}, however, indicate that a single-component evolving BB does not match the data adequately, especially in the blue and red ends of the wavelength range. In the UV we observe two features: a drop in the UVM2 band in epochs past maximum that becomes stronger with time and come from increasing absorption in Fe-lines at that wavelength. Furthermore, there is an excess in the UVW2 band which, however, is likely an instrumental feature: The UVW1 and 2 bands have a leak in the red that becomes more pronounced for redder sources \citep{2010ApJ...721.1608B}, reflected by the increase as the SN BB cools. In the IR we have a consistent small excess in the $J$ and $H$ bands, starting already from the time of the precursor, but which do not change significantly in strength compared to the BB. An additional component in the IR is likely due to dust emission, which is usually observed to increase with time \citep[see e.g.][]{2014Natur.511..326G}. An additional IR component has also been observed for 2009ip \citep{2012ATel.4454....1G, 2014ApJ...780...21M} already at times before maximum, which was attributed to emission from a shell expelled long before the main event. Such an explanation is also appealing to explain this feature of SN 2015bh. Unfortunately, we lack a good coverage in the IR to better assess a possible additional component due to pre-existing dust or dust production after the main event.

\subsection{Spectral evolution and line profiles}\label{Sec:specs}
\begin{figure}
   \centering
   \includegraphics[width=\columnwidth]{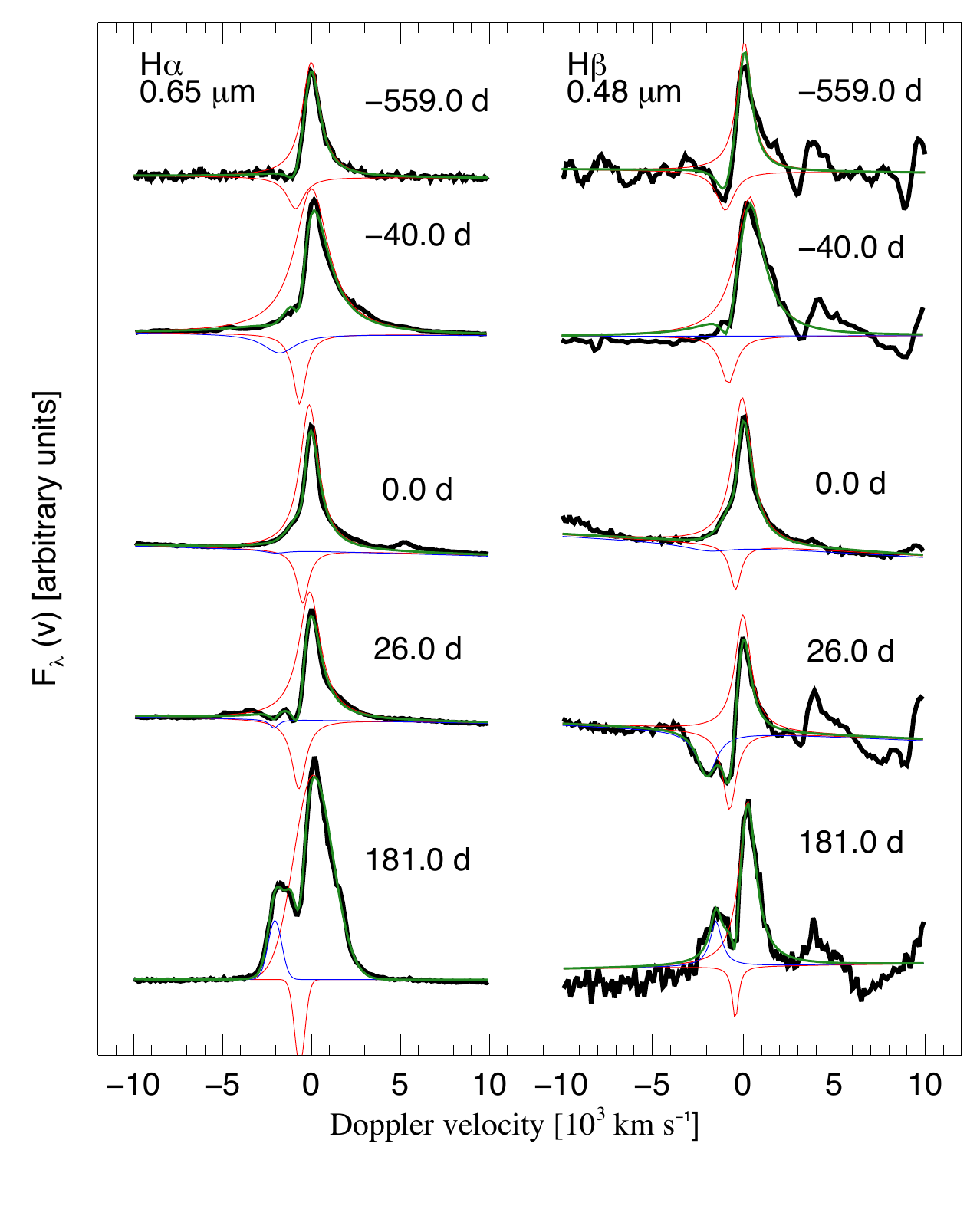}
         \caption{Components fitted to the H$\alpha$ and H$\beta$ line at different representative epochs. The colors of the components are identical to the ones in Fig.\ref{Fig:specfitsresults}: red/orange are the narrow P-Cygni profile emission and absorption components, blue/turquoise the broad components and green the total fitted profile.}
         \label{Fig:specfits}
   \end{figure}

\begin{figure}
   \centering
   \includegraphics[width=\columnwidth]{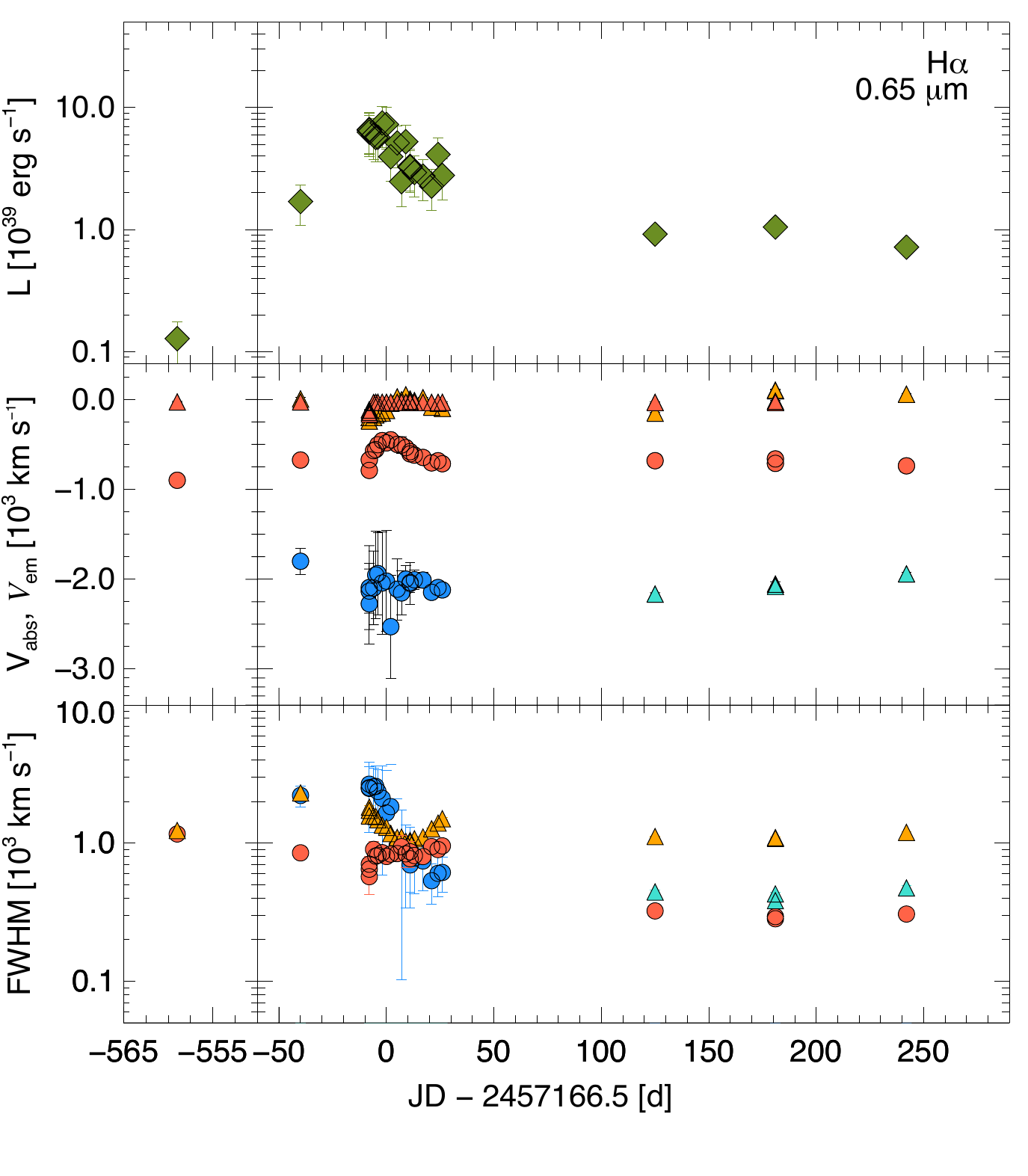}
      \includegraphics[width=\columnwidth]{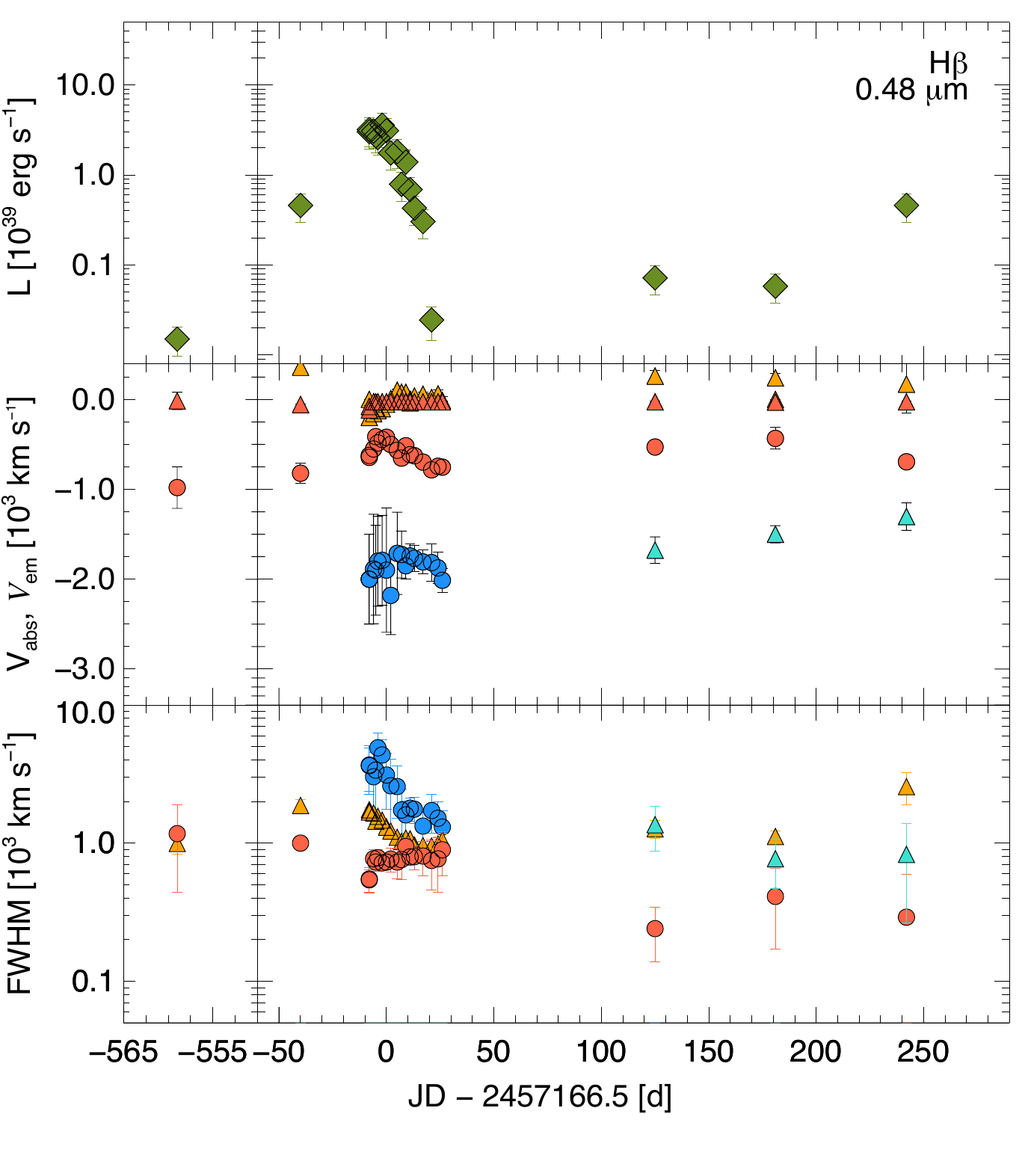}
       \caption{Properties of H$\alpha$ and H$\beta$ derived from the component fits during the LBV phase, precursor and the main event. From top to bottom: Total luminosity, velocity shift and FWHM and flux of the different components. Round symbols correspond to absorption components, triangles to emission components. Colors correspond to the components fitted as shown in Fig.\ref{Fig:specfits}}
         \label{Fig:specfitsresults}
   \end{figure}

We modeled the different components in H$\alpha$ and H$\beta$ in our complete set of spectra from --559 to 240\,d. The lines are fitted well with a single P-Cygni profile during the outburst in 2013 (see above), a Lorentzian profile in emission (as it is typical for IIn SNe) plus two absorption components from the precursor until +26\,d and a double emission together with an absorption component at late times ($>$126\,d, see Sect.~\ref{sec:late}). In Fig. \ref{Fig:specfits} and \ref{Fig:specfitsresults} we plot the H$\alpha$ and H$\beta$ profiles at different representative epochs and the fitted components.

Our observations contain a single epoch during the precursor in April 2015 around the time of its maximum, which was taken after a new outburst (episode 2015A) was reported by SN hunt and confirmed as a SN impostor/LBV outburst \citep{2015ATel.7042....1E}. The spectrum resembles very much the one obtained during episode 2013A. In the blue wing of the absorption part of the single P-Cygni profile a second velocity component starts to appear which could be the onset of the double absorption profile appearing post maximum of the main event. This would imply that this faster material (v$\sim$2000~km\,s$^{-1}$ is directly related to the precursor (2015A) event, a possibility also noted by \cite{2016arXiv160502450O}.

During the main event until maximum, the line profiles start to become rather smooth and almost symmetric with broad wings and very little (visible) absorption, although still present. HeI lines get more pronounced while the Fe lines disappear. This could be due to flash ionization of the material and time dependent recombination, affecting different transitions at a different timescale. The velocity of the narrow absorption component (and to a smaller degree the emission component) slightly decreases until maximum after which is increases again, although only by $\sim$150--200~km\,s$^{-1}$. There is possibly a similar tendency in the faster absorption component but fitting around maximum is difficult and hence imply large errors. The FWHM of the broad absorption component decreases past maximum while the one of the narrow component slightly increases. These effects are probably associated to the increase in radiation field around maximum, ionizing the inner part of the emitting and absorbing shells. 

In this phase, we also observe some broad ``bump'' blue wards of H$\beta$. It is not clear whether this bump is also present blue wards of H$\alpha$. If associated to the Balmer lines, this would mean some material ejected at very high velocities ($\sim$13,000 km\,s$^{-1}$) Alternatively it could be a blend of NIII $\lambda\lambda$ 4634, 4640, C III $\lambda$ 4647, 4650  and He II $\lambda$4686, very similar to the one observed in the Ibn SNe 2013cu \citep{2014Natur.509..471G} and 2015U \citep{2016arXiv160304866S}, as well as the type IIn SN 1998S \citep{2015ApJ...806..213S}. At later times (26\,d and beyond) there is some conspicuous bump in absorption blue wards of H$\beta$ and possibly H$\alpha$, albeit at somewhat lower velocities ($\sim$7,000km\,s$^{-1}$).  \cite{2016arXiv160502450O} observed a 

Past maximum, the spectrum starts to show clearly the double absorption profile with an increasing strength in absorption in time. It is important to note that the double absorption system is visible in all line, implying that the first and second shell corresponding to the two components have very similar abundance patterns.

\subsection{Late evolution}\label{sec:late}
After the object appeared behind the Sun again in late September 2015, the LC had dropped to $\sim$--12.5\,mag, a similar brightness to the object during the outbursts before the main event. The LC after September 2015 shows only a very shallow decline.

The spectra after the sun gap changed dramatically: The emission lines showed a FWHM of several thousand km\,s$^{-1}$. H$\alpha$ had now a double peaked profile that can be modeled with the original emission component at the same velocity and an additional component at $\sim$--2000~km\,s$^{-1}$, the same velocity as the second absorption component observed before the observational gap. This could be an indication that those two components are related (see Sect. \ref{sec:geometry}). 

HeI $\lambda$5889 and $\lambda$7065, which had disappeared in late June, show up as broad emission lines with the same broad profile as the Balmer lines. A strong [CaII] $\lambda\lambda$7291, 7323 doublet is also visible in those spectra. Fe II becomes visible again in emission, but shows a smaller FWHM than the other emission lines. The spectra thus resemble that of a nebular SN, with the exception that there are still absorption components present. One possible interpretation is that the ejecta have become optically thin and we are seeing deeper into the material with a wide range of velocities. 

Together with the change in spectral features, we also observe a large increase in the H$\alpha$ EW (see Fig. \ref{Fig:EWs}). During the outbursts and the precursor, it had an EW of around 200\,\AA{}, during the main event it dropped to values of 50\,\AA{} after which it rose again to finally of values of more than 1000\,\AA{} at $>$ 126 days. A similar behavior has been observed for other IIn SNe and also SN 2009ip \citep{2014MNRAS.438.1191S}. Our object shows a remarkable similarity to the EW evolution of SN 1998S, a genuine type IIn SN. SN 2009ip had values of close to 1000\,\AA{} during the pre-explosion outbursts years before the main event, and increased again to even somewhat higher levels after the main event. \cite{2014MNRAS.438.1191S} interpret the large increase in EW as evidence for a terminal explosion of SN 2009ip as this takes place when the interacting material from the SN becomes optically thin, causing the continuum to drop and the H$\alpha$ photons from the shock escape more easily.

  \begin{figure}
   \centering
   \includegraphics[width=\columnwidth]{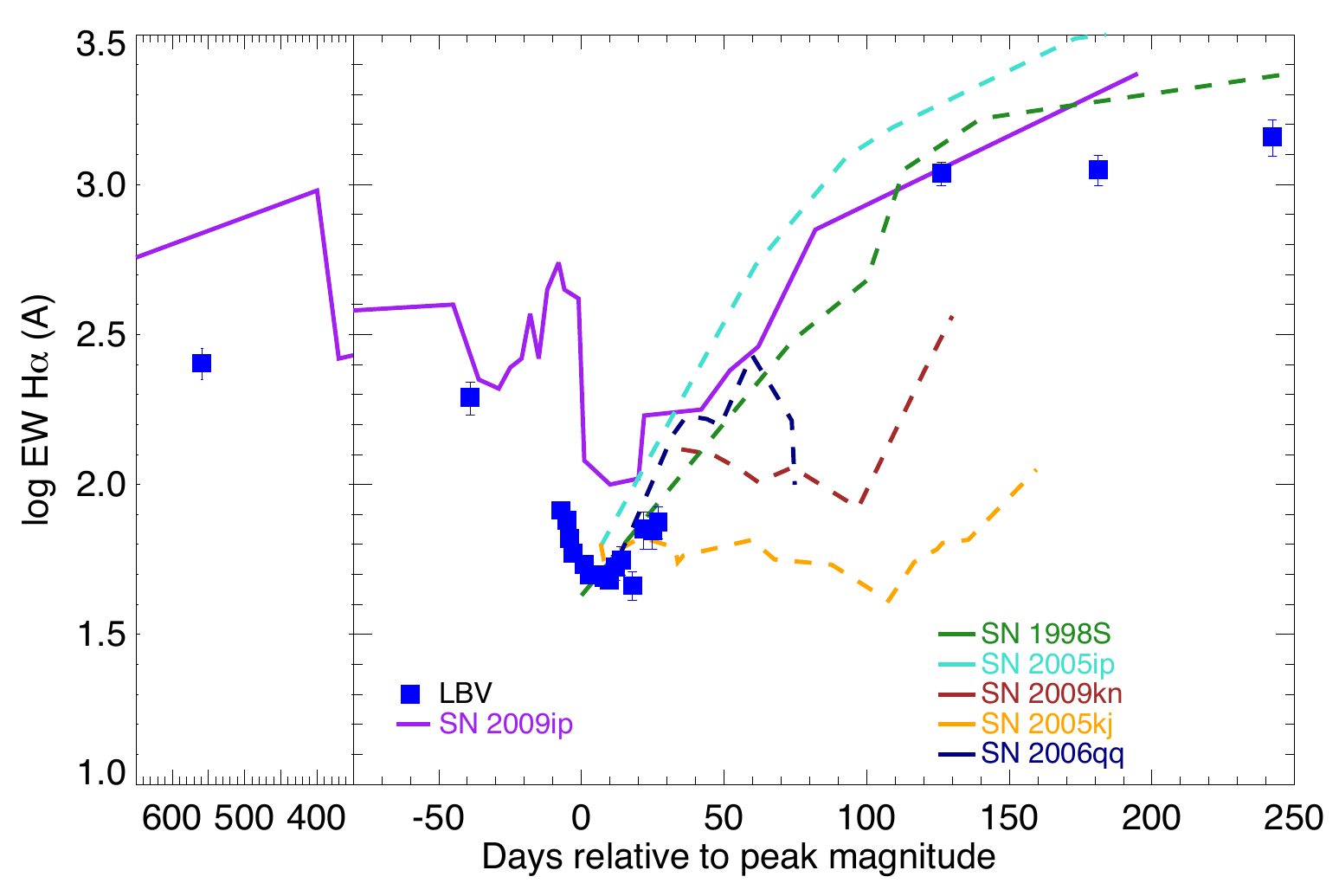}
      \caption{H$\alpha$ EW evolution from the LBV phase to the possible SN explosion. For comparison we also show the EW evolution for SN 2009ip and several there type IIn SNe (data from \citealt{2014MNRAS.438.1191S} and F. Taddia priv. comm.) }
         \label{Fig:EWs}
   \end{figure}

Another interesting aspect of the late spectra is the value of the Balmer decrement H$\alpha$/H$\beta$, which in Case B recombination has a value of 2.73 in absence of extinction. While it showed relatively high values during the outburst and precursor (8.5 and 3.5), it dropped to values below Case B recombination around maximum. From $\sim$ day 10 on it rises again reaching up to values of 6--10 at late times, depending on the method used (i.e. taking the ratio of the peak fluxes or the ratio of the line fluxes). SN 2009ip showed a relatively normal Balmer decrement of 3$\pm$0.2 during the precursor but then dropped to a value of 1.1--1.3 during the peak of the main event \citep{2014AJ....147...23L}. The authors explain this with a very dense (n$_e > 10^{13}$ cm$^{-3}$) plasma in the ejecta which thermalizes the emission due to collisions. At late times of $\sim$ 200 days, however, the Balmer decrement of SN 2009ip increases to a value of 12 \cite{2015arXiv150206033F}. This is typical for nebular phases of outbursts from different objects when they cool and become optically thin and is caused by hydrogen self-absorption. Similar high Balmer decrement values have been observed in a number of other objects such as expanding shells of novae \citep{2003A&A...404..997I} and the outflow from an X-ray binary outburst at late times \citep{2016arXiv160502358M}.

     \begin{figure}
   \centering
   \includegraphics[width=\columnwidth]{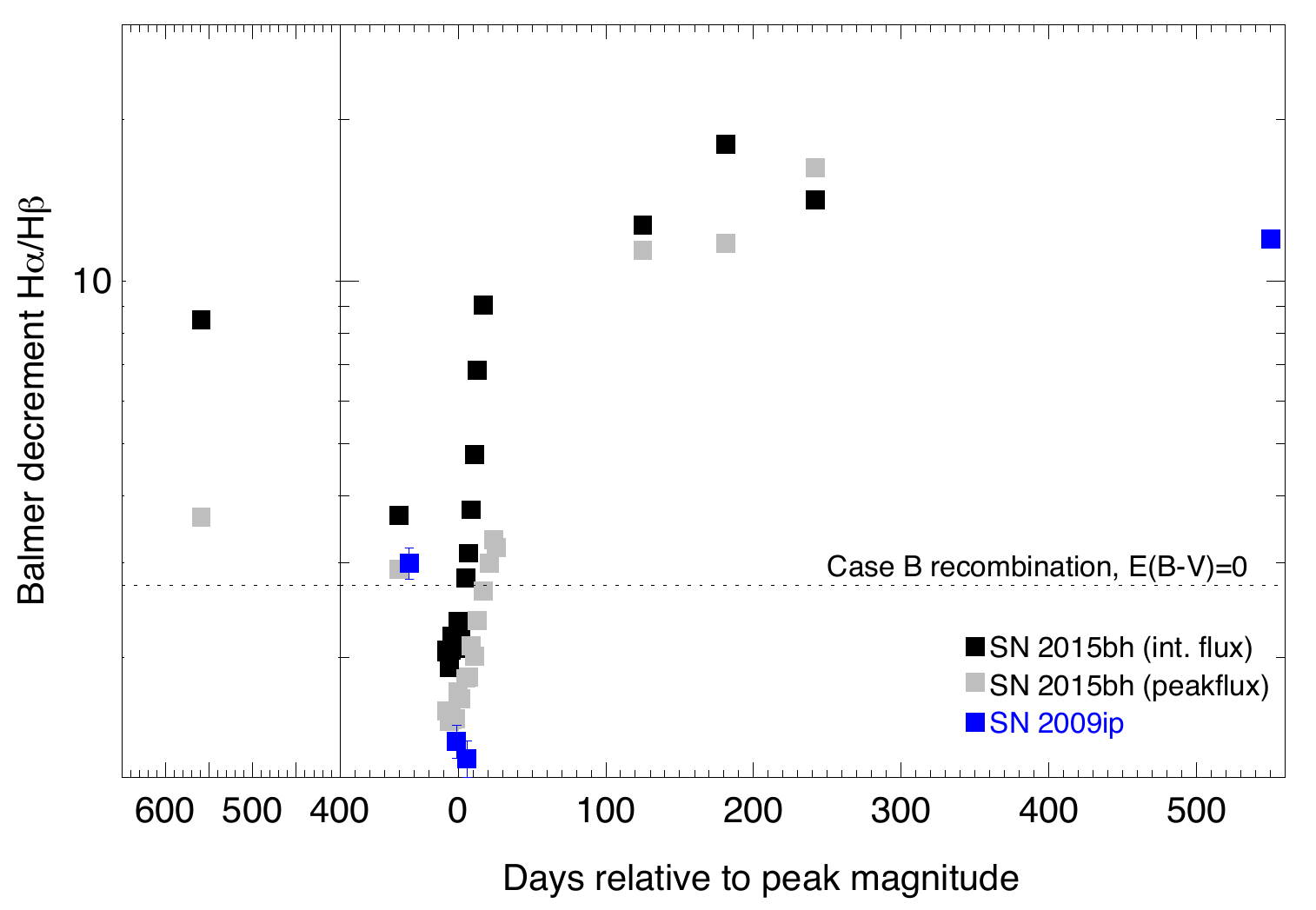}
      \caption{Balmer decrement evolution using total integrated and peak fluxes of H$\alpha$ and H$\beta$ and comparison with SN 2009ip (data from \citealt{2014AJ....147...23L}, who uses peak flux values, and \citealt{2015arXiv150206033F}).}
         \label{Fig:EWs}
   \end{figure}

\subsection{A possible geometry of the event}\label{sec:geometry}
\begin{figure}
   \centering
   \includegraphics[width=\columnwidth]{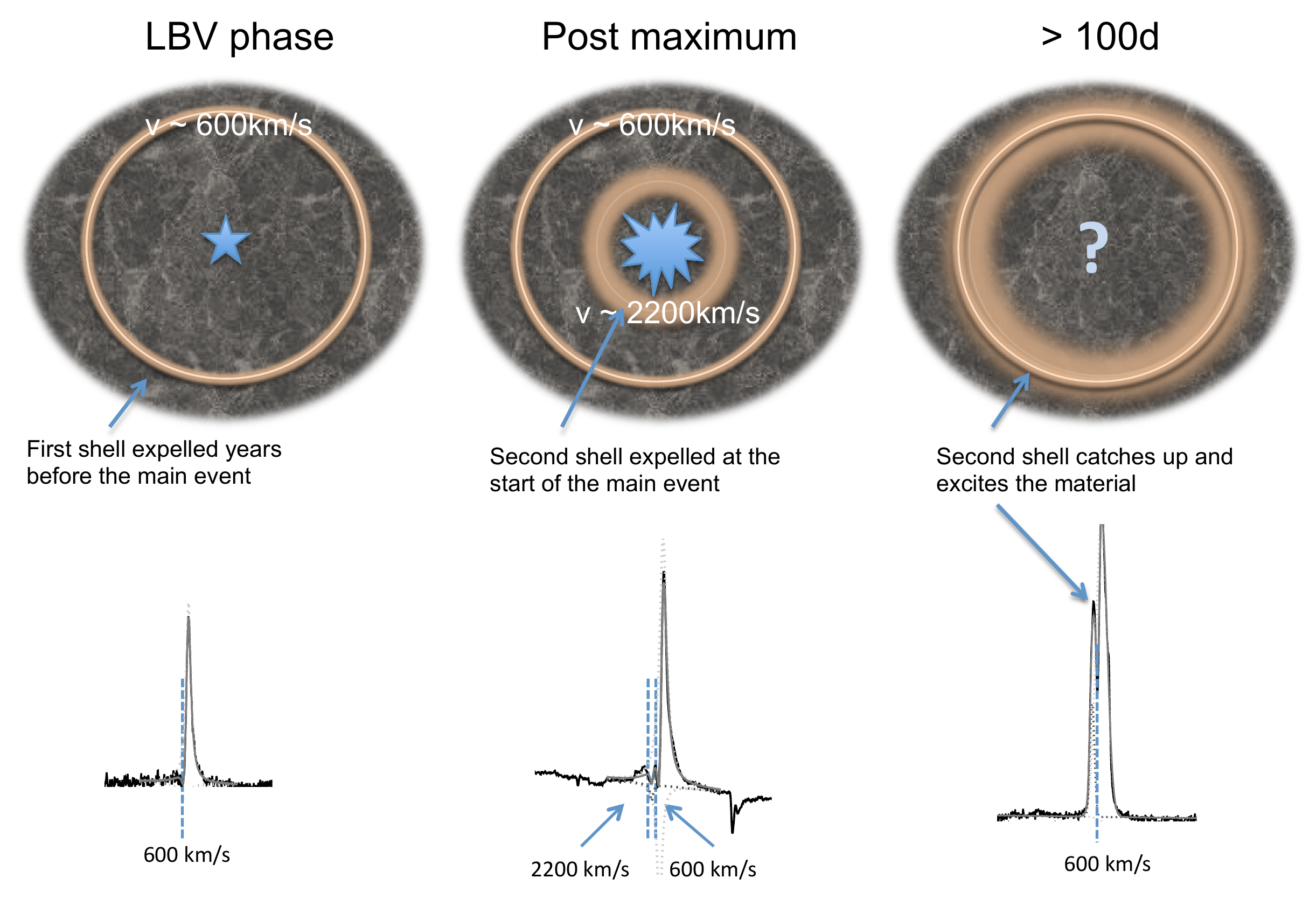}
      \caption{Scetch of the event during the LBV phase, after maximum brightness and at late times ($>$ 120d) together with the corresponding profiles of H$\alpha$. The LBV is embedded in a dense CSM indicated by the grey shaded area. During the LBV phase, only one absorption component is present while post maximum, after the emission starts to fade, a second component becomes visible in all emission lines which we therefore attribute to a shocked region associated to the shell ejection or SN explosion. At late times, the second component shows up in emission, either implying that the second shell has interacted with CSM (probably the earlier ejected shell) or that is has become optically thin.}
         \label{Fig:scetch}
   \end{figure}

The evolution of the spectral lines give us some clues as to what happened and how it happened (see Fig.\ref{Fig:scetch}) The narrow component both in absorption and emission remains rather unaffected since the LBV outburst in November 2013. This implies that the material must have been so far from the star that it remained fairly unaffected even during the main event. We associate this to a thin shell at a large distance from the star moving with a moderate speed of v~$\sim$~--700km\,s$^{-1}$ that was expelled many years back in the past.

Already during the precursor a broad absorption component with v~$\sim$\,--2000\,km\,s$^{-1}$ appears and stays up to the observational gap while becoming more pronounced. At late times, $>$126\,d of the main event, this absorption component is replaced by a second component in emission at the same velocity. One possible scenario is that this second component is associated with a shell expelled at the time of the precursor. At late times this v~$\sim$\,--2000\,km\,s$^{-1}$ shell catches up with the v\,$\sim$\,--700km\,s$^{-1}$ shell and shock excitation leads to an emission component at --2000~km\,s$^{-1}$. Assuming that the second shell was expelled around --50\,d, at $\sim$100\,d when the fast emission component appears, the shell would have reached a distance of $\sim$\,2.5$\times$10$^{15}$cm, which is at the same order of magnitude as the extrapolation of the BB radius evolution (see Fig.~\ref{Fig:Tevol}) out to 100\,d. Alternatively, the material becomes optically thin, with the material that had first shown up in absorption now appearing as emission component. 

SN 2009ip is thought to have a complicated geometry. Based on polarization measurements \cite{2014MNRAS.442.1166M} propose a toroidal CSM with which the ejecta interact. The explosion or main event itself is also asymmetric but orthogonal to the toroidal CSM implying a bipolar shape of the photosphere of the SN or ejected shell at the main event.  A similar conclusion was reached by \cite{2014AJ....147...23L} who propose a disk-like geometry of the ejecta as the more likely option considering the observed (low) Balmer decrement around maximum which would require very special conditions in a shell-like structure.

SN 2015bh has a similar behavior of the Balmer decrement with values below 2.73 during maximum albeit less extreme than SN 2009ip. One interesting feature is the presence of a small additional emission component in the red wing of H$\alpha$ at late times with a velocity shift of $\sim$ +2000~km\,s$^{-1}$ which would be the receding counterpart of the second emission component showing up in the late time spectra. A curious case is SN 1998S which showed a triple-peaked H$\alpha$ profile at late times, which could imply a similar scenario but very asymmetric or bipolar ejecta or a different viewing angle.

\subsection{Modeling of the main event}
\begin{figure}
   \centering
   \includegraphics[width=7.2cm]{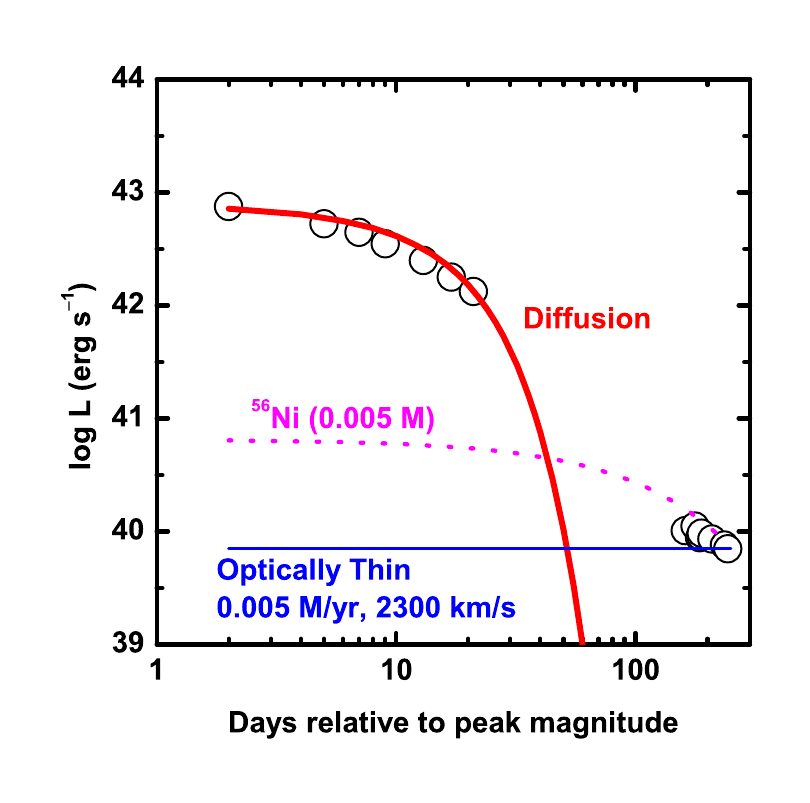}
      \includegraphics[width=7.6cm]{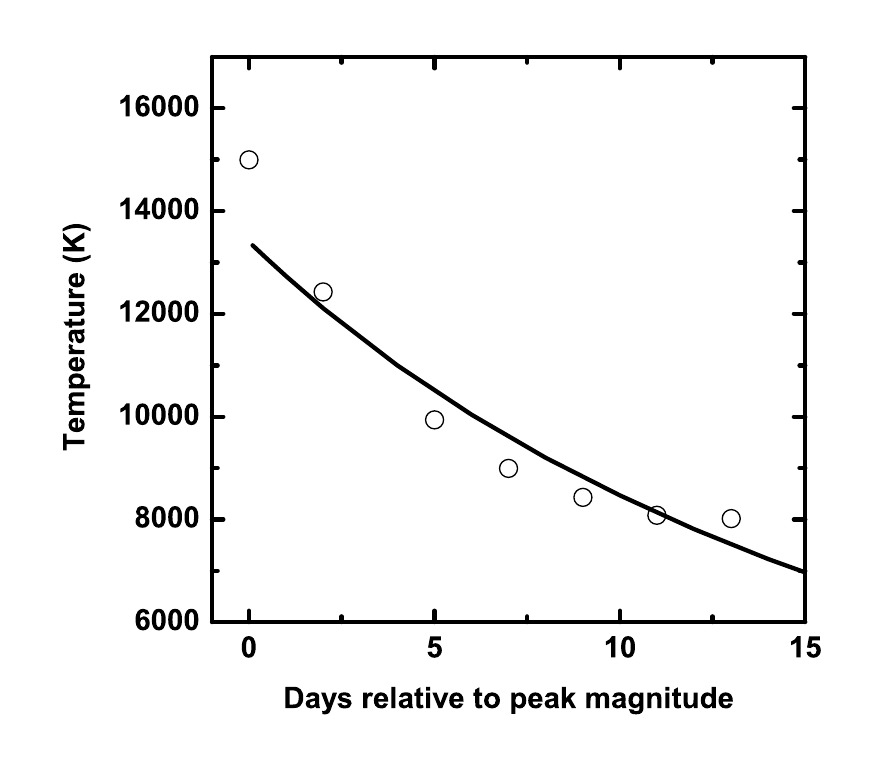}
      \caption{Model fit to the bolometric LC and temperature evolution.}
         \label{Fig:model}
\end{figure}

We estimated the physical quantities related to the 2015B event, using the bolometric LC and other observational constraints. Note that the quantities given below are only order-of-magnitude estimates. First, we fit the bolometric LC within the context of the ``shell-diffusion'' model \citep[e.g.][]{1982ApJ...253..785A, 1996sunu.book.....A, 2007ApJ...671L..17S}. This model assumes that the ejecta crashes into the dense CSM (a shell in this case), creating an optically thick, dense shell. This shell then cools down through adiabatic losses while the optical emission is created through diffusion. The fit to the bolometric LC in the first month after the peak is shown in Fig. \ref{Fig:model}. For the model in this figure, the diffusion time scale is set to be 20 days. This requires $M_0 \sim 0.5 M_{\odot}$ as the total mass of the ejecta and the dense shell, if the distance to the shell (created by the most recent eruption) is $R_0 \sim 5 \times 10^{14}$ cm (see below). The LC decline is well reproduced by the diffusion model, which supports the main physical assumption of the model. We note that adopting the similar formalism for SN 2009ip, a similar value for $M_0$ ($0.1\,M_{\odot}$ or larger) has been derived \citep{2014ApJ...780...21M,2015arXiv150206033F, 2015ApJ...803L..26M}.

The physical parameters are further constrained by the temperature evolution, assuming a BB emission. Figure \ref{Fig:model} shows the temporal evolution of the photospheric temperature based on the diffusion model (Arnett 1996), as compared to the one obtained from the BB fit to the SED sequence. This particular (best) model has $M_0 = 0.005\,M_{\odot}$, $R_0=4.5\times10^{14}$\,cm, and $v=2,300$\,km\,s$^{-1}$. The last quantity is the characteristic velocity of the shell after the crash, i.e. the sum of the ejecta and the dense shell. This estimate shows a discrepancy as compared to that through the diffusion time scale. However, our model is very simplified and therefore we simply regard this as an uncertainty intrinsic to the model presented here. Therefore, our estimates are $M_0 \sim 0.005-0.5~M_{\odot}$, $R_0 \sim 5\times 10^{14}$~cm, and $E \sim 10^{47}-10^{49}$ erg. A more detailed model will be presented in a forthcoming paper. We do note, however, that the upper limits on the ejecta properties we derive are robust, $M_0 \lesssim 0.5~M_{\odot}$ and $E \lesssim 10^{49}$~erg. This shell radius is fully consistent with the idea that it was created by the pre-burst event just before 2015B, about  two months before. With a velocity of $\sim 2,000$ km s$^{-1}$, the shell would be at $\sim 1 \times 10^{15}$ cm, roughly consistent with the value we estimated. With these parameters, not only the initial temperature, but also its evolution are reproduce nicely.

Finally, the late-time LC is interesting as it shows nearly constant luminosity after $\sim$100 days. This suggests that the ejecta (i.e., ejecta plus the dense shell initially at $R_0$) are not decelerated any more \citep[see also][]{2015ApJ...803L..26M}. Adopting $2,300$ km s$^{-1}$ as the constant shell velocity (i.e., free-expansion) expanding within a CSM (distributed outside the dense shell at $R_0$, as $r^{-2}$ for simplicity), we estimate that the required mass loss rate in $\sim$1--2 years before the 2015A event (assuming a mass loss velocity of $800$ km s$^{-1}$) is quite high, $\sim$ 0.005~M$_{\odot}$ yr$^{-1}$. If this would have been created by a steady-state mass loss, then the CSM mass swept up by the ejecta (plus the dense shell inside this CSM) would be $0.008~ M_{\odot}$. Given the need for free expansion ($M_0 > 0.008~M_{\odot}$), this favors the large value for $M_{0}$ within our estimate through the diffusion time scale and temperature evolution. It is likely that this CSM would indeed be better described by another shell than the smooth CSM, and in this case the required CSM mass would be reduced. The required distance to this second CSM shell would be then $\sim 4 \times 10^{15}$ cm. This suggests an eruption activity about one and half year before the 2015A event, corresponding to the 2013A event.

Under the assumption that SN 2015bh was indeed a CC event, we can calculate the produced $^{56}$Ni mass in the 2015A event. We derive an upper limit of $\sim 0.003 M_{\odot}$ if we assume the full deposition of energy from $\gamma$-rays into the SN ejecta. This small amount also agrees with our energy estimate, $\sim$10$^{48}-10^{49}$ erg, which is not large enough to lead to explosive nucleosynthesis of $^{56}$Ni. The very small upper limit could also mean that we actually do not have a CC event here. 

\section{The progenitor star and its evolution}\label{sec:prog}
Deriving any conclusion about the progenitor star properties is hampered by the fact that the LBV has been in outburst as far back as 1994 until its possible SN explosion. The two epochs with the lowest luminosities during the 21 years of observations were in early February 2008 and a year later in January/February 2009: at both times it had an absolute magnitude of $\sim$ --8 in R-band. Whether these epochs correspond to a genuinely ``quiescent'' state is debatable, infact, it is more likely that the LBV was in a hot phase of the S Dor variability cycle or another type of outburst state in these epochs. 

\begin{figure*}
   \centering
   \includegraphics[width=15cm]{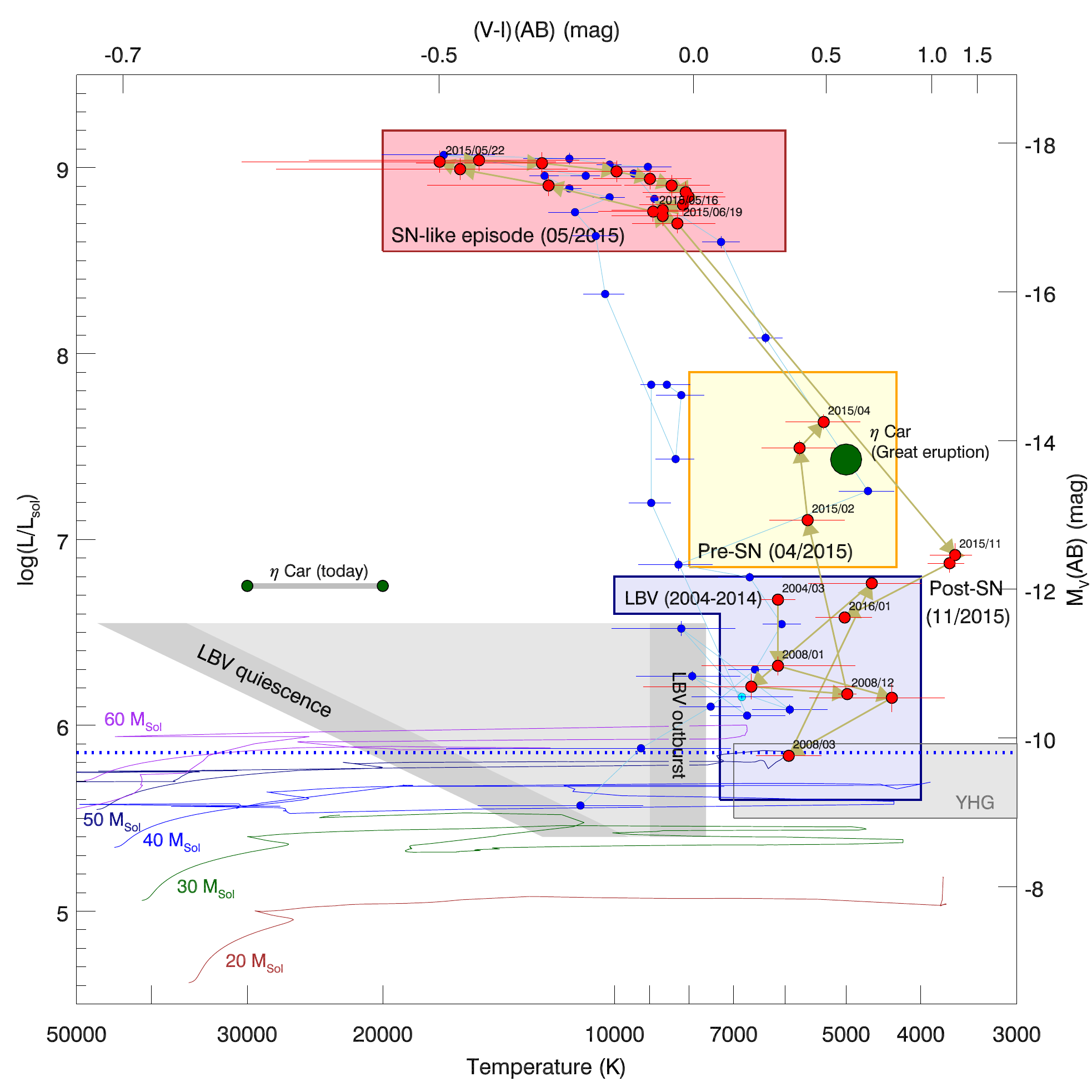}
      \caption{HR diagram showing the evolution of SN\,2015bh during the early outbursts, throughout the precursor, main event and decline (red dots). Blue dots show the same for SN\,2009ip from 2009 to 2015 \citep[data from][]{2013ApJ...767....1P, 2015arXiv150206033F}. The last observation at --9.1\,mag ($R$-band) is below the proposed progenitor level from HST observations in 1999 (blue dotted line, no color information is available for this epoch). We also plot data for Eta Carinae \citep{2012Natur.482..375R} today and during the main eruption as well as the range of LBVs between quiescence and outburst.}
         \label{Fig:HR}
   \end{figure*}

In Fig. \ref{Fig:HR} we plot the evolution of the object throughout the different states of outbursts (``LBV phase''), the precursor, the main 2015B event, and the subsequent post-maximum decline. We take the luminosity in R-band while the temperature is derived from V--I colors. We also plot stellar evolution tracks from the Geneva models \citep{2012A&A...537A.146E} for progenitors of 20 -- 60 M$_\odot$ at solar metallicity (although the true metallicity is closer to half solar, see Sect.~\ref{sec:env}) and with zero rotation. Furthermore we obtain data from the literature on Eta Carinae \citep{2012Natur.482..375R} and SN 2009ip \citep{2013ApJ...767....1P, 2015arXiv150206033F}.

\subsection{Evolution through the HR diagram}
Since 2004, the LBV was redder than S-Dor like LBVs during outburst, even during the periods of lowest luminosity luminosity: such states can be considered (perhaps falsely) as quiescent. LBVs typically have much higher temperatures and bluer colors during quiescence and become redder during outburst while remaining constant in bolometric luminosity. Our object never reaches the colors of LBVs in quiescent state, and instead are rather similar than the color of Eta Carinae during its giant eruption in the 1800s, albeit at a lower luminosity.  

A possible interpretation is that the outbursts observed from 2004 to 2015 more resemble the giant outburst of Eta Car than S-Dor type variability, followed by a much more violent outburst or SN that Eta Car has not experienced (yet). The difference in luminosity might simply be a consequence of the higher mass of Eta Car compared to the LBV progenitor of SN 2015bh. Another possibility is that some of the observations, e.g. in April 2008, actually do correspond to the quiescent state of the star. In this case, the object would be a cool, yellow hyper giant (YHG) such as $\rho$ Cas which are probably evolved red giant stars. YHGs also show variations in luminosity and occasionally suffer larger eruptions, e.g. every $\sim$ 50 years for $\rho$ Cas; hence they are usually enshrouded by dust and material from their frequent eruptions. YHGs have been classified through the ``Keenan-Smolinski'' criterion which requires one or several broad components in H$\alpha$ and broad absorption lines. In 2013, SN\,2015bh showed emission lines with narrow P-Cygni profiles which is a characteristic of LBVs in outburst. However, YHGs and LBVs can not always be clearly distinguished based purely on their spectra. 

After the main event the object returned to the same luminosity as it had during the LBV outbursts, but showing lower temperatures. Also, the emission that we observed at 200 days post maximum might still be a radiating shell from the shock interaction and not from the photosphere of the star in case it is still alive. From its current position in the HR diagram the object can either return to a bluer and fainter state (as is the case for SN 2009ip, see below) or simply continue cooling and dropping out of the HR diagram as predicted for a terminal explosion. Future observations in a few years from now might be able to make a more concrete statement on the final fate of the star.

\subsection{The possible progenitor star}
The progenitor type and mass also depends on the question whether we only observed the LBV in outburst since 2004 or whether some of the observations reflect the quiescent state. Assuming that we did observe occasionally a quiescent state of a yellow hyper giant (YHG), stellar evolution tracks indicate that the progenitor is a very massive star of $\sim$ 50 M$_\odot$. In case the star is in an Eta Car like giant eruption, the true quiescent state of the star should be lower and hence its mass; though how much lower is difficult to estimate. Considering the spectra during outburst resemble those of LBVs rather than YHGs, the progenitor in (unobserved) quiescence may be bluer and possibly fainter, though this is only speculation. The progenitor of the Type IIb SN in fact was proposed to be a possible YHG but with lower wind velocities than SN 2015bh \citep{2014A&A...572L..11G}. Another case could be SN 1998S \citep[see a recent paper by][]{} whose spectra shows a lot of similarities with SN 2015bh, where indications for pre-explosion mass losses have been found. 

Only a handful of SN IIn progenitors have been identified in the past: SN 2005gl was proposed to be an LBV of $>$50 M$_\odot$ \citep{2007ApJ...656..372G}, seen to have exploded due to its absence in post-explosion \emph{HST} images \citep{2009Natur.458..865G} which first put in doubt the traditional model that LBVs would not directly explode as SNe. SN 2010jl, a very luminous SN IIn in a dusty region, had a $>$30 M$_\odot$ progenitor \citep{2011ApJ...732...63S} . An LBV progenitor was also suggested for the ultra luminous SN IIn SN 1978K \citep{2007ApJ...669.1130S} although no direct pre-explosion imaging is available for this SN. The progenitor of SN 2009ip is poorly constrained as pre imaging was only obtained in one filter in 1999 and consistent with a very massive star of at least 60 M$_\odot$\citep{2011ApJ...732...32F, 2014MNRAS.438.1191S}. Also indirect evidence for mass-loss episodes such as complex absorption line profiles \citep[e.g.][]{2008A&A...483L..47T} indicate the possibility of LBVs as progenitor stars. SNe IIn show in general a large variety in explosion properties \citep{2013A&A...555A..10T} and hence might come from a range of progenitors masses.

SN impostors might be an even more diverse group: Events such as SN 2008S and 2008 NGC 300-OT1 seem to originate from cool, red, so-called ``extreme AGB'' stars of relatively low masses ($\sim$ 9\,M$_\odot$;  \citealt{0004-637X-741-1-37}). UGC 2773 OT2009-1 was identified as an LBV with a mass of $>$25\,M$_\odot$ \citep{2011ApJ...732...32F} but showed strong absorption lines during its outburst event, contrary to SN 2015bh, and is possibly an Eta Car giant outburst analogue \citep{2016MNRAS.455.3546S}. Several other SN impostors have also been determined to be LBV stars, e.g. SN 2002kg \citep{2006astro.ph..3025V}, SN 2000ch \citep{2004PASP..116..326W} and SN 1961V \citep{1989ApJ...342..908G, 2011ApJ...737...76K}.

\subsection{SN 2009ip in the HR diagram and late-time observations}

SN 2009ip has been observed now for more than 3 years since the main event in 2012 \citep{2015arXiv150206033F}. During the outburst phase from its discovery in 2009 to the main event in 2012 it occupied a very similar region as SN\,2015bh did between 2004--2014 (see Fig.\ref{Fig:HR}). Also the subsequent precursor and the main event are found in very similar locations in both cases, although our object might have shown a bit bluer colors during the main event \citep[data taken from][]{2013ApJ...767....1P}. After the main event, SN 2009ip went back to the region of the precursor and, after a short turn towards cooler temperatures, now turned hotter and bluer again \citep[data taken from][]{2015arXiv150206033F}. In 2014 it reached nearly the same magnitude ($\sim$0.1 mag brighter) than in the HST pre-explosion images (the color cannot be determined, see above). 

On Nov. 30, 2015 we obtained observations in $g',r','i'$ with OSIRIS/GTC of SN 2009ip \citep{2015ATel.8417....1T} which revealed that the object has now dropped to a luminosity below the level in the \emph{HST} images in 1999 \citep{2011ApJ...732...32F, 2014MNRAS.438.1191S} and has become again bluer (see Fig. \ref{Fig:HR}). Several authors \citep[e.g.][]{2014ApJ...787..163G, 2014ApJ...780...21M} have already noted this blue-turn but most agree that a BB fit might not be appropriate at this stage and that the temperature was actually not increasing. We also obtained spectra of SN 2009ip on Dec. 23 using LDSS3C on the 6.5m Magellan-Clay telescope (Las Campanas Observatory, Chile, see Fig.\ref{Fig:2009ip} in the Appendix). The spectrum shows little continuum and a huge, nearly symmetric H$\alpha$ line with an EW of $>$3000~\AA{} well as weak emission lines of Fe{\sc ii} and NaD possibly blended with HeI. H$\beta$ is only marginally detected implying a Balmer decrement of $\gtrsim$50. In comparison to spectra of SN 2009ip from May 6, 2014 (as part of the PESSTO Data release), the hydrogen lines except for H$\alpha$ have basically vanished while the Fe\,II emission lines are still visible. 

The new observations below the ``progenitor level'' have several possible implications: (1) SN 2009ip was in a quiescent state in 1999, but after the 2012 events, it lost a large amount of mass such that its future quiescent state will be below the level it had before the events, hence the progenitor mass estimate would still be valid. (2) The star was already in outburst in 1999 and its quiescent state (and hence the progenitor mass) is actually lower. Considering that the LBV progenitor of SN 2015bh had definitely been in outburst state 21 years before the main event, this scenario seems likely also for SN 2009ip 13 years before the main event. We consider the second scenario more likely, implying that there is a class of objects that shows Eta-Car like outbursts lasting for more than a decade before an even more violent event that Eta Car has not shown (yet) or an actual terminal explosion.

The late time spectra also seem to be different from ``typical'' SN IIn such as SN 1998S \citep{2012MNRAS.424.2659M} or SN 2010jl \citep{2014ApJ...797..118F} which, at those times, still show the characteristic, broad nebular emission lines. This could imply that the dense ejecta visible at late times are actually not present here. In either case, if the progenitor of SN 2009ip survived those events as a kind of ``zombie'' star, it would reach its current end-state in only a few more years from now. Deep high-resolution imaging might settle the question about the nature of SN 2009ip soon. The spectra indicate that there is furthermore very little emission left from the CSM interaction at this stage, facilitating those observations.

\section{The progenitor environment}\label{sec:env}

Studying H$\alpha$ maps of SN hosts, \cite{2012MNRAS.424.1372A} argued that SNe~IIn were less associated with HII regions than what would be expected if they (all) originated from very massive stars such as LBVs, and that they are even less associated with SF regions than the low-mass progenitors of SN type II-P. \cite{2015MNRAS.447..598S} studied the association of LBV stars with O-stars and found that LBV stars are usually isolated from star formation, which lead them to propose that they are mass gainers in binary systems rather than very massive stars. \cite{2015arXiv150504719T} studied the environments of 60 interacting transients including SNe IIn, Ibn and SN impostors, and found that SN impostors and long lasting (SN 1988Z-like) SNe~IIn are typically found at lower metallicities than SNe~IIn that resemble SN~1998S. They proposed that the former class originates from LBV progenitors, while the latter can result from the explosion of Red Supergiants (RSG) in a CSM. \citep{2014MNRAS.441.2230H} studied the distribution of SN IIn and SN impostors within their hosts, finding that they are not associated to ongoing SF traced by H$\alpha$ emission. They also find that ``SN 2008S-like'' impostors (those with low mass progenitors) fall on regions without H$\alpha$ emission while ``Eta-Car like'' impostors (possible LBV progenitors) do have underlying H$\alpha$ emission. 

Our H$\alpha$ tunable filter map of NGC~2770 indicates that SN 2015bh is probably not lying within a bright SF region. It is, however, within the spiral arm where SN 2007uy exploded within a row of smaller SF regions spanning from the region of SN 2007uy along the spiral arm. A definite answer as to whether it is associated to a SF region or not is hard to give since any image available of NGC 2770 is contaminated by the LBV and there might well be a faint underlying SF region. However, in contrast to SN 2009ip, it is clearly hosted within the galaxy in one of the star-forming outer spiral arms. 

\begin{figure}
   \centering
   \includegraphics[width=\columnwidth]{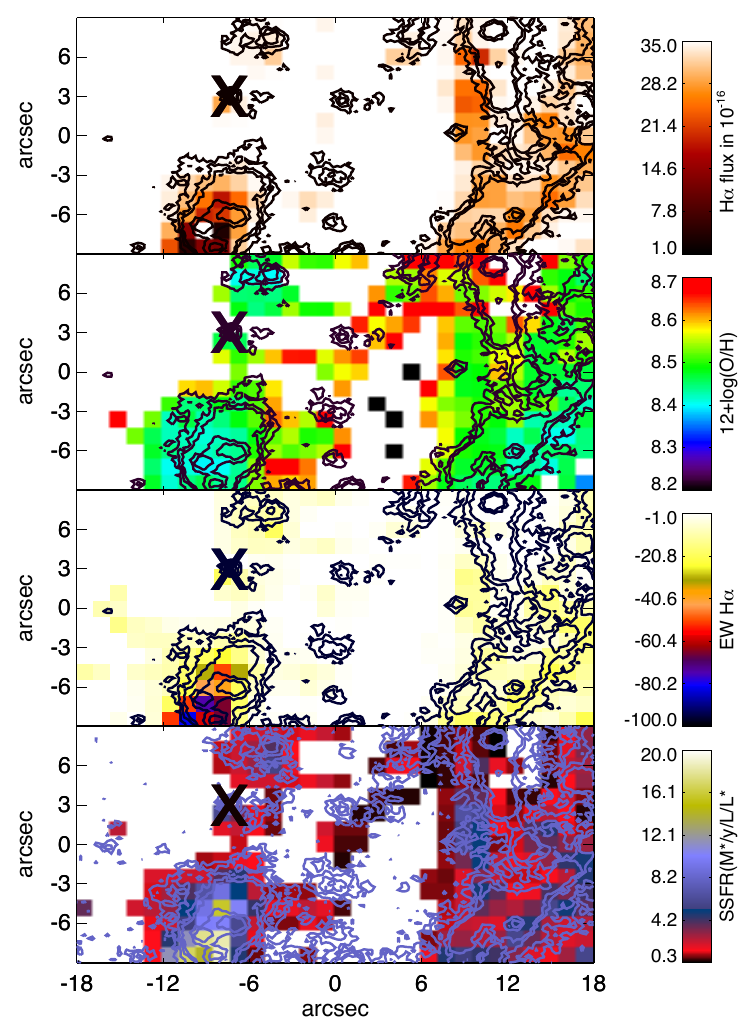}
      \caption{The galactic environment of SN\,2015bh  in H$\alpha$, metallicity, H$\alpha$ EW and specific SFR as observed with OSIRIS/GTC and drift-scan spectroscopy. The area covers about 5$\times$2.5 kpc. Note that the properties at the explosion site cannot be considered as the data are contaminated by emission from the LBV during outburst, marked by a cross.}
         \label{Fig:3Dplots}
   \end{figure}

A map of the region around the LBV site corresponding to half of the data cube constructed from the drift-scan spectroscopy dataset ($\sim$5$\times$2.5\,kpc) is shown in Fig. \ref{Fig:3Dplots}. Since we cannot determine properties of the environment at the actual location of SN 2015bh (in all available spectra we have contamination from the LBV star), we study the properties at the SF regions in the same spiral arm SE and NW of the LBV location. The metallicity is derived from the N2 parameter in the calibration of \cite{2013A&A...559A.114M}, the SSFR is obtained from scaling the H$\alpha$ luminosity with the magnitude in the region of 3980--4920\,\AA{} of the spectrum, corresponding to the coverage of a B-band broad-band filter.

The HII region to the SE is part of a cluster of several HII regions in one of which SN 2007uy was located (which is just outside the range covered by the driftscan cube). The SE and NW SF region have both a metallicity of 12+log(O/H)$=$8.46, H$\alpha$ EWs of $\sim$ --40 and --10\AA{} and a specific SFR (SSFR) of $\sim$ 3.1 and 2.1 M$_\odot$/y/L/L*. The metallicity of both neighboring SF regions is below the mean of the metallicity gradient in NGC 2770 by about 0.06 dex and similarly low as for SN 2007uy and SN 1999eh (\citet{2009ApJ...698.1307T}, Th\"one et al. in prep.). Compared to other SN IIn site metallicities, SN\,2015bh  lies at the mean of the distribution \citep{2015arXiv150504719T}. The EW is rather low which would imply an age of the SF region of more than 7 and 10 Myr or the lifetime of a 25 or 17 M$_\odot$ star for the two regions respectively. In line with this, the SSFR at both sites is rather low both within the galaxy and lower than any of the SN Ib sites (Th\"one et al. in prep.). The fact that, in particular the region NW of the LBV site is neither very young nor particularly star-forming is supported by the absence of H$\beta$ in emission, which is the case in many HII regions within NGC 2770, the HII cluster around SN 2007uy, however does show H$\beta$ in emission. 

\section{Comparison with SN impostors and SN type IIn}\label{sec:comp}

     \begin{figure}
   \centering
   \includegraphics[width=\columnwidth]{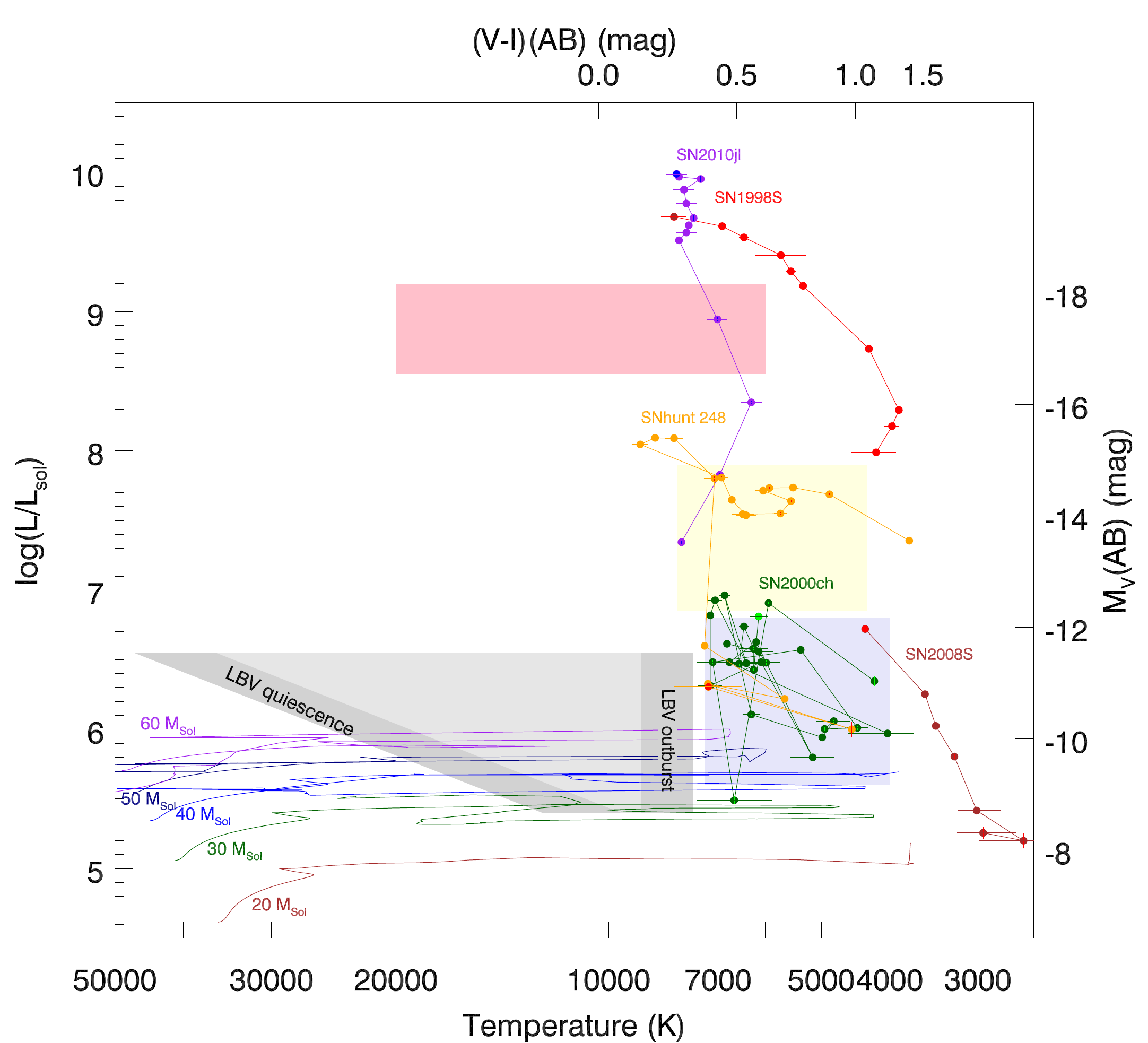}
      \caption{Evolution throughout the HR diagram for several other events: The SN impostors SN 2008S, SN 2000ch and SNhunt248 and the type IIn SNe 1998S and 2010jl. SN 2000ch occupies a very similar region to SN 2009ip and SN 2015bh in the LBV phase and could be a similar star in this pre-main event period. SN 2008S was always redder than SN 2009ip and SN 2015bh at much lower luminosities. SN 2010jl and SN 1998S reached higher magnitudes than the main events for SN 2009ip and SN 2015bh. SNhunt248 occupies a similar region during the outburst phase but has a lower luminosity at peak. The small late blue turn corresponds to a bump in the LC past maximum. }
         \label{Fig:HRothers}
   \end{figure}
   
 \subsection{SN 2015bh and SN type IIn}

  \begin{figure*}
   \centering
   \includegraphics[width=12cm]{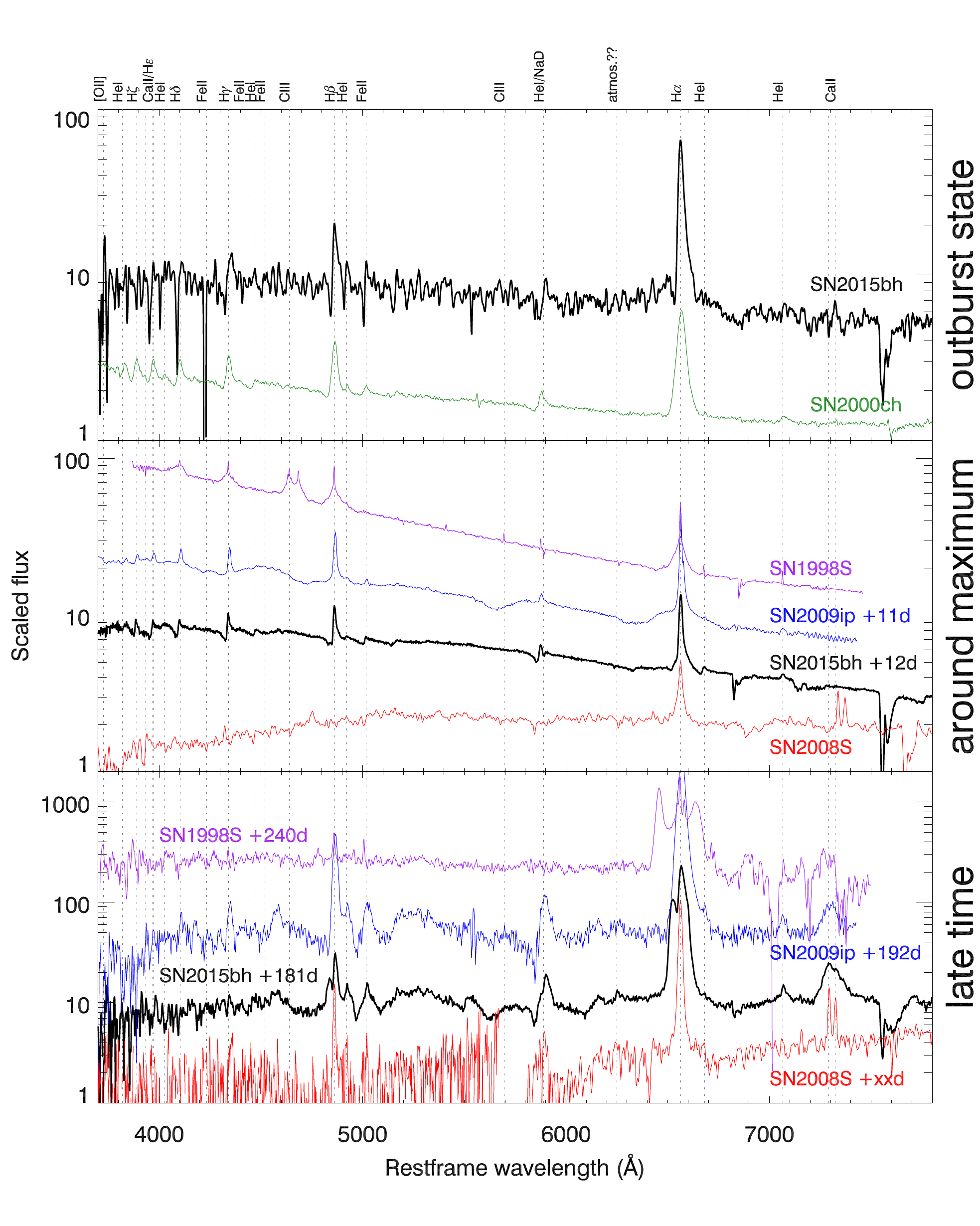}
      \caption{Comparison between the spectra of SN 2015bh (red), SN 2009ip (black), SN 1998S (a genuine type IIn), SN 2008S (an impostor) and SN  at  different epochs. Unfortunately, there are no spectra publicly available on SN 2009ip during the outburst phase, the same applies to the very similar event SNhunt 248 \citep{2015A&A...581L...4K, 2015MNRAS.447.1922M}}
         \label{Fig:specsIIn}
   \end{figure*} 
 
SNe IIn are actually a rather mixed bag of events: as the classification is made through signs of shock interaction of some sort of ejecta with a thick CSM, the central engine can be very different and includes examples of hydrogen-poor events (called SN Ibn) and even thermonuclear explosions (SN Ian). Here we only concentrate on H-rich CC SNe IIn. Circumstellar interaction has also been one of the proposed mechanisms for super-luminous SNe (SLSNe) \citep{2012Sci...337..927G}. SN 2006gy \citep{2007ApJ...666.1116S} and SN 2010jl  \citep{2011ApJ...730...34S} with peak magnitudes of --22\,mag and --20.5\,mag, respectively, were classified as a SN IIn but had a much higher luminosity than typical SNe IIn. It seems that SNe IIn span a wide range of luminosities from SLSNe to more ``normal'' peak luminosities or maybe even down to sub-luminous events. Most SN IIn show indications for massive progenitors, hence it might be a question of mass-loss through winds, eruptions and the structure of the CSM which defines the luminosity of the final explosion.

Several classes have been established in the past of ``real'' Type IIn SNe: SN 1994W, 1998S, 1988Z-like and ``generic'' SNe IIn \citep[see e.g.][]{2013A&A...555A..10T, 2012ApJ...744...10K}. 1994W-like IIn (other members are e.g. SN 2009kn \citep{2012MNRAS.424..855K} and SN 2006bo \citep{2013A&A...555A..10T}) have a characteristic plateau post maximum and have also been called SN IIP-n, which is not the case for SN\,2015bh. They have, however, a similar set of emission lines as we observe here, but in contrast they never show the dramatic change to broad late-time emission lines as we observed. SN 1998S and 2008fq \citep{2013A&A...555A..10T} showed broad line profiles at $>$50\,d and the Balmer line profiles actually resemble very much those of SN 2009ip. SN 1988Z-like SN such as SN 2006jd and SN 2006qq \citep{2013A&A...555A..10T} have very asymmetric emission lines with a broad component that can even be detached from the narrow component, and a suppression in the red wing due to dust, which we do not observe here. Spectra of some type IIn SNe are shown in Fig. \ref{Fig:specsIIn}.

Some SNe IIn have also shown a precursor directly prior to the main event, as we observed for SN 2015bh, well-studied examples of which are SN 2010mc \citep{2013Natur.494...65O} and LSQ13zm \citep{2016MNRAS.tmp..462T}. Several authors have suggested that those precursors are directly related to the main event/core-collapse or possibly even being the core-collapse itself followed by a strong interaction with the CSM producing the main event. Outbursts in the years prior to explosion such as observed for SN 2009ip and SN 2015bh have only been detected in a few cases and unfortunately consist of mostly single detections and no spectroscopy. Examples are SN 2011ht \citep{2013ApJ...779L...8F}, SN 2006jc \citep{2007ApJ...657L.105F}, a SN Ibn and PTF12cxj \citep{2014ApJ...789..104O}.\cite{2014ApJ...789..104O} argue that since those outbursts and precursors are so common in type IIn, they should be directly related to the end state of very massive stars.

\subsection{SN 2015bh and SN impostors}
SN impostors are classified as having lower peak luminosities ($>$--16\,mag) than actual SNe \citep{2011MNRAS.415..773S} but are otherwise probably separated into different types of eruptions stemming from very different processes. Rather than their peak luminosity, the shape of the LCs and the duration of the outburst point to different types of events within the SN impostor class. Several authors have tried to put SN impostors into different sub-classes based on various properties \citep[see e.g.][]{2011MNRAS.415..773S, 2012ApJ...758..142K, 2014MNRAS.441.2230H}, but the nature of those transients remains unclear. As a rough distinction we can divide SN impostors into three classes:

{\it Large outbursts (Eta Car or S-Dor type)}. This class of impostors shows lightcurves with very broad peaks of only a few magnitudes above the ``quiescent'' state that can last several years and with rather erratic behavior, the prototype of which are the eruptions of Eta Carinae in 1843 and 1890. They usually show narrow emission lines similar to SNe IIn but usually without high velocity components or major absorption lines. Other members of this class are SN 2002kg/V37 in NGC 2403 \citep{2005A&A...429L..13W}, an LBV outburst ,and U2773-OT \citep{2016MNRAS.455.3546S}, a possible Eta Car giant eruption analogue. HD 5980 in the SMC is another interesting event in this respect, however,  this system is thought to be a binary consisting of an LBV and a Wolf-Rayet (WR) star, the latter of which is a variable star which changed to a B-supergiant during outburst \citep{2010AJ....139.2600K}. In all of these cases, a star is clearly present after the outburst and hence those are genuine impostors. 

{\it Short-term outbursts}. This class comprises shorter eruptions (timescale of a few weeks) that do not belong to the type of outbursts describe above, but are longer than micro variations of massive stars ($<$0.5 mag). Their spectra are similar to LBV outbursts showing narrow emission lines. In this class are SN 2000ch \citep{2004PASP..116..326W} (and its subsequent outbursts in 2008--2009 \citet{2010MNRAS.408..181P} ) as well as the pre-explosion outbursts of SN 1964V, SN 2009ip, SNhunt 248 and SN 2015bh. All of those short outbursts lead to a CC type IIn SN or a hyper-eruption except for SN 2000ch, which we will describe further below (Sect. \ref{Sect:newclass}). There might be many more of those outbursts but they are often missed due to their faintness and the low cadence of current surveys. 

{\it Impostors with SN-type LCs}. This class is probably the most difficult to distinguish from CC SNe IIn and some of them might actually be real CC events. Their LCs resemble ``genuine'' SNe with fast rises and slow decays that are either linear or have an early plateau similar to type IIP SNe but with peak luminosities lower than for typical type IIn SNe although there is a continuum of peak luminosities and any cut-off is rather arbitrary. The spectra of those events often look similar to type IIn SNe, some even showing high velocity components indicative of outflows more typical of SNe, hence a distinction on the spectra is often also hard to make. The definite classification of a SN impostor is a surviving star, but even this is still disputed for many SN impostors.

There are numerous examples in this class and probably a range of progenitors. SN 2008S-like impostors (which includes SN 2008S, NGC300-OT and maybe SN 2010dn \citep{2011MNRAS.415..773S}) have relatively faint and red progenitors \citep{0004-637X-741-1-37} and occupy a very different region in the HR diagram (see Fig. \ref{Fig:HRothers}). However, recent observations \citep{2015arXiv151107393A} reveal that those two objects are now fainter than the progenitor level in the IR. If they were indeed SN impostors, the surviving star has to be enshrouded by significant amounts of dust corresponding to $>$1\,M$_\odot$ of ejected mass, which seems unlikely for a $<$14\,M$_\odot$ progenitor. Other impostors such as SN 1997bs \citep{2000PASP..112.1532V}, SN 2002bu \citep{2012ApJ...760...20S} and SN 2007sv \citep{2015MNRAS.447..117T} could have either an LBV or a red giant progenitor but also here the final fate of the star is sometimes disputed \citep[see e.g. the discussion on SN 1997bs][]{2000PASP..112.1532V, 2002PASP..114..403L}. There are many more events in this class but not always with very good observational coverage \citep[for further examples see e.g.][]{2011MNRAS.415..773S}.

Finally, bright events with peak luminosities between SN impostors and Type IIn SNe are highly disputed about their nature and final state. The prototype of those is SN 1961V whose final fate is still highly discussed \citep{2011ApJ...737...76K, 2012ApJ...746..179V}. More recently, SNhunt 248 \citep{2015A&A...581L...4K}  seemed to show a very similar behavior but at lower peak luminosity. SN 2009ip and SN 2015bh would also be members of this class if we consider them impostors. All of these events have shown precursors which might either be an intrinsic property or simply an observational bias as corresponding precursors of the fainter SN-like class are very likely to be missed.

\subsection{A new class of event?}\label{Sect:newclass}
Looking at the events described above there is evidence that SN 2015bh and a few other events (SN 2009ip, SNhunt 248 and possibly SN 1961V) are members of a new class that either might or might not be core-collapse SNe, but that nevertheless share a number of common properties: (1) Short-term outbursts of $\sim$2 mag during at  least a few decades before the main event (2) A precursor 40-80 days before the main event that is brighter than the outbursts before. (3) A main event with a SN-like LC (fast rise, linear decay and late flattening), (4) LBV type spectra with strong, narrow Balmer, He and Fe lines with (narrow) P-Cygni profiles during the outbursts and until after the maximum of the main event. After this a more complex absorption profile develops some of which is likely associated to the material expelled during the main event. (5) Despite the LBV type spectra, the objects are found red wards of the LBV outburst region during the pre-eruption outburst phase, in the region of YHGs. During the precursor, the temperature stays at similar values, and only the luminosity increases. The main event finally resembles a normal SN evolution moving first towards higher temperatures and subsequently cools down. At late times there is another shift to higher temperatures observed. (6) There is no indication for major dust production in the main event. 

There seems to be some variation in the maximum luminosity, the shape of the LC and the shape of the emission and absorption lines of the object. The latter two are likely only a product of slightly different CSM and different viewing angles. The variance in maximum luminosity might be either due to the mass of the progenitor, the amount of mass expelled (if it was not a core-collapse event) or also due to the density of the CSM. Also, the location inside the host seems to vary to a certain degree: SN 2009ip was far away from the main disk while SN 2015bh and SNhunt 248 were in an outer spiral arm but not within any large SF region. 

Of all the possible members of this class, {\it SN 2009ip} is the most similar to SN 2015bh, in fact, it almost looks like a carbon copy of it. The LC in all stages looks surprisingly similar (see Fig.~\ref{Fig:LC}), with SN 2009ip showing a small shoulder in the decline after the main event, but a similar feature could have been missed in SN 2015bh due to the sun gap in the observations. The absolute luminosities of both events also match rather well, the fact that the LBV is fainter is mostly due to foreground extinction in the NGC 2770 while SN 2009ip exploded at a large distance from the center and hence has negligible local extinction. The spectral features and evolution are equally similar (see Fig.~\ref{Fig:specsIIn}) with SN 2009ip showing a somewhat richer absorption structure and no double profile at late times. Both events also behave extremely similar in their path through the HR diagram. 

Ample observations of {\it SNhunt 248} including a decade-spanning LC have recently be published by \cite{2015A&A...581L...4K}, but see also an earlier paper by \cite{2015MNRAS.447.1922M}. They note a series of small outbursts  over more than a decade and a ``triple peak'' during the main events 2014A, B, and C. Event 2014A looks very much like the pre-explosion bumps of 2009ip, 2015bh and other IIns, 2014B is the main event and 2014C likely just a rebrightening due to some further interaction with the CSM, similar to the smaller bump or shoulder observed in the declining LC of 2009ip main event. The spectra show a very similar set of lines including the rich forest of Fe lines, but somewhat weaker He lines than 2009ip and 2015bh. H$\alpha$ consists of a broad (> 1100 km\,s$^{-1}$) and a narrow component as well as at least two absorption components. \cite{2015MNRAS.447.1922M} note that the object was in the YHG region during the pre-explosion outbursts but that it does not show any IR excess as common for the more dusty YHGs, which is exactly where SN 2015bh and 2009ip are found during outburst. The only major difference between SNhunt 248 and SN 2015bh is the peak luminosity which is $\sim$2.5 mag lower than for SN 2015\,bh. Nevertheless, there are so many similarities that SNhunt 248 can be considered a member of this class, but either with a lower mass progenitor or a weaker interaction with the CSM. 

Finally the famous {\it SN 1961V} might be another member of this class. \cite{1964ApJ...139..514Z} searched archival plates for pre-explosion observations and found that it showed frequent variations over 20 years before the main event in 1961. In fact, the magnitude of the variations and the final outburst are similar in scale to SN 2009ip and SN 2015bh. As there is no color information and no spectra available of 1961V, a final assessment of this event is hard to make. \cite{2012ApJ...746..179V} also noted that the proposed ``quiescent state'' of SN 1961V in the 30ies might have actually been an outburst state, similar to the outburst phase of SN 2009ip. As mentioned above, even more than 50 years after the event there is a large discussion about whether this star exploded or whether there is a survivor star present, partly also due to the lack of a very precise location of the event. In any case, if there is a survivor, it has not shown any further outbursts. 

{\it SN 2010mc} has been frequently suggested as another carbon copy of SN 2009ip \citep{2014MNRAS.438.1191S, 2014ApJ...780...21M} showing a similar pre-explosion event, total energy release, LC and spectral properties. However, there are no records of outbursts in the years before the explosion. Also, the spectra of SN 2010mc are somewhat different, e.g. they do not show Fe II and the absorption components of the P-Cygni profiles are less pronounces. Unfortunately, there are only a few spectra of SN 2010mc available around the peak of the main event when e.g. the Fe II lines are also nearly absent in SN 2009ip and SN 2015bh, and one late-time spectrum \citep{2014MNRAS.438.1191S} that resembles our latest spectrum of SN 2009ip but has equally little features. SN 2010mc might or might not be a member of the same class but lacks observations for a final conclusion. It is very likely that this new class has more members than the 3-4 established above, but lack observations to meet all of the 6 criteria that we used to define the ``zombie-star'' class. \cite{2014ApJ...789..104O} determined that about 98\% of all SN IIn show precursors in the years before explosion, so it is likely that many more examples of this class exist.

A conspicuous candidate in this respect is {\it SN 2000ch}, a (clear) impostor, with several outbursts observed in 2000 \citep{2004PASP..116..326W} and 2008-2009 \citep{2010MNRAS.408..181P}. In the HR diagram (see Fig. \ref{Fig:HRothers}) the behavior of SN 2000ch is strikingly similar to the pre-explosion activity of SN 2009ip and SN 2015bh. As already noted by \cite{2011MNRAS.415..773S}, SN 2000ch is an object that should be carefully monitored and we suggest that it might experience a similar destiny as SN 2015bh, although over what time-frame is difficult to say.

\subsection{The ultimate question: was it a terminal explosion or is the star still "alive"?}
SN 2015bh is obviously not a unique event, but it is still unclear why the events described above are so similar: The objects in this class are probably LBVs that are starting to experience some major instabilities and mass loss during the end of their lives, or evolutionary stage. While smaller variations in LBV stars such S-Dor outbursts are probably driven by pulsations causing irregular changes in radius, temperature and luminosity but do not seem supersede the Eddington limit \citep{2011ApJ...736...46G}, giant eruptions need much more violent processes and probably involve deeper layers of the star. The physical processes for such major shedding of the outer layers of the star is still not fully explained by theoretical models. The two main physical mechanisms currently favored ar super-eddington winds or explosions \citep{2014ARA&A..52..487S}. Super-eddington winds rely on an increase of the starÕs bolometric luminosity for reasons that are not entirely clear, but further drive a massive outflow \citep[see e.g.]{1994PASP..106.1025H, 1999PASP..111.1124H, 2006ApJ...645L..45S}. In the explosion model, an underlying mechanism needs to inject significants amounts of energies in deep stellar layers. Proposed mechanisms have invoked envelope instabilities \citep[see e.g.]{1993MNRAS.263..375G, 2012ARA&A..50..107L}, unstable nuclear burning \citep{2014ApJ...785...82S}, pulsational pair instability \citep{2007Natur.450..390W}, explosive shell burning instabilities \citep{2010MNRAS.405.2113D}, wave-driven mass loss \citep{2007ApJ...667..448M, 2013MNRAS.430.1736S}, stellar collisions or mergers in  binary systems \citep{2010NewAR..54...39P, 2014ApJ...785...82S}, and extreme proximity to the Eddington limit \citep{2014A&A...564A..83M}.

The current picture is that LBV stars are a transition state of a massive O-star on its way to becoming a WR star. There are two problems with this picture: First LBV stars have been observed to directly explode as CC-SNe. Second, there is no mechanism securely established on how to remove the outer H and probably He layers of the star. Steady mass loss by winds of $\sim$ 10$^{-4}$M$_\odot$/yr might not to be sufficient to shed enough mass in a short enough time. It has been proposed in the past that a series of giant eruptions could be a way to remove the outer layers \citep{2006ApJ...645L..45S} in order to create a WR star. Other possible mechanisms are binary interactions or mergers. However, both main-sequence and especially post-MS time-averaged mass-loss rates remain poorly understood, which directly impact our understanding of whether massive stars explode as LBVs or WRs, and their underlying properties. An interesting example of this is SN 2006jc (a SN Ibn) that had an outburst 1.5 years before the final explosion. H and He had different velocity structures so they were possibly ejected at different eruptions. \cite{2007ApJ...657L.105F} speculate that the progenitor had to be a WNR star that recently ($\sim$ years) transitioned from an LBV phase (which ejected the H envelope) to a WR phase. 

A single giant eruption might not be sufficient to transition an LBV star to a WR star, e.g. Eta Carinae survived already at least two of these eruptions and still returned to its LBV state. Detailed spectroscopic modeling \citep{2001ApJ...553..837H, 2012MNRAS.423.1623G} indicates that it has retained a significant hydrogen envelope after those events. The possible survivors of the SN 2015bh of ``zombie star'' class, however, as well as some other SN-like SN impostors have not shown any further variability like the one exhibited before the main event \citep[see e.g. late time data of SN 2009ip][]{2015arXiv150206033F}. Also, in all cases with a potential survivor, the star has to be less luminous and probably hotter than the progenitor star before the main event. So, either, those stars actually did explode as a type IIn SN and hence we are just observing deeper layers of the SN or those surviving ``zombie'' stars have actually become WR stars, explaining their luminosity below the progenitor level and the bluer colors. If the latter is the case, such a hyper-eruption as we observed for SN 2015bh and others might be an attractive way to create WR stars from an LBV star within on a few years to decades. If this is the case we would expect to observe a late time spectrum showing high excitation lines of C, N or O depending in the WR star type. However, the lower luminosity of those stars makes these observations very challenging (e.g. the potential survivor stars of SN 1961V have magnitude of I$\sim$24 mag, see \citealt{2002PASP..114..700V, 2012ApJ...746..179V}). A last option is that the star lost only part of its H-envelope and still remains as an LBV, albeit at lower luminosity. While it is currently very difficult to determine whether those zombie-star events are terminal or not, they might offer an appealing explanation of very fast mass loss to produce a WR star.

\section{Conclusions}\label{sec:concl}
We monitored the progenitor of SN 2015bh/iPTF13evf/SNhunt 275 over more than 20 years from a phase of LBV-like eruptions and instabilities to a possible terminal explosion or hyper-eruption in May 2015. The key observations and results are the following: 

\begin{enumerate}
\item The progenitor of SN\,2015bh has been in an active state at least 21 years prior to the main event with variations of $\sim$ 2\,mag together with small color/temperature changes. \cite{2014ApJ...789..104O} found that SNe IIn frequently show precursors several months to years prior to the (possible) explosion and that this might be a feature inherent to very massive stars at the end of their lives. SN\,2009ip, the possible twin of SN\,2015bh , showed similar precursors \citep{2013ApJ...767....1P} and several known SN impostors might be actually precursors of future SNe, with the best example being SN\,2000ch \citep{2004PASP..116..326W}.
\item The main event (the core-collapse or hyper-eruption) was preceded by a precursor event peaking about 30 days before the main event, something which might be the case for as much as 50\% of all SNe IIn \citep{2014ApJ...789..104O}. It has been argued that those precursors might be the actual explosion and the main event the interaction of the ejected photosphere with the CSM. 
\item The main event had a peak magnitude of M$_V$ = --17.5 which is higher than the peak luminosity of SN impostors, but on the lower end of ``true'' SN IIn peak luminosities \citep[see e.g.][]{2013A&A...555A..10T}. There are now several events known such as SN 1961V and SN 2009ip that had similar peak luminosities but their terminal nature is still disputed. The total energy release of the main event was $\sim$ 1.8$\times$10$^{49}$\,erg. 
\item The shape of the LC of the main event of SN 2015bh resembles a typical SN LC with fast raise and linear decay before flattening out at late times ($>$120\,d). The behavior of the LC can be modeled with the following quantities: The emission of the main event is created by a shell of $\lesssim$0.5~M$_\odot$ at $\sim$ 5$\times$10$^{14}$cm crashing into a thick CSM and the optical emission is created by diffusion. The shallow decay at late times indicates that the eject are now optically thin and that only a very small amount ($\sim$0.005\,M$_\odot$) of Ni$^{56}$ had been created.
\item The object was not detected in radio during the LBV phase with an upper limit of 8.7$\times$10$^{26}$ erg\,s$^{-1}$\,Hz$^{-1}$ at 4.8\,GHz. The main event was not detected in X-rays at any epoch down to limits of 3$\times$10$^{40}$\,erg\,s$^{-1}$. SNe IIn have been frequently detected in X-ray with peak luminosities up to 3$\times$10$^{41}$\,erg\,s$^{-1}$ in the case of SN 2010jl \citep{2015ApJ...810...32C} but have been elusive at radio wavelengths. Nevertheless, SNe IIn with luminosities above our limit during the LBV phase have been detected \citep[see e.g.][]{2014ApJ...780...21M} albeit radio LCs are peaking at late time ($>$50 days post maximum). SN 2009ip showed some X-ray emission at peak with low significance at $\sim$2.5$\times$10$^{39}$\,erg\,s$^{-1}$ and was detected at 9\,GHz with $\sim$5$\times$10$^{25}$  erg\,s$^{-1}$\,Hz$^{-1}$ \citep{2014ApJ...780...21M}.
\item The UV-nIR SED is fitted reasonably well with an expanding and cooling BB starting from a temperature of 1.7$\times$10$^4$ K and a radius of $\sim$ 2$\times$10$^{14}$\,cm at --7 days down to 6000 K and 2$\times$10$^{14}$\,cm 26 days after maximum. The SED shows a small excess in flux both in the UV and in the nIR, the latter could be due to dust. However, there is no significant increase in the nIR excess and no obvious sign of dust production at least until 180 days post maximum.
\item The spectra during the LBV phase show the typical strong Balmer and Fe II emission lines with P-Cygni profiles as observed for other LBVs in outburst and SN impostors. Little change happens during the pre-cursor event apart from a second absorption component in the P-Cygni profile appearing. The spectra during the main event are typical for a type IIn SN and has a similar set of emission lines as SN 2009kn (a SN 1994W-like SN IIn), but showing late broad nebular-type emission profiles such as seen for SN 1998S. At the onset of the main event, the absorption part of the lines basically disappears and reappears as a clear double profile about 2 weeks post maximum. At about a month past maximum the spectra resemble again those before the main event, although with a more complex absorption structure. Past 100 days after maximum, the spectrum shows nebular-type emission and a few new lines such as [CaII] start to appear. 
\item The narrow peak of the Balmer emission lines changes very little over time, the same applies for the lower velocity absorption component at $\sim$ --700 km~s$^{-1}$. The second component at v$\sim$ --2000 km~s$^{-1}$ might already be visible during the precursor, shows up as broad absorption component past maximum and finally in emission past 100 days. A small additional emission is also visible in the red wing of the H$\alpha$ line at roughly +2000 km~s$^{-1}$ indicative of an asymmetric geometry of the second shell.
\item The Balmer decrement and H$\alpha$ EW show large changes during the main event. The Balmer decrement first drops to values below Case B recombination and then reaches values of more than 10. The first can be observed in a disk-like geometry where H$\alpha$ and H$\beta$ are coming from different regions, the high values at late times are characteristic of nebulae that become optically thin. A similar behavior is observed for the EW reaching values of $>$1000\AA{} past 100 days, also observed in a number of type IIn SNe. 
\item The ejecta of SN 2015bh likely have an assymmetric and disk-like geometry. Post 100 days we observe an interaction between an earlier shell and the one probably ejected during the precursor event. The distance of the faster shell at this time is roughly consistent with the extrapolated radius of the BB at that time.
\item The long time scale of observations available for SN 2015bh allowed us to trace the evolution of the star throughout the HR diagram for almost two decades. Despite its spectra resembling an LBV in outburst, the star has actually never been observed in the region typically spanned for LBVs before and during outbursts, but red wards of the instability strip. The pre-explosion event only shifts the event to higher luminosities while remaining at similar temperatures. During the main event SN 2015bh behaved similar to other SNe type IIn. At late times, SN 2015bh started to turn again towards hotter temperatures, which can also be observed in other SNe IIn, and might simply reflect that we start to see deeper into the ejecta. SN 2009ip shows a very similar path across all its evolution, but has by now (1176 days post maximum) dropped below its progenitor level and showing high temperatures. 
\item SN\,2015bh is situated within the outer spiral arm of NGC 2770, close to several smaller SF regions and possible within a small SF region itself, although this is difficult to determine as any historical observation of NGC 2770 is contaminated by progenitor outbursts. The metallicity of surrounding regions are 12+log(O/H)$=$8.46 dex and the SSFRs 2.1 -- 3.1 M$_\odot$/y/L/L*, which are both close to the average in NGC 2770 at that distance from the centre. The age of the population of neither of them is very young and hence does not directly support a massive star origin. 
\end{enumerate}

Despite the wealth of data collected for SN\,2015bh  it is difficult at this time to make a definite statement whether the object actually exploded or not. However, it appears that SN  2015bh represents a new class of events that share a lot of properties with SN IIn but that could equally be some new kind of very luminous outburst of a massive star at the end of its life or transitionary state, more massive than an Eta Car type eruption. In case the star survives such a hyper-eruption, this has probably a large impact on the further life (and final death) of such a ``zombie'' star. SN 2009ip has not shown any further outbursts since the main event in 2012 despite having reached a luminosity below the outburst phase, hence if the star is still alive, the 2012 event has significantly altered its properties. A very speculative idea is that those hyper-eruptions could be one way for massive stars to shed most of their envelope in very short time in contrast to a long-time mass loss via winds, hence those surviving ``zombie stars'' would actually be WR stars.

Very few SNe have been monitored and available pre-explosion data for more than 10 years which is the case here (and at a shorter timescale for SN 2009ip). There might be many more of these objects that have either been missed so far because of the faintness of the outbursts before the main event, the objects being hidden by extinction and/or the lack of frequent monitoring. If at least a part of SN impostors are a similar kind of event before a hyper-eruption such as SN 2000ch, such objects should be monitored on a regular basis. Future deep high-cadence large-scale surveys such as the LSST will allow us to detect and follow many more of these objects over a long time to determine whether such eruptions might be common in the end stages of massive stars or whether they can actually survive those hyper eruptions.

\section*{Acknowledgements}

CT and AdUP acknowledge support from Ram\'on y Cajal fellowships and AYA2014-58381P. C. G. acknowledges funding provided by the Danish Agency for Science and Technology and Innovation. SSchulze and FEB acknowledge support from Basal-CATA PFB-06/2007, Iniciativa Cientifica Milenio grant P10-064-F (Millennium Center for Supernova Science), by Project IC120009 "Millennium Institute of Astrophysics (MAS)" of Iniciativa Cient\'ifica Milenio del Ministerio de Econom\'ia, Fomento y Turismo de Chile, with input from "Fondo de Innovaci\'{o}n para la Competitividad, del Ministerio de Econom\'{\i}a, Fomento y Turismo de Chile" and CONICYT-Chile FONDECYT 1101024. ZC is funded by a Project Grant from the Icelandic Research Fund. K.W. acknowledges funding from the UK STFC. The work by K.M. is partly supported by Japan Society for the Promotion of Science KAKENHI Grant (26800100), by World Premier International Research Center Initiative, MEXT, Japan and the JSPS Open Partnership Bilateral Joint Research Project between Japan and Chile.

Ground based observations were collected at the Gran Telescopio Canarias (GTC), the 0.9 and 1.5m telescopes of Sierra Nevada Observatory, the 2.2m and 3.5m telescopes of Calar Alto Observatory, the 0.6m REM telescope, and the 0.5m telescope of the University of Leicester. Further observations were obtained with the UVOT and XRT telescopes onboard {\it Swift}. We thank the staff at the different observatories for performing the observations. KW thanks R. McErlean for making the transients programme at the 0.5m UoL observatory possible. We thank M. Phillips for devoting observing time on the Magellan Baade Telescope to obtain a late spectrum of SN 2009ip. This paper makes use of data obtained from the Isaac Newton Group Archive which is maintained as part of the CASU Astronomical Data Centre at the Institute of Astronomy, Cambridge. This work made use of the ``Weizmann Supernova repository'' (WiseREP) \url{http://wiserep.weizmann.ac.il} \citep{wiserep}.

\bibliographystyle{mnras}  
\bibliography{NGC2770_LBV_astroph}

\newpage



\appendix

\begin{table}
\caption{Log of the photometric observations}
\begin{tabular}{c c c c}

\label{Tab:photlog}\\
\hline\hline
Date 		& Telescope & Band & Magnitude \\
\hline
19940409		& JKT		& $B$			& $>$21.88		\\
19940409		& JKT		& $V$			& 21.39$\pm$0.34	\\
19951224		& INT		& $R$			& 21.66$\pm$0.28	\\
19970503		& WHT		& $B$			& $>$20.62		\\
19970603		& WHT		& $B$			& $>$20.70		\\
19970503		& WHT		& $R$			& $>$19.89		\\
20020322.881	& INT		& $R$			& 20.61$\pm$0.25	\\
20040318.141	& SDSS		& $u'$			& 23.02$\pm$0.46 	\\
20040318.141	& SDSS		& $g'$			& 21.17$\pm$0.04 	\\
20040318.141	& SDSS		& $r'$			& 20.74$\pm$0.03 	\\
20040318.141	& SDSS		& $i'$			& 20.58$\pm$0.05 	\\
20040318.141	& SDSS		& $z'$			& 20.35$\pm$0.14 	\\			
20080110.004	& ALFOSC		& $R$			& 21.75$\pm$0.16	\\
20080110.016	& ALFOSC		& $V$			& 22.06$\pm$0.18	\\
20080110.028	& ALFOSC		& $I$				& 21.47$\pm$0.23	\\
20080110.038	& ALFOSC		& $B$			& 22.62$\pm$0.22	\\
20080110.053	& ALFOSC		& $U$			& $>$ 23.57		\\
20080110.252	& ALFOSC		& $R$			& 21.64$\pm$0.13	\\
20080110.261	& ALFOSC		& $V$			& 22.15$\pm$0.19	\\
20080110.304	& ALFOSC		& $R$			& 20.93$\pm$0.17	\\
20080111.304	& ALFOSC		& $R$			& 21.19$\pm$0.18	\\
20080112.012 	& ALFOSC 		& $U$			& 23.29$\pm$0.24	\\
20080112.018	& ALFOSC		& $B$			& 22.70$\pm$0.22	\\
20080112.023	& ALFOSC		& $R$			& 21.72$\pm$0.16	\\
20080112.026	& ALFOSC		& $V$			& 22.21$\pm$0.20	\\
20080113.023	& ALFOSC		& $R$			& 21.78$\pm$0.15	\\
20080114.222	& ALFOSC		& $U$			& 23.82$\pm$0.37	\\
20080114.248	& ALFOSC		& $I$				& 21.45$\pm$0.18	\\
20080114.238	& ALFOSC		& $V$			& 22.49$\pm$0.28	\\
20080114.243	& ALFOSC		& $R$			& 22.12$\pm$0.22	\\
20080115.243	& ALFOSC		& $R$			& 21.21$\pm$0.19	\\
20080116.245	& ALFOSC		& $U$    			& $>$ 22.57		\\
20080116.252	& ALFOSC		& $B$			& 22.71$\pm$0.33	\\
20080116.257	& ALFOSC		& $V$			& 22.40$\pm$0.37	\\
20080116.261	& ALFOSC		& $R$			& 22.17$\pm$0.34	\\
20080116.266	& ALFOSC		& $I$				& 21.76$\pm$0.32	\\
20080117.261	& ALFOSC		& $R$			& 22.08$\pm$0.35	\\
20080125.197  	& VLT		& $B$			& $>$ 21.50		\\
20080125.200  	& VLT		& $V$			& 21.17$\pm$0.27	\\ 
20080125.201  	& VLT		& $I$				& $>$ 21.29		\\
20080125.203  	& VLT		& $R$			& $>$ 20.53		\\
20080125.899	& ALFOSC		& $B$			& $>$ 22.04		\\
20080128.045	& ALFOSC		& $B$			& 23.11$\pm$0.35	\\
20080128.052	& ALFOSC		& $V$			& 22.68$\pm$0.34	\\
20080128.067	& ALFOSC		& $R$			& 22.13$\pm$0.26	\\
20080128.071	& ALFOSC		& $I$				& $>$ 22.16		\\
20080129.067	& ALFOSC		& $R$			& 22.09$\pm$0.23	\\
20080129.209	& ALFOSC		& $V$			& $>$ 21.81		\\
20080129.217	& ALFOSC		& $R$			& $>$ 21.66		\\
20080129.220	& ALFOSC		& $I$				& $>$ 21.34		\\
20080130.215	& ALFOSC		& $R$			& $>$ 22.63		\\
20080131.062	& ALFOSC		& $U$			& 23.77$\pm$0.38	\\
20080131.076	& ALFOSC		& $B$			& 23.26$\pm$0.37	\\
20080131.083	& ALFOSC		& $V$			& $>$ 21.78		\\
20080131.086	& ALFOSC		& $R$			& 22.14$\pm$0.25	\\
20080131.089	& ALFOSC		& $I$				& $>$ 22.27		\\
20080201.009	& ALFOSC		& $U$			& $>$ 23.66		\\
20080201.014	& ALFOSC		& $B$			& 22.87$\pm$0.24	\\
20080201.016	& ALFOSC		& $V$			& 22.76$\pm$0.34	\\
20080201.017	& ALFOSC		& $R$			& 22.34$\pm$0.31	\\
20080201.019	& ALFOSC		& $I$				& $>$ 22.23		\\
20080201.086	& ALFOSC		& $R$			& 22.26$\pm$0.24	
\end{tabular}
\end{table}

\begin{table}
\caption{Log of the photometric observations - continued}
\begin{tabular}{c c c c}

\label{Tab:photlog}\\
\hline\hline
Date 		& Telescope & Band & Magnitude \\
\hline
20080202.017	& ALFOSC		& $R$			& 22.60$\pm$0.34	\\
20080203.038	& ALFOSC		& $U$			& $>$ 23.52		\\
20080203.042	& ALFOSC		& $B$			& 23.11$\pm$0.31	\\
20080203.044	& ALFOSC		& $V$			& $>$ 22.85		\\
20080203.046	& ALFOSC		& $R	$			& 22.08$\pm$0.23	\\
20080203.047	& ALFOSC		& $I$				& $>$ 22.32		\\
20080204.046	& ALFOSC		& $R$			& 22.35$\pm$0.25	\\
20080209.51 	& HST    		& $F606W$		& 22.8 			\\
20080210.044	& 2.5m NOT	& $R$			& 22.40$\pm$0.31	\\
20080210.046	& 2.5m NOT		& $I$				& 21.95$\pm$0.31	\\
20080211.044	& 2.5m NOT		& $R$			& 22.56$\pm$0.31	\\
20080218.975	& 2.5m NOT		& $B$			& 22.45$\pm$0.36	\\
20080218.982	& 2.5m NOT		& $V$			& $>$ 22.35		\\
20080218.988	& 2.5m NOT		& $R$			& $>$ 22.16		\\
20080218.993	& 2.5m NOT		& $I$				& $>$ 21.80		\\
20080222.191  	& VLT		& $B$			& 21.72$\pm$0.31	\\ 
20080222.193  	& VLT		& $V$			& $>$ 21.83		\\
20080222.195  	& VLT		& $I$			& $>$ 21.62		\\
20080222.196  	& VLT		& $R$			& $>$ 21.75		\\
20080229.078	& 2.5m NOT		& $I$				& $>$ 21.95		\\
20080229.081	& 2.5m NOT		& $U$			& $>$ 23.42		\\
20080229.097	& 2.5m NOT		& $V$			& $>$ 22.68		\\
20080229.101	&2.5m NOT		& $R$			& $>$ 22.48		\\
20080306.091  	& VLT		& $B$			& 22.82$\pm$0.36	\\
20080306.099  	& VLT		& $V$			& 22.15$\pm$0.35	\\
20080306.103  	& VLT		& $R$			& 21.88$\pm$0.36	\\
20080306.113  	& VLT		& $I$			& 21.38$\pm$0.26	\\
20080306.113	& 2.5m NOT		& $I$				& 21.06$\pm$0.23	\\
20080310.879	& 2.5m NOT		& $B$			& $>$  22.54		\\
20080310.884	& 2.5m NOT		& $V$			& 21.81$\pm$0.39	\\
20080310.887	& 2.5m NOT		& $R$			& $>$ 21.93		\\
20080310.889	& 2.5m NOT		& $I$				& $>$ 21.53		\\
20080314.061 	& VLT		& $U$			& $>$ 23.48	\\
20080314.076 	& VLT		& $B$			& $>$ 22.92	\\
20080314.082 	& VLT		& $V$			& 22.17$\pm$0.33	\\
20080314.085 	& VLT		& $I$			& 21.65$\pm$0.36	\\
20080314.087 	& VLT		& $R$			& 21.78$\pm$0.30	\\
20080317.836	& 2.5m NOT		& $B$			& $>$ 22.02		\\
20080317.855	& 2.5m NOT		& $V$			& $>$ 21.89		\\
20080317.862	& 2.5m NOT		& $R$			& $>$ 21.83		\\
20080317.867	& 2.5m NOT		& $I$				& $>$ 21.32		\\
20080330.45   	& HST    		& $F606W$   		& 21.5			\\
20080402.930	& 2.5m NOT		& $B$			& 20.81$\pm$0.05	\\
20080402.946	& 2.5m NOT		& $V$			& 20.30$\pm$0.06	\\
20080402.952	& 2.5m NOT		& $R$			& 20.36$\pm$0.09	\\
20080402.956	& 2.5m NOT		& $I$				& 20.60$\pm$0.24	\\
20080416.948	& 2.5m NOT		& $V$			& 22.11$\pm$0.25	\\
20080416.944	& 2.5m NOT		& $R$			& 22.49$\pm$0.33	\\
20080416.955	& 2.5m NOT		& $I$				& 22.05$\pm$0.30	\\
20080427.030 	& VLT		& $B$			& 22.28$\pm$0.19	\\
20080427.033 	& VLT		& $V$			& 21.90$\pm$0.30	\\
20080427.039 	& VLT		& $R$			& 21.30$\pm$0.17	\\
20080427.040 	& VLT		& $I$			& 21.06$\pm$0.19	\\
20081219.05   	& HST    		& $F505W$   		& 22.436$\pm$0.029	\\ 
20081219.05   	& HST    		& $F814W$   		& 21.586$\pm$0.024	\\
20081220.50   	& HST    		& $F450W$   		& 22.021$\pm$0.022	\\
20081220.50   	& HST    		& $F675W$   		& 20.847$\pm$0.017	\\
20090118.178  	& VLT		& $V$			& 21.76$\pm$0.25	\\
20090118.189  	& VLT		& $R$			& 21.40$\pm$0.22	\\
20090118.197  	& VLT		& $I$			& 21.15$\pm$0.24	\\
20090120.68   	& HST    		& $F450W$   		& 23.729$\pm$0.067	\\
20090120.68   	& HST    		& $F675W$   		& 22.417$\pm$0.047	\\
\end{tabular}
\end{table}

\begin{table}
\caption{Log of the photometric observations - continued}
\begin{tabular}{c c c c}
\label{Tab:photlog}\\
\hline\hline
Date 		& Telescope & Band & Magnitude \\
\hline\hline
20090219.144  	& VLT		& $V$			& 21.86$\pm$0.21	\\
20090219.169  	& VLT		& $R$			& 21.58$\pm$0.23	\\
20090219.194  	& VLT		& $I$			& 21.79$\pm$0.35	\\
20090225.57	& HST    		& $F606W$    		& 22.5			\\
20120201.026	& INT		& $R$			& 19.84$\pm$0.10	\\
20131107.164	& 10.4m GTC	& $r'$			& 22.07$\pm$0.16	\\
20131107.165	& 10.4m GTC	& $r'$			& 22.20$\pm$0.16	\\
20131111.169	& 10.4m GTC	& $r'$			& 21.92$\pm$0.12	\\
20131126.146	& 10.4m GTC	& $r'$			& 20.08$\pm$0.12	\\
20131128.133	& 10.4m GTC	& $r'$			& 20.27$\pm$0.14	\\
20150209.93   	& Asiago 		& $u'$            	  	& 21.2			\\
20150209.93 	& Asiago 		& $g'$            	 	& 20.1			\\
20150209.93    	& Asiago 		& $r'$            		& 19.5			\\
20150209.93    	& Asiago 		& $i'$            	  	& 19.4			\\
20150211.504	& UVOT/Swift	& $UVW2$		& $>$20.90		\\
20150211.504	& UVOT/Swift	& $UVM2$			& $>$20.92		\\
20150211.504	& UVOT/Swift	& $UVW1$		& $>$20.25		\\
20150211.504	& UVOT/Swift	& $U$			& $>$19.64		\\
20150211.504	& UVOT/Swift	& $B$			& $>$18.97		\\
20150211.504	& UVOT/Swift	& $V$			& $>$18.08		\\
20150218.821	& UVOT/Swift	& $UVW2$		& $>$20.95		\\
20150218.821	& UVOT/Swift	& $UVM2$			& $>$20.87		\\
20150218.821	& UVOT/Swift	& $UVW1$		& $>$20.32		\\
20150218.821	& UVOT/Swift	& $U$			& $>$19.77		\\
20150218.821	& UVOT/Swift	& $B$			& $>$19.08		\\
20150218.821	& UVOT/Swift	& $V$			& $>$18.20		\\
20150327.00    	& 10.4m GTC    	& $f723/45$ ($\sim i'$)   	& 18.46			\\
20150327.01    	& 10.4m GTC    	& $f627/24$ ($\sim r'$)   	& 18.75$\pm$0.10	\\
20150327.02    	& 10.4m GTC    	& $f458/13$ ($\sim g'$)  	& 19.13			\\
20150409.02    	& 10.4m GTC    	& $f723/45$ ($\sim i'$)   	& 18.02			\\
20150409.02    	& 10.4m GTC    	& $f627/24$ ($\sim r'$)   	& 18.48$\pm$0.10	\\
20150409.03    	& 10.4m GTC    	& $f458/13$ ($\sim g'$) 	& 18.78			\\
20150414.953  	& 10.4m GTC    	& $r'$              	& 18.18$\pm$0.05	\\
20150420.989  	& 1.5m OSN    		& $R$ 	         	& 17.43$\pm$0.03	\\
20150430.854 	& 1.5m OSN    		& $R$  	 	       	& 17.74$\pm$0.02	\\
20150506.852 	& 1.5m OSN    		& $R$	         	& 17.85$\pm$0.02	\\
20150507.857 & OMEGA2000/CAHA	&$J$		&17.52$\pm$0.03 \\
20150507.869 & OMEGA2000/CAHA	&$H$	&17.25$\pm$0.04 \\
20150507.882 & OMEGA2000/CAHA	&$Ks$	& 16.97$\pm$0.06 \\
20150514.30	& ASASSN	& $V$			& 16.78$\pm$0.2	\\ 
20150515.864 	& 0.9m OSN 	& $R $	         	& 15.86$\pm$0.02	\\
20150516.685	& UVOT/Swift	& $UVW2$		& 17.25$\pm$0.04	\\
20150516.685	& UVOT/Swift	& $UVM2$		& 16.81$\pm$0.03	\\
20150516.685	& UVOT/Swift	& $UVW1$		& 16.48$\pm$0.04	\\
20150516.685	& UVOT/Swift	& $U$			& 15.99$\pm$0.05	\\
20150516.685	& UVOT/Swift	& $B$			& 16.00$\pm$0.06	\\
20150516.685	& UVOT/Swift	& $V$			& 16.09$\pm$0.08	\\
20150516.874  	& 1.5m OSN 	& $B$		        	& 15.96$\pm$0.03	\\
20150516.877  	& 1.5m OSN 	& $V$		        	& 15.81$\pm$0.02	\\
20150516.881  	& 1.5m OSN	& $R$	         	& 15.80$\pm$0.02	\\
20150516.884 	& 1.5m OSN 	& $I$		        	& 15.87$\pm$0.03	\\
20150516.910	& 0.5m UoL	& $B$			& 15.95$\pm$0.11	\\ 
20150516.910	& 0.5m UoL	& $V$			& 15.80$\pm$0.09	\\ 
20150516.910	& 0.5m UoL	& $R$			& 15.75$\pm$0.13	\\ 
20150516.910	& 0.5m UoL	& $I$			& 15.89$\pm$0.10	\\ 
20150517.853	& UVOT/Swift	& $UVW2$		& 16.88$\pm$0.03	\\
20150517.853	& UVOT/Swift	& $UVM2$		& 16.49$\pm$0.03	\\
20150517.853	& UVOT/Swift	& $UVW1$		& 16.22$\pm$0.04	\\
20150517.853	& UVOT/Swift	& $U$			& 15.71$\pm$0.05	\\
20150517.853	& UVOT/Swift	& $B$			& 15.75$\pm$0.05	\\
20150517.853	& UVOT/Swift	& $V$			& 15.83$\pm$0.07	\\
20150517.853 	& 0.9m OSN 	& $U$		        	& 15.67$\pm$0.06	\\
\end{tabular}
\end{table}

\begin{table}
\caption{Log of the photometric observations - continued }
\begin{tabular}{c c c c}
\label{Tab:photlog}\\
\hline\hline
Date 		& Telescope & Band & Magnitude \\
\hline\hline
20150517.867 	& 0.9m OSN 	& $V$		        	& 15.45$\pm$0.05	\\
20150517.874  	& 0.9m OSN 	& $R	$	        	& 15.58$\pm$0.04	\\
20150517.881  	& 0.9m OSN 	& $I	$	        	& 15.69$\pm$0.05	\\
20150517.988	& 0.6m REM	& $g'$			& 15.63$\pm$0.04	\\
20150517.986	& 0.6m REM	& $r'$			& 15.39$\pm$0.09	\\
20150517.986	& 0.6m REM	& $i'	$		& 15.38$\pm$0.03	\\
20150517.986	& 0.6m REM	& $H$			& 16.05$\pm$0.10	\\
20150517.988	& 0.6m REM	& $z'$			& 15.38$\pm$0.04	\\
20150517.991	& 0.6m REM	& $J$			& 16.07$\pm$0.08	\\
20150518.000	& 0.6m REM	& $H$			& 16.29$\pm$0.19	\\
20150518.055	& UVOT/Swift	& $UVW2$		& 16.72$\pm$0.03	\\
20150518.055	& UVOT/Swift	& $UVM2$		& 16.46$\pm$0.10	\\
20150518.055	& UVOT/Swift	& $UVW1$		& 16.17$\pm$0.03	\\
20150518.055	& UVOT/Swift	& $U$			& 15.75$\pm$0.05	\\
20150518.055	& UVOT/Swift	& $B$			& 15.76$\pm$0.05	\\
20150518.055	& UVOT/Swift	& $V$			& 15.71$\pm$0.07	\\
20150518.895  	& 10.4m GTC    		& $r'$              	& 15.62$\pm$0.02	\\
20150519.388	& UVOT/Swift	& $UVW2$		& 16.45$\pm$0.03	\\
20150519.388	& UVOT/Swift	& $UVM2$		& 16.27$\pm$0.03	\\
20150519.388	& UVOT/Swift	& $UVW1$		& 16.01$\pm$0.03	\\
20150519.388	& UVOT/Swift	& $U$			& 15.64$\pm$0.06	\\
20150519.388	& UVOT/Swift	& $B$			& 15.47$\pm$0.06	\\
20150519.388	& UVOT/Swift	& $V$			& 15.57$\pm$0.08	\\
20150519.887  	& 10.4m GTC    		& $r' $             	& 15.56$\pm$0.03	\\
20150519.917	& 0.5m UoL	& $B$			& 15.57$\pm$0.08	\\
20150519.917	& 0.5m UoL	& $V$			& 15.47$\pm$0.09	\\
20150519.917	& 0.5m UoL	& $R$			& 15.40$\pm$0.09	\\
20150519.917	& 0.5m UoL	& $I	$		& 15.58$\pm$0.09	\\
20150519.927 	& 0.9m OSN 	& $B$		        	& 15.52$\pm$0.04	\\
20150519.931 	& 0.9m OSN 	& $V$		        	& 15.49$\pm$0.04	\\
20150519.934 	& 0.9m OSN 	& $R$		        	& 15.45$\pm$0.03	\\
20150519.986	& 0.6m REM	& $g'$			& 15.40$\pm$0.05	\\
20150519.986	& 0.6m REM	& $r'$			& 15.24$\pm$0.08	\\
20150519.986	& 0.6m REM	& $i'	$		& 15.30$\pm$0.02	\\
20150519.987	& 0.6m REM	& $H$			& 16.09$\pm$0.05	\\
20150519.988	& 0.6m REM	& $z'$			& 15.36$\pm$0.04	\\
20150520.002	& UVOT/Swift	& $UVW2$		& 16.69$\pm$0.04	\\
20150520.002	& UVOT/Swift	& $UVM2$		& 16.36$\pm$0.04	\\
20150520.002	& UVOT/Swift	& $UVW1$		& 16.09$\pm$0.04	\\
20150520.002	& UVOT/Swift	& $U$			& 15.68$\pm$0.07	\\
20150520.002	& UVOT/Swift	& $B$			& 15.58$\pm$0.07	\\
20150520.002	& UVOT/Swift	& $V$			& 15.77$\pm$0.12	\\
20150520.002	& 0.6m REM	& $g'$			& 15.49$\pm$0.03	\\
20150520.002	& 0.6m REM	& $r'$			& 15.29$\pm$0.03	\\
20150520.002	& 0.6m REM	& $i'	$		& 15.28$\pm$0.03	\\
20150520.005	& 0.6m REM	& $J$			& 15.86$\pm$0.05	\\
20150520.007	& 0.6m REM	& $H$			& 16.11$\pm$0.07	\\
20150520.129	& UVOT/Swift	& $UVW1$		& 16.14$\pm$0.04	\\
20150520.131	& UVOT/Swift	& $UVW2$		& 16.40$\pm$0.04	\\
20150520.899  	& 10.4m GTC    		& $r'$              	& 15.48$\pm$0.02	\\
20150521.059	& UVOT/Swift	& $UVW1$		& 16.04$\pm$0.06	\\
20150521.059	& UVOT/Swift	& $U$			& 15.50$\pm$0.06	\\
20150521.059	& UVOT/Swift	& $B$			& 15.42$\pm$0.04	\\
20150522.007	& 0.6m REM	& $g'$			& 15.35$\pm$0.09	\\
20150522.007	& 0.6m REM	& $r'$			& 15.39$\pm$0.12	\\
20150522.007	& 0.6m REM	& $J$			& 16.07$\pm$0.13	\\
20150522.010	& 0.6m REM	& $H$			& 16.06$\pm$0.08	\\
20150522.450	& UVOT/Swift	& $UVW2$		& 16.40$\pm$0.03	\\
20150522.450	& UVOT/Swift	& $UVM2$		& 16.27$\pm$0.04	\\
20150522.450	& UVOT/Swift	& $UVW1$		& 16.04$\pm$0.04	\\
20150522.450	& UVOT/Swift	& $U	$		& 15.46$\pm$0.08	\\
20150522.450	& UVOT/Swift	& $B$			& 15.41$\pm$0.06	\\
20150522.450	& UVOT/Swift	& $V$			& 15.46$\pm$0.08	\\
\end{tabular}
\end{table}

\begin{table}
\caption{Log of the photometric observations - continued}
\begin{tabular}{c c c c}
\label{Tab:photlog}\\
\hline\hline
Date 		& Telescope & Band & Magnitude \\
\hline\hline
20150522.921  	& 10.4m GTC    		& $r' $             	& 15.35$\pm$0.01	\\
20150523.986	& 0.6m REM	& $g'$			& 15.38$\pm$0.09	\\
20150523.986	& 0.6m REM	& $r'$			& 15.09$\pm$0.08	\\
20150523.986	& 0.6m REM	& $i'$			& 15.18$\pm$0.06	\\
20150523.986	& 0.6m REM	& $H$			& 16.02$\pm$0.04	\\
20150524.012	& 0.6m REM	& $J$			& 15.55$\pm$0.11	\\
20150524.015	& 0.6m REM	& $H$			& 15.79$\pm$0.11	\\
20150524.780	& UVOT/Swift	& $UVW2$		& 16.49$\pm$0.03	\\
20150524.780	& UVOT/Swift	& $UVM2$		& 16.33$\pm$0.03	\\
20150524.780	& UVOT/Swift	& $UVW1$		& 16.11$\pm$0.03	\\
20150524.780	& UVOT/Swift	& $U$			& 15.48$\pm$0.05	\\
20150524.780	& UVOT/Swift	& $B$			& 15.39$\pm$0.05	\\
20150524.780	& UVOT/Swift	& $V$			& 13.34$\pm$0.06	\\
20150524.888  	& 10.4m GTC    		& $r' $             	& 15.42$\pm$0.03	\\
20150524.910	& 0.5m UoL	& $B$			& 15.45$\pm$0.08	\\
20150524.910	& 0.5m UoL	& $V$			& 15.37$\pm$0.09	\\
20150524.910	& 0.5m UoL	& $R$			& 15.30$\pm$0.09	\\
20150524.910	& 0.5m UoL	& $I	$		& 15.49$\pm$0.09	\\
20150525.445	& UVOT/Swift	& $UVW2$		& 16.82$\pm$0.03	\\
20150525.445	& UVOT/Swift	& $UVM2$		& 16.49$\pm$0.03	\\
20150525.445	& UVOT/Swift	& $UVW1$		& 16.15$\pm$0.03	\\
20150525.445	& UVOT/Swift	& $U$			& 15.53$\pm$0.05	\\
20150525.445	& UVOT/Swift	& $B$			& 15.45$\pm$0.05	\\
20150525.445	& UVOT/Swift	& $V$			& 15.39$\pm$0.07	\\
20150526.312	& UVOT/Swift	& $UVW2$		& 17.03$\pm$0.03	\\
20150526.312	& UVOT/Swift	& $UVM2$		& 16.70$\pm$0.03	\\
20150526.312	& UVOT/Swift	& $UVW1$		& 16.23$\pm$0.03	\\
20150526.312	& UVOT/Swift	& $U$			& 15.59$\pm$0.05	\\
20150526.312	& UVOT/Swift	& $B$			& 15.46$\pm$0.05	\\
20150526.312	& UVOT/Swift	& $V$			& 15.41$\pm$0.06	\\
20150526.874 	& 0.9m OSN 	& $U$		        	& 15.60$\pm$0.02	\\
20150526.881  	& 0.9m OSN 	& $B$		        	& 15.47$\pm$0.02	\\
20150526.888 	& 0.9m OSN 	& $V$		        	& 15.27$\pm$0.05	\\
20150526.895  	& 0.9m OSN 	& $R$		        	& 15.14$\pm$0.07	\\
20150526.915  	& 10.4m GTC    		& $r' $             	& 15.37$\pm$0.01	\\
20150527.864  	& 0.9m OSN 	& $U$		        	& 15.67$\pm$0.04	\\
20150527.871 	& 0.9m OSN 	& $B$		        	& 15.54$\pm$0.02	\\
20150527.878 	& 0.9m OSN 	& $V$		        	& 15.31$\pm$0.03	\\
20150527.885  	& 0.9m OSN 	& $R$		        	& 15.23$\pm$0.02	\\
20150527.973	& 0.6m REM	& $g'$			& 15.58$\pm$0.10	\\
20150527.973	& 0.6m REM	& $r'$			& 15.27$\pm$0.06	\\
20150527.973	& 0.6m REM	& $i'	$		& 15.37$\pm$0.03	\\
20150527.973	& 0.6m REM	& $J$			& 15.73$\pm$0.06	\\
20150527.978	& 0.6m REM	& $H$			& 16.07$\pm$0.05	\\
20150529.101	& UVOT/Swift	& $UVW2$		& 17.47$\pm$0.05	\\
20150529.101	& UVOT/Swift	& $UVM2$		& 17.12$\pm$0.04	\\
20150529.101	& UVOT/Swift	& $UVW1$		& 16.58$\pm$0.04	\\
20150529.101	& UVOT/Swift	& $U$			& 15.80$\pm$0.06	\\
20150529.101	& UVOT/Swift	& $B$			& 15.55$\pm$0.06	\\
20150529.101	& UVOT/Swift	& $V$			& 15.68$\pm$0.09	\\
20150529.430	& UVOT/Swift	& $UVW2$		& 17.69$\pm$0.07	\\
20150529.430	& UVOT/Swift	& $UVW1$		& 16.67$\pm$0.05	\\
20150529.430	& UVOT/Swift	& $U	$		& 15.90$\pm$0.05	\\
20150529.430	& UVOT/Swift	& $B$			& 15.55$\pm$0.06	\\
20150529.909	& 10.4m GTC		& $r'$			& 15.49$\pm$0.02	\\
20150531.091	& UVOT/Swift	& $UVW2$		& 17.77$\pm$0.04	\\
20150531.091	& UVOT/Swift	& $UVM2$		& 17.40$\pm$0.04	\\
20150531.091	& UVOT/Swift	& $UVW1$		& 16.88$\pm$0.04	\\
20150531.091	& UVOT/Swift	& $U$			& 16.05$\pm$0.05	\\
20150531.091	& UVOT/Swift	& $B$			& 15.57$\pm$0.07	\\
20150531.091	& UVOT/Swift	& $V$		& 15.57$\pm$0.07	\\
20150531.879	& 10.4m GTC		& $r'$			& 15.58$\pm$0.05	\\
\end{tabular}
\end{table}

\begin{table}
\caption{Log of the photometric observations - continued}
\begin{tabular}{c c c c}
\label{Tab:photlog}\\
\hline\hline
Date 		& Telescope & Band & Magnitude \\
\hline\hline
20150602.427	& UVOT/Swift	& $UVW2$		& 18.19$\pm$0.05	\\
20150602.427	& UVOT/Swift	& $UVW1$		& 17.17$\pm$0.04	\\
20150602.427	& UVOT/Swift	& $U$			& 16.17$\pm$0.06	\\
20150602.427	& UVOT/Swift	& $B$			& 15.86$\pm$0.05	\\
20150602.427	& UVOT/Swift	& $V$			& 15.74$\pm$0.07	\\
20150603.870  	& 0.9m OSN 	& $B$		        	& 15.89$\pm$0.02	\\
20150603.888  	& 0.9m OSN 	& $U$		        	& 16.23$\pm$0.04	\\
20150603.902  	& 0.9m OSN 	& $V$		        	& 15.63$\pm$0.05	\\
20150603.909  	& 0.9m OSN 	& $R$		        	& 15.44$\pm$0.03	\\
20150603.916  	& 0.9m OSN 	& $I$			        	& 15.24$\pm$0.03	\\
20150604.862  	& 0.9m OSN 	& $U$	        	& 16.27$\pm$0.04	\\
20150604.877  	& 0.9m OSN 	& $V$		        	& 15.57$\pm$0.03	\\
20150604.884  	& 0.9m OSN 	& $R$		        	& 15.48$\pm$0.06	\\
20150604.896  	& 0.9m OSN 	& $I$		        	& 15.41$\pm$0.04	\\
20150604.926	& 0.5m UoL	& $B$			& 15.99$\pm$0.09	\\
20150604.926	& 0.5m UoL	& $R$			& 15.64$\pm$0.09	\\
20150604.926	& 0.5m UoL	& $I$			& 15.65$\pm$0.09	\\
20150606.078	& UVOT/Swift	& $UVW2$		& 19.16$\pm$0.10	\\
20150606.078	& UVOT/Swift	& $UVM2$		& 18.58$\pm$0.11	\\
20150606.078	& UVOT/Swift	& $UVW1$		& 17.98$\pm$0.10	\\
20150606.078	& UVOT/Swift	& $U$			& 16.53$\pm$0.08	\\
20150606.078	& UVOT/Swift	& $B$			& 15.94$\pm$0.07	\\
20150606.078	& UVOT/Swift	& $V$			& 15.82$\pm$0.11	\\
20150606.881	& 10.4m GTC		& $r'$			& 15.73$\pm$0.04	\\
20150607.083	& UVOT/Swift	& $UVW2$		& 19.58$\pm$0.12	\\
20150607.083	& UVOT/Swift	& $UVM2$		& 19.67$\pm$0.12	\\
20150607.083	& UVOT/Swift	& $UVW1$		& 18.28$\pm$0.06	\\
20150607.083	& UVOT/Swift	& $U$			& 16.97$\pm$0.07	\\
20150607.083	& UVOT/Swift	& $B$			& 16.18$\pm$0.06	\\
20150607.083	& UVOT/Swift	& $V$			& 15.90$\pm$0.07	\\
20150610.855  	& 0.9m OSN 	& $U$		        	& 16.76$\pm$0.04	\\
20150610.856  	& 0.9m OSN 	& $B$		        	& 16.32$\pm$0.07	\\
20150610.857  	& 0.9m OSN 	& $V$		        	& 15.90$\pm$0.03	\\
20150610.858  	& 0.9m OSN 	& $R$		        	& 15.71$\pm$0.03	\\
20150610.859  	& 0.9m OSN 	& $I$		        		& 15.68$\pm$0.03	\\
20150610.870	& UVOT/Swift	& $UVW2$	& 20.09$\pm$0.26	\\
20150610.870	& UVOT/Swift	& $UVW1$	& 18.81$\pm$0.09	\\
20150610.870	& UVOT/Swift	& $U$		& 17.04$\pm$0.07	\\
20150610.870	& UVOT/Swift	& $B$		& 16.42$\pm$0.06	\\
20150610.888	& 10.4m GTC		& $r'$		& 15.78$\pm$0.02	\\
20150612.002	& UVOT/Swift	& $UVW2$	& 20.18$\pm$0.16	\\
20150612.002	& UVOT/Swift	& $UVM2$		& 20.73$\pm$0.26	\\
20150612.002	& UVOT/Swift	& $UVW1$	& 18.94$\pm$0.10	\\
20150612.002	& UVOT/Swift	& $U$		& 17.35$\pm$0.07	\\
20150612.002	& UVOT/Swift	& $B$		& 16.52$\pm$0.06	\\
20150612.002	& UVOT/Swift	& $V$		& 15.99$\pm$0.08	\\
20150613.864  	& 0.9m OSN 	& $U$	        	& 17.11$\pm$0.05	\\
20150613.865  	& 0.9m OSN 	& $B$	        	& 16.38$\pm$0.07	\\
20150613.868  	& 0.9m OSN 	& $R$	        	& 15.76$\pm$0.03	\\
20150613.869  	& 0.9m OSN 	& $I$		        	& 15.68$\pm$0.03	\\
20150614.002	& UVOT/Swift	& $UVW2$	& 20.55$\pm$0.16	\\
20150614.002	& UVOT/Swift	& $UVM2$		& 21.49$\pm$0.36	\\
20150614.002	& UVOT/Swift	& $UVW1$	& 19.29$\pm$0.10	\\
20150614.002	& UVOT/Swift	& $U$		& 17.45$\pm$0.06	\\
20150614.002	& UVOT/Swift	& $B$		& 16.48$\pm$0.05	\\
20150614.002	& UVOT/Swift	& $V$		& 16.19$\pm$0.07	\\
20150614.882	& 10.4m GTC		& $r'$		& 15.92$\pm$0.02	\\
20150616.129	& UVOT/Swift	& $UVW2$	& 21.52$\pm$0.34	\\
20150616.129	& UVOT/Swift	& $UVW1$	& 19.88$\pm$0.16	\\
20150616.129	& UVOT/Swift	& $U$		& 17.85$\pm$0.11	\\
20150616.129	& UVOT/Swift	& $B$		& 16.67$\pm$0.08	\\
20150616.129	& UVOT/Swift	& $V$		& 16.33$\pm$0.10	\\
\end{tabular}
\end{table}

\begin{table}
\caption{Log of the photometric observations - continued}
\begin{tabular}{c c c c}
\label{Tab:photlog}\\
\hline\hline
Date 		& Telescope & Band & Magnitude \\
\hline\hline
20150617.897	& 10.4m GTC		& $r'$		& 16.02$\pm$0.02	\\
20150618.870  	& 0.9m OSN 	& $R$	        	& 16.00$\pm$0.02	\\
20150613.867  	& 0.9m OSN 	& $V$	        	& 15.92$\pm$0.04	\\
20150619.883	& 10.4m GTC		& $r'$		& 16.09$\pm$0.02	\\
20150620.865  	& 0.9m OSN 	& $V$	        	& 16.16$\pm$0.03	\\
20150620.867  	& 0.9m OSN 	& $R$	        	& 16.04$\pm$0.03	\\
20150620.872  	& 0.9m OSN 	& $B$	        	& 16.84$\pm$0.04	\\
20150620.870  	& 0.9m OSN 	& $I$		        	& 15.92$\pm$0.04	\\
20150620.928  	& 0.5m UoL	& $R$	        	& 16.07$\pm$0.12	\\
20150925.180 	& 1.5m OSN	& $R$		& 19.63$\pm$0.12	\\
20150925.193	& 1.5m OSN	& $B$		& 21.68$\pm$0.29	\\
20151008.161	& 1.5m OSN	& $R$		& 19.54$\pm$0.13	\\
20151008.159	& 1.5m OSN	& $B$		& 21.38$\pm$0.24	\\
20151008.183 	& 1.5m OSN	& $I$			& 19.17$\pm$0.21	\\
20151016.164	& 1.5m OSN	& $R$		& 19.84$\pm$0.21	\\
20151030.189 	& 1.5m OSN	& $R$		& 19.58$\pm$0.09	\\
20151030.198	& 1.5m OSN	& $I$			& 19.20$\pm$0.12	\\
20151101.188   	& 1.5m OSN	& $V$		& 20.46$\pm$0.21	\\
20151101.195   	& 1.5m OSN	& $R$		& 19.65$\pm$0.13	\\
20151101.198 	& 1.5m OSN	& $I$			& 19.13$\pm$0.15	\\
20151111.237   	& 1.5m OSN	& $R$		& 19.64$\pm$0.11	\\
20151118.211   	& 1.5m OSN	& $R$		& 19.72$\pm$0.10 	\\
20151118.214   	& 1.5m OSN	& $I$			& 19.31$\pm$0.12 	\\
20151118.230  	& 1.5m OSN	& $V$		& 20.59$\pm$0.15 	\\
20151125.191   	& 1.5m OSN	& $R$		& 19.79$\pm$0.15 	\\
20151125.200   	& 1.5m OSN	& $I$			& 19.58$\pm$0.25 	\\%
20151127.167   	& 1.5m OSN	& $R$		& 19.72$\pm$0.13 	\\
20151127.175 	& 1.5m OSN	& $I$			& 19.19$\pm$0.16 	\\
20151126.221		&PANIC/CAHA	&$J$		&18.44$\pm$0.17	\\
20151128.175   	& 1.5m OSN	& $R$		& 19.61$\pm$0.11 	\\
20151128.185   	& 1.5m OSN	& $I$			& 19.51$\pm$0.17 	\\
20160118.092  	& 0.9m OSN	& $V$		& 20.97$\pm$0.26 	\\
20160118.107   	& 0.9m OSN	& $R$		& 19.87$\pm$0.15 	\\
20160118.123   	& 0.9m OSN	& $I$			& 19.46$\pm$0.36 	\\
20160113.212   	& 0.9m OSN	& $R$		& 19.99$\pm$0.35 	\\
20160121.179	& 10.4m GTC	& $g'$		& 21.41$\pm$0.10	\\
20160121.182	& 10.4m GTC	& $r'$		& 20.22$\pm$0.10	\\
20160121.184	& 10.4m GTC	& $i'$		& 20.57$\pm$0.05	\\
\end{tabular}
\end{table}


  \begin{figure*}
   \centering
   \includegraphics[width=17cm]{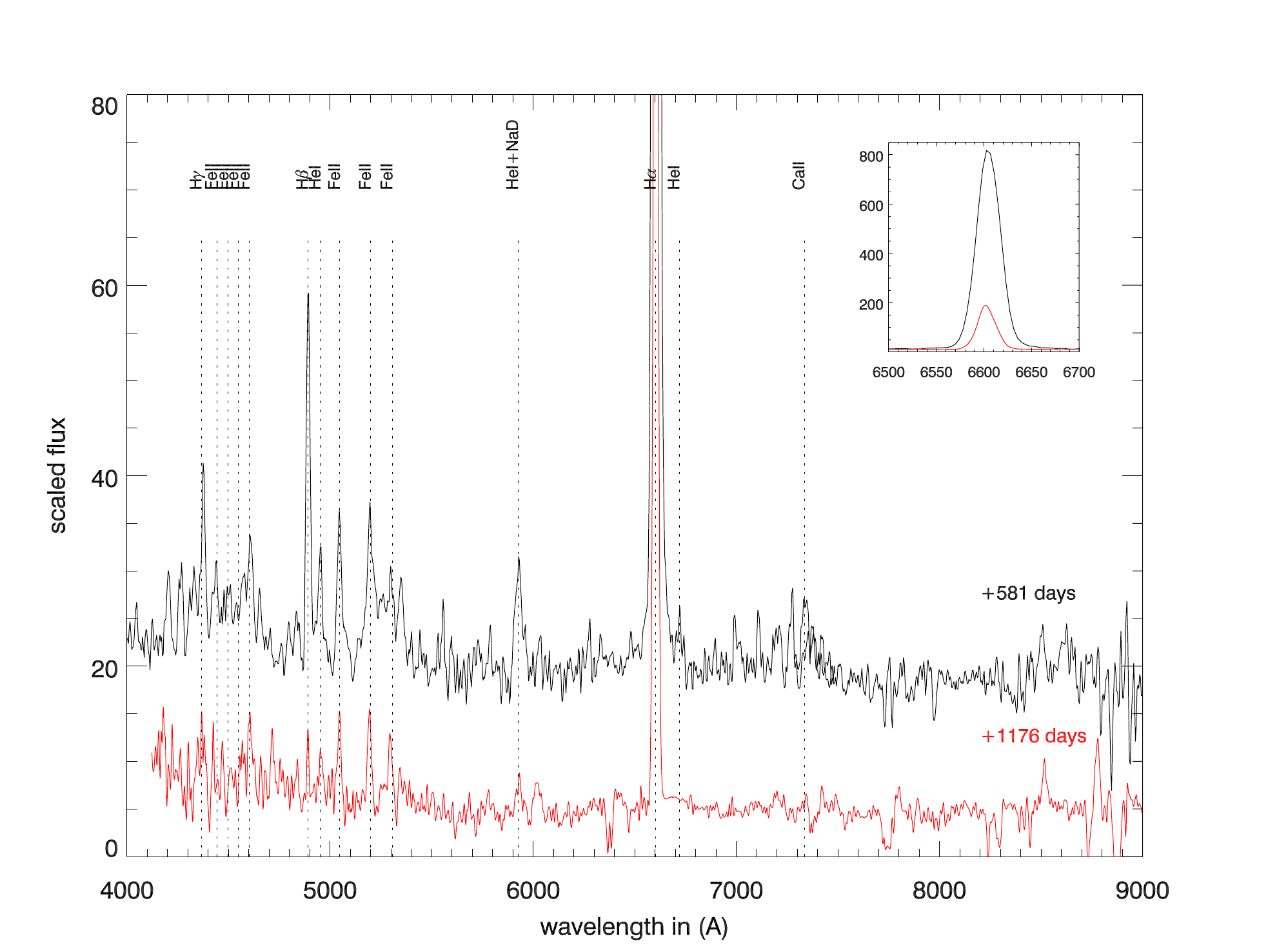}
      \caption{Late time spectra of SN 2009ip: At +581d (May 6, 2014) taken with EFOSC at the NTT (PESSTO data release, black) and at +1176d (Dec. 23, 2015) taken with Magellan-Clay (red). The inset shows the H$\alpha$ line at the two epochs.}
         \label{Fig:2009ip}
   \end{figure*}

  \begin{figure*}
   \centering
   \includegraphics[width=17cm]{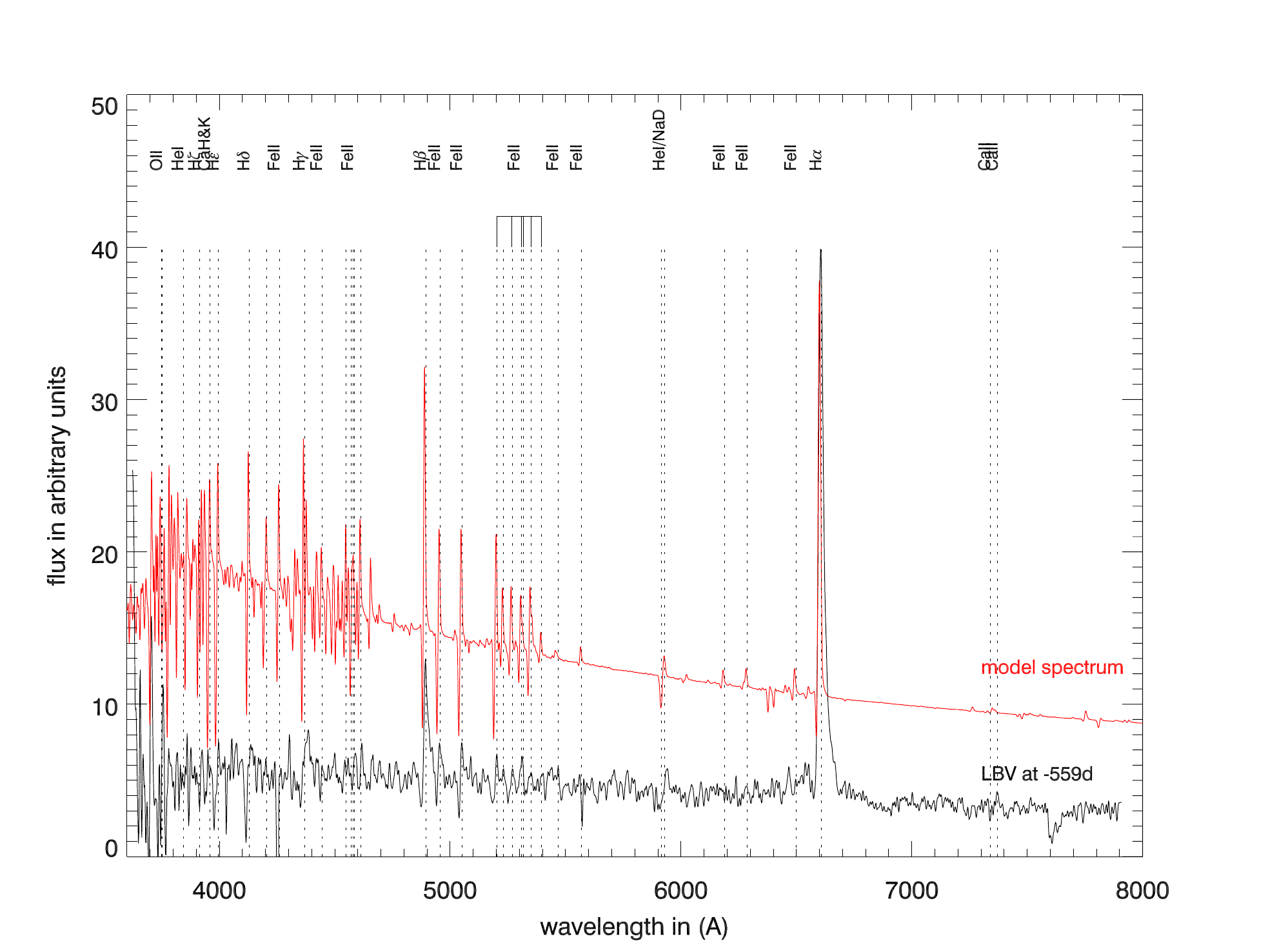}
      \caption{Comparsion between our Nov. 11, 2013 spectra at the onset of episode 2013A and a model spectrum of an LBV in outburst with L$=$1$\times$10$^{6}$ L$_\odot$, T$_\mathrm{eff}$=8500\,K, a mass loss rate of $\dot{M}=$5$\times$10$^{-4}$M$_\odot$\,y$^{-1}$ and a wind terminal speed of 600 km\,s$^{-1}$. The lines present are mainly Balmer lines, NaD, CaII and FeII.}
         \label{Fig:specsmodel}
   \end{figure*}

\bsp	
\label{lastpage}
\end{document}